\documentclass[aps,prx,reprint,twocolumn,floatfix,superscriptaddress,longbibliography]{revtex4-2}
\usepackage[english]{babel}
\usepackage{standalone}
\usepackage{amsthm}
\usepackage{amsfonts, amsmath, amssymb}
\usepackage{graphicx}
\usepackage{dcolumn}
\usepackage{bm}
\usepackage{dsfont}
\usepackage[utf8]{inputenc}
\usepackage[colorlinks,citecolor=blue,linkcolor=blue,urlcolor=blue]{hyperref}
\usepackage{physics}
\usepackage{soul}
\usepackage{mathtools}
\usepackage{float}
\usepackage{stmaryrd} 
\usepackage{enumerate}
\newcommand{\Ecal}{\mathcal{E}}
\newcommand{\Vcal}{\mathcal{V}}

\newcommand{\bea}{\begin{equation}\begin{aligned}}
\newcommand{\eea}{\end{aligned}\end{equation}}

\newcommand{\bean}{\begin{equation*}\begin{aligned}}
\newcommand{\eean}{\end{aligned}\end{equation*}}

\newcommand{\be}{\begin{equation}}
\newcommand{\ee}{\end{equation}}
\newcommand{\nqubits}{n}

\DeclarePairedDelimiterX{\mynorm}[1]{\lVert}{\rVert}{#1}
\renewcommand{\norm}[1]{\mynorm[\small]{#1}} % The original \norm is too ugly. Let us redefine one whose height can be fixed by hand once for all. \small can be replaced with for example \big or \bigg to give other sizes. 

\newcommand{\expectt}[2]{\mathbb{E}_{#1}\left[ #2\right]}
\renewcommand{\(}{\left(}
\renewcommand{\)}{\right)}
\renewcommand{\[}{\left[}
\renewcommand{\]}{\right]}
\def\rme{{\rm {e}}}
\def\rmi{{\mathrm {i}}}

\renewcommand{\tr}[1]{{\rm{Tr}}\left[#1\right]}
\newcommand{\U}{\hat{U}}
\newcommand{\V}{\hat{V}}
\newcommand{\W}{\hat{W}}
\renewcommand{\P}{\hat{P}}
\renewcommand{\O}{\hat{O}}
\newcommand{\Ucal}{\mathcal{U}}
\newcommand{\Wcal}{\mathcal{W}}
\newcommand{\Ud}{\hat{U}^{\dag}}
\newcommand{\Vd}{\hat{V}^{\dag}}

\newcommand{\thetav}{\vb*{\theta}}
\newcommand{\variance}[2]{\mathrm{Var}_{#1}\left[#2\right]}
\newcommand{\multicomp}{\mathop{\bigcirc}}

\newcommand{\rhohat}{\hat{\rho}}

\newcommand{\id}{\mathds{1}}

\newcommand{\Shat}{\hat{S}}
\newcommand{\Zhat}{\hat{Z}}
\newcommand{\Xhat}{\hat{X}}

\newcommand{\CN}{\widehat{CNOT}}

\newcommand{\CNx}{\CN_X}

\newcommand{\ptil}{\tilde{p}}
\newcommand{\sgn}{\mathop{\mathrm{sgn}}}

\newtheorem{theorem}{Theorem}[section]

\theoremstyle{definition}
\newtheorem{definition}{Definition}[section]
\theoremstyle{remark}
\newtheorem*{remark}{Remark}
  {\left\lbrace\begin{array}{@{}l@{}}}%
  {\end{array}\right.}

\newcommand{\Xtil}{\tilde{X}}
\newcommand{\vbXtil}{\tilde{\vb*{X}}}
\newcommand{\Util}{\tilde{\Ucal}}
\newcommand{\ftil}{\tilde{f}}

\begin{document}

\newcommand{\upcite}{\affiliation{Universit\'{e} Paris Cit\'{e}, CNRS,  Mat\'{e}riaux et Ph\'{e}nom\`{e}nes Quantiques (MPQ), F-75013 Paris, France}}

\newcommand{\ictp}{\affiliation{The Abdus Salam International Center for Theoretical Physics (ICTP), Strada Costiera 11, I-34151 Trieste, Italy}}

\title{Efficient estimation of trainability for variational quantum circuits}

\author{Valentin Heyraud}
\upcite
\author{Zejian Li}
\upcite
\ictp
\author{Kaelan Donatella}
\upcite
\author{Alexandre Le Boit\'{e}}
\upcite
\author{Cristiano Ciuti}
\upcite

\begin{abstract}
Parameterized quantum circuits used as variational ansätze are emerging as promising tools to tackle complex problems ranging from quantum chemistry to combinatorial optimization. These variational quantum circuits can suffer from the well-known curse of barren plateaus, which is characterized by an exponential vanishing of the cost-function gradient with the system size, making training unfeasible for practical applications. Since a generic quantum circuit cannot be simulated efficiently, the determination of its trainability is an important problem. Here we find an efficient method to compute the gradient of the cost function and its variance for a wide class of variational quantum circuits. Our scheme relies on our proof of an exact mapping from randomly initialized circuits to a set of Clifford circuits that can be efficiently simulated on a classical computer by virtue of the celebrated Gottesmann-Knill theorem. This method is scalable and can be used to certify trainability for variational quantum circuits and explore design strategies that can overcome the barren plateau problem. As illustrative examples, we show results with up to 100 qubits.
\end{abstract}

\date{\today}
\maketitle

\section{Introduction}
\label{section:introduction}

Inspired by the success of machine-learning methods, variational quantum algorithms~\cite{cerezo2021d, carleo2019, cerezo2022a} have emerged as a promising way to harness the power of quantum computing in various domains ranging from quantum chemistry \cite{peruzzo2014, kandala2017, googleaiquantumandcollaborators2020a} to combinatorial optimization problems \cite{farhi2014, lacroix2020, harrigan2021}. These algorithms use the output of parameterized quantum circuits as variational ansätze, whose parameters are classically optimized through gradient-based methods. 

Variational quantum circuits can suffer from trainability issues caused by the existence of barren plateaus \cite{mcclean2018}, a limitation that has been extensively studied in the recent literature \cite{holmes2022a, wang2021c, ortizmarrero2021a, uvarov2021a, cerezo2021c, patti2021a, wiersema2021, kim2022, kim2022a, sack2022, friedrich2022, grant2019a, liu2022, mitarai2022a, ravi2022, kim2021, kim2022b, cheng2022, dborin2022, pesah2021, schatzki2022, holmes2021a, arrasmith2021a, arrasmith2022,  wang2021d, du2022a, sharma2022, depalma2023}. It is characterized by an exponential vanishing of the cost function's gradient with the system size that makes training variational quantum circuits impossible for a large number of qubits. Barren plateaus can originate from various and fundamentally different phenomena. Their emergence was first shown in Ref.~\cite{mcclean2018} for 2-designs (random unitary transformation matching the Haar distribution up to the second moment). Recent works linked barren plateaus to the expressibility of the ansatz \cite{holmes2022a} as well as noise \cite{wang2021c} and entanglement. In particular, the authors of Ref.~\cite{ortizmarrero2021a} showed that for architectures that can be split into a hidden and a visible subsystem, such as quantum Boltzmann machines or feed-forward quantum neural networks, an excess of entanglement between the two subsystems would result in a highly mixed state for the visible subsystem. This can lead to a flat landscape for the cost function. The effect of the structure of the cost function on the appearance of barren plateaus was also investigated in other works \cite{uvarov2021a, cerezo2021c}, and it was shown that global cost functions are more prone to exhibit barren plateaus. Note that shallow models such as quantum kernel machines \cite{heyraud2022, jerbi2023, li2022, schuld2021a, rebentrost2014a} and reservoir computing models \cite{mujal2021, denis2022a, marcucci2020, pierangeli2021}, while often easier to train than variational quantum algorithms, might also suffer from trainability issues of a similar nature \cite{thanasilp2022}.

Numerous investigations have proposed strategies to address the barren-plateau issue. In the context of entanglement-induced barren plateaus, most strategies rely on limiting the amount of entanglement~\cite{patti2021a, wiersema2021, kim2022, kim2022a, sack2022}. Other methods make use of tailored distributions of the initial circuits parameters and carefully designed circuits architectures~\cite{friedrich2022, grant2019a, liu2022,  dborin2022, mitarai2022a,ravi2022, cheng2022, kim2021, kim2022b}. Yet, only a handful of configurations offer trainability guarantees and robustness against barren plateaus \cite{pesah2021, schatzki2022}. Tackling this fundamental issue remains an important theoretical challenge. 

In this paper, we propose an alternative approach to the problem by providing an efficient method to estimate the average gradients and their variance  for a wide class of variational quantum algorithms. By studying the quantum channel associated to a random single-qubit rotation, we prove that under some simple conditions the first and second moments can be expressed as mixed-unitary channels \cite{watrous2018} composed of Clifford gates \cite{nielsen2010}. Upon some additional general assumptions for the random angles distribution, we demonstrate that this allows us to exactly map randomly initialized circuits composed of Clifford gates and parametrized rotations to an ensemble of Clifford circuits. Moreover, we prove that the obtained ensemble can be efficiently sampled to compute quantities of interest, such as the variance of the gradient or the average of the cost function over the initial random parameters. Making use of the celebrated Gottesman-Knill theorem \cite{gottesman1998,aaronson2004}, we analytically prove the efficiency of our method that can be implemented on a classical computer with a complexity scaling polynomially in both the number of variational parameters and the system size. In addition, we show some numerical experiments to illustrate our method on examples of random circuits and faithfully reproduce the exponential suppression of the variance first found in Ref.~\cite{mcclean2018} with polynomial resources and for circuits acting on up to 100 qubits. 

\begin{figure*}[ht]
    \includegraphics[width=\textwidth]{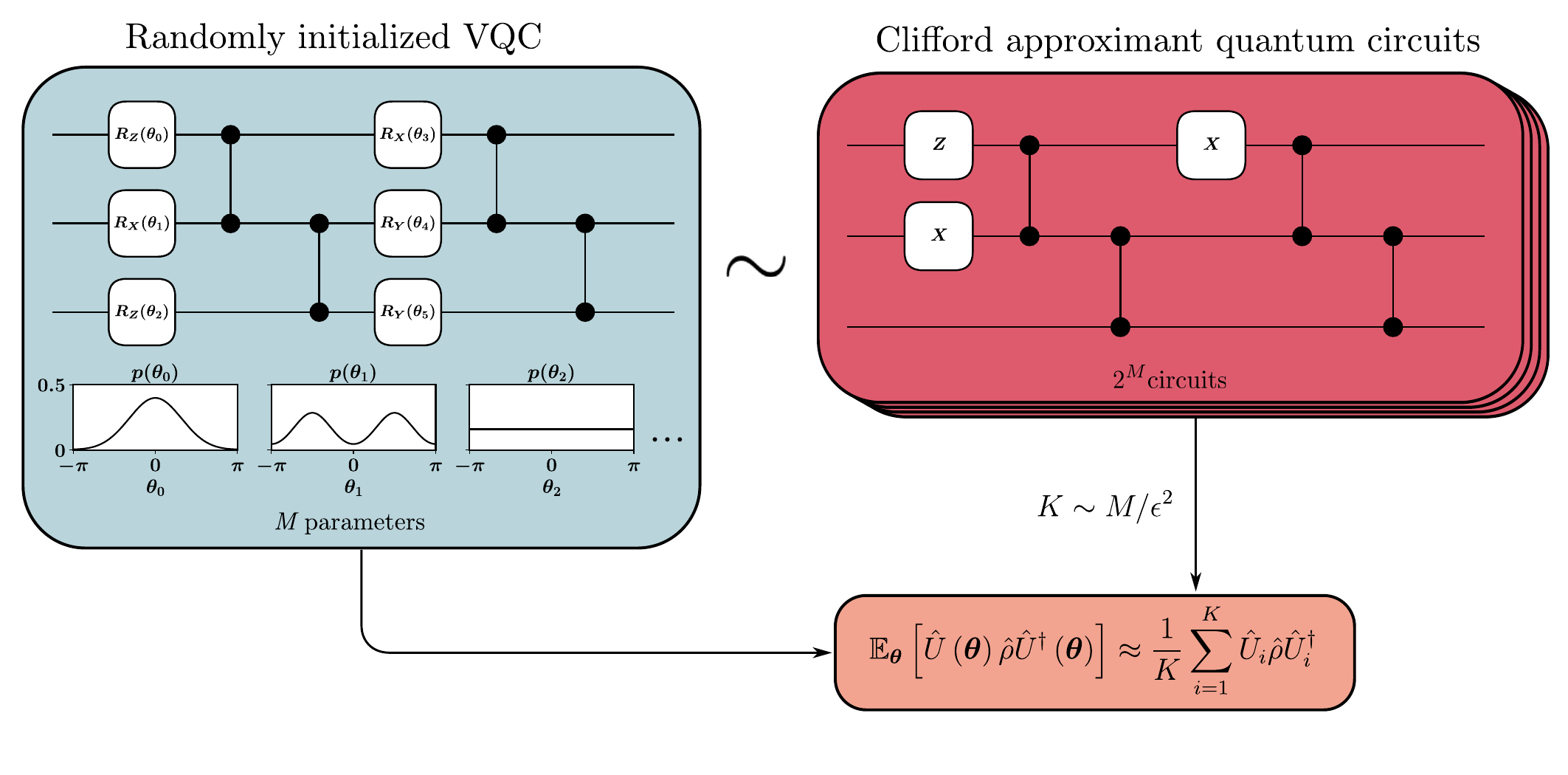}
    
    \caption{A schematic representation of the mapping from a parameterized quantum circuit with random parameters to Clifford approximant circuits for first-order quantities (quantities that requires only the knowledge of the  rotation $1$-fold channels to be computed, see App. \ref{appendix:first-and-second-order-quantities}). For a circuit with $M$ parameters, a sample size of the order of $M/\epsilon^2$ is enough to get an approximation of the average on the initial circuit with a precision $\epsilon$ on the observables mean values (see App. \ref{appendix:sampling_efficiency}).}
    \label{fig:vqc_to_clifford_mapping}
    
\end{figure*}

\section{Theoretical framework}
\label{section:theoretical_framework}

\subsection{Variational problem}
\label{section:variational_problem}
In variational quantum algorithms, a parameterized unitary transformation $\U(\thetav)$ acting on $\nqubits$ qubits is used as a variational ansatz to achieve a task expressed as the minimization of a cost function 
\bea
C(\thetav) = \tr{\U(\thetav)\rhohat\Ud(\thetav)\O}
\label{eq:cost_function}
\eea
for some observable $\O$ and some initial $n$-qubit state $\rhohat$. This formulation is general and encompasses typical tasks, such as the preparation of a target state $\ket{\psi}$ (setting $\O=-\dyad{\psi}$) or ground state search for some Hamiltonian $\hat{\mathcal{H}}$ (setting $\O=\hat{\mathcal{H}}$). The considered parameterized unitaries are typically composed of a succession of parameterized gates and fixed layers. Here we consider a generic ansatz of the form
\bea
\U (\thetav) = \prod_{i=1}^{M}\U_{i}(\theta_i)\W_i,
\label{eq:layered_ansatz}
\eea
where each unitary $\U_i(\theta_i) = e^{-\mathrm{i}\frac{\theta_i}{2}\P_i}$ is a single qubit rotation associated to a given Pauli operator $\P_i\in\{\hat{X},\hat{Y},\hat{Z}\}$, while the $\W_k$ are fixed layers composed of a sequence of unparameterized gates that can act on multiple qubits. Upon absorbing Clifford gates in the fixed layers, we will consider the rather general class of circuits based on $Z$ rotations \footnote{We denote $\hat{H}$ the Hadamard gate and $\hat{S}$ the phase gate, which both belong to the Clifford group. For $X$-rotations we have that $\hat{X} = \hat{H}^{\dag}\hat{Z}\hat{H}$ and hence $e^{-\mathrm{i}\frac{\theta_i}{2}\hat{X}} = \hat{H}^{\dag}e^{-\mathrm{i}\frac{\theta_i}{2}\hat{Z}}\hat{H}$ and we can replace $\W_i$ and $\W_{i+1}$ respectively by $\hat{H}\W_i$ and $\W_{i+1}\hat{H}$ to get another ansatz with the same form as the original one and with only $Y$ and $Z$ rotations. We proceed likewise for $Y$-rotations using the fact that $\hat{Y} = (\hat{S}\hat{H})\hat{Z}(\hat{S}\hat{H})^{\dag}$. Note that in the case of the last layer one of the extra gates must be absorbed in the cost function observable to get the same ansatz structure.}. The unitary transformation $\U (\thetav)$ depends on $M$ continuous parameters gathered in the vector $\thetav = \(\theta_0, \dots, \theta_{M-1}\)$. These rotation parameters can be optimized using classical gradient-descent techniques. The gradient of the cost function with respect to the $k$-th parameter can be conveniently estimated using the parameter-shift rule \cite{mitarai2018, schuld2019a}:
\bea
\partial_k C(\thetav)=\frac{1}{2}\(C(\thetav+\frac{\pi}{2}\vb*{e}_k)-C(\thetav-\frac{\pi}{2}\vb*{e}_k)\),
\label{eq:parameter_shift_rule}
\eea
where $\vb*{e}_k$ is the canonical vector along the component $k$. It is worth noticing that the $\pm\pi/2$ shifts in the parameter $\theta_k$ can be factored out and seen as an extra Clifford gate added to the fixed layer $\W_k$. In fact, remarking that the phase gate $\hat{S}$ can be written $\hat{S} = e^{\mathrm{i}\frac{\pi}{4}}e^{-\mathrm{i}\frac{\pi}{2}\hat{Z}}$ and assuming $\hat{P}_k = \hat{Z}$, we have:
\begin{equation}
\begin{aligned}
	\U_k(\theta_k + \pi/2)\W_k &= e^{-\mathrm{i}\frac{\theta_k}{2}\hat{Z}}e^{-\mathrm{i}\frac{\pi}{2}\hat{Z}}\W_k\\ &= e^{-\mathrm{i}\frac{\pi}{4}} \U_k(\theta_k)\hat{S}\W_k \,.
\end{aligned}
\end{equation}
We define $\W_{k,\pm} = e^{\mp\mathrm{i}\frac{\pi}{4}}\hat{S}\W_k$ such that we can write
\begin{equation}
	\U_k(\theta_k \pm \pi/2)\W_k = \U_k(\theta_k)\W_{k,\pm}\,.
\end{equation}
We denote
\begin{equation}
\begin{aligned}
\U_{\pm}(\thetav) = \U\(\thetav \pm \frac{\pi}{2}\vb*{e}_k\)
\label{eq:shifted_ansatz}
\end{aligned}
\end{equation}
the shifted unitaries appearing in the parameter-shift-rule. From what precedes we have 
\begin{equation}
\U_{\pm}(\thetav) = \prod_{i=1}^{M}\U_{i}(\theta_i)\V_{i,\pm}\,,
\end{equation}
with
\begin{equation}
\V_{i,\pm} = \begin{dcases*}
	\,\W_{k,\pm} & if $i=k$ \,,\\
	\,\W_i & otherwise \,.
\end{dcases*}
\label{chap_3:eq:shifted-unitaries-decomposition-layers}
\end{equation}

\subsection{Unitary ensembles and \texorpdfstring{$\boldsymbol{t}$}{t}-fold channels}

To start the optimization process the rotation angles are randomly initialized according to some probability distribution $ p(\thetav)$. The initialized circuit can then be represented by a unitary ensemble $\mathbb{U} = \{\U, \mathbb{P}(\U)\}$, where $\mathbb{P}$ is a probability measure on $\mathbb{U}$. One is often interested in computing averages of quantities that are polynomial of a given order $t$ in the entries of $\U$. Such quantities can be completely determined by the knowledge of the $t$-fold channel \cite{roberts2017a}
\bea
\Phi^{(t)}_{\mathbb{U}}\(\rhohat\) = \int_{\mathbb{U}}\U^{\otimes t}\rhohat\U^{\dagger \otimes t}\mathrm{d}\mathbb{P}(\U),
\label{eq:$t$-fold-channel}
\eea
where $\rhohat$ is an initial state of $t$ copies of the original $\nqubits$-qubit system. As an example, the expected value of the square of the cost function at the initialization can be evaluated using  $\expectt{\thetav}{C(\thetav)^2} = \tr{\Phi_{\thetav}^{(2)}(\rhohat^{\otimes 2})\O^{\otimes 2}}$, where we denote $\Phi_{\thetav}^{(t)}$ the $t$-fold channel associated to the unitary ensemble $\{\U(\thetav), \; \mathbb{R}^{M} \ni \thetav  \sim p(\thetav)\}$. In App.~\ref{appendix:first-and-second-order-quantities} we define \emph{second-order quantities} (respectively \emph{first-order quantities}) as quantities that can be obtained from the knowledge of the $2$-fold (respectively $1$-fold) channel. To give a concrete example, $\expectt{\thetav}{C(\thetav)^2}$ is a second-order quantity. 

More generally, one can characterize the expressivity of a given ansatz by comparing its $t$-fold channels to the ones obtained for a Haar (uniform) distribution over the whole unitary group \cite{sim2019a, nakaji2021b, holmes2022a}. Unitary ensembles whose $t$-fold channels match the $t$-fold channels for the Haar measure, the so-called $t$-designs \cite{gross2007, iosue2022a}, have played a crucial role in the original discovery of the barren plateaus phenomenon \cite{mcclean2018}. Moreover, in multiple cases random quantum circuits are approximate $t$-designs \cite{harrow2009, brandao2016, haferkamp2022}. For instance, the authors of 
\cite{harrow2009} showed that quantum circuit composed of a polynomial number of gates randomly
drawn from a universal set of two-qubit gates and applied to random pairs of qubits
are approximate 2-designs. This result have been extended to cases where the gates
are applied to nearest-neighbor qubits in \cite{brandao2016} and \cite{haferkamp2022}.

\subsection{Barren Plateaus}

For a unitary ensemble $\mathbb{U}$ that describes parameterized ansätze $\U(\thetav)$ with random continuous parameters $\thetav$ and a possibly random architecture, a cost function $C(\U(\thetav))$ is said to exhibit a barren plateau if the probability of obtaining a gradient that deviates from zero by some $\epsilon>0$ vanishes exponentially with the system size $\nqubits$. More precisely, $\mathbb{P}_{\mathbb{U}}(\abs{\partial_k C}>\epsilon)\leq \mathcal{O}(\exp(-\alpha\nqubits))$ for some $\alpha>0$ \cite{holmes2022a}. As mentionned earlier, barren plateaus were first found for unitary ensembles forming 2-designs \cite{mcclean2018} and connections to expressivity \cite{holmes2022a}, noise \cite{wang2021c}, entanglement \cite{ortizmarrero2021a, patti2021a, wiersema2021, kim2022, kim2022a, sack2022} and the degree of locality of the cost function \cite{uvarov2021a, cerezo2021c} were later discovered. In many cases, the average value of the gradient vanishes exactly, for instance when the rotation parameters are initialized uniformly in $\[-\pi, \pi\]$. However, this does not imply a vanishing of the gradient amplitude on average, and thus does not tell much about the trainability of the model. In this unbiased case, the variance is a relevant quantity. Due to the Chebyshev inequality, one has $\mathbb{P}_{\mathbb{U}}(\abs{\partial_k C}>\epsilon)\leq \variance{}{\partial_k C}/\epsilon^2$, so that a vanishing variance implies the existence of a barren plateau. On the other hand, a non-vanishing variance guarantees large fluctuations of the initial gradient and thus a good initial trainability, independently of the gradient bias. 

For variational quantum algorithms, the gradient is to be estimated through measurements
realised on an hardware platform. As argued in Ref.~\cite{mcclean2016}, probing an
exponentially small gradient requires an exponential precision on the measurements
and thus an exponential number of experiments, which is prohibitive. Barren
plateaus are often seen as a quantum version of the vanishing gradient phenomenon
in classical machine learning. The impact of vanishing gradients on the trainability of
classical deep neural network is discussed in Ref.~\cite{goodfellow2016}. The exact nature of barren plateaus was investigated further by the authors of Ref.~\cite{liu2022d} in
a quantum machine learning context, using a least square loss function and relying
on a neural tangent kernel formalism. In particular, the authors discussed the fundamental
differences between barren plateaus and classical vanishing gradient, and they
show that in a certain over-parameterized regime the training procedure might be
robust to noise. Let us notice that variational quantum algorithms can suffer from other
issues beyond barren plateaus, related for instance to the number of local minima of
the loss landscape \cite{anschuetz2022a} or
to the complexity associated with the classical optimization procedure \cite{bittel2021}. 
Also, higher-order moments may help diagnose the trainability of variational quantum algorithms~\cite{cerezo2021f}.

\section{Results and discussion}
\label{section:result_and_discussion}

The main finding of this theoretical work is that under some rather general assumptions on the distribution of the rotation parameter $\theta$, it is possible to map the $1$-fold and $2$-fold channels of a random rotation $\hat{R}_Z(\theta)$ to a finite unitary ensemble of Clifford gates. Moreover, we prove that such mapping allows us to estimate quantities of interest such as the gradient variance using only Clifford circuits. Finally, we illustrate our rigorous proofs through numerical experiments. The detailed mathematical proofs are presented in the Appendix. 

\begin{figure*}[htb!]
    \includegraphics[width=\textwidth]{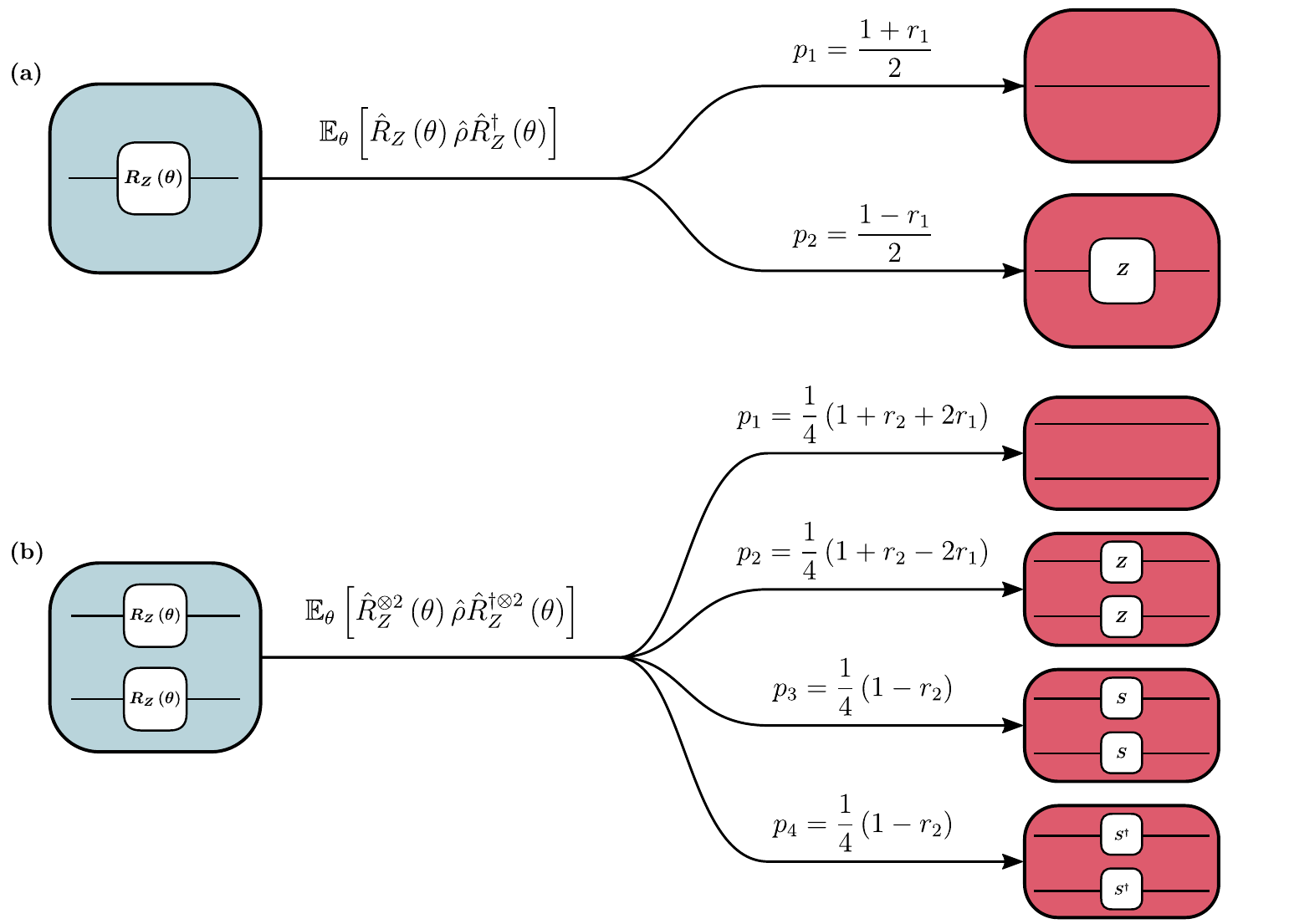}
    \caption{Schematic representation of the mapping rules from $Z$-rotations with a random parameter to unitary ensembles composed of Clifford gates. Panels (a) and (b) respectively are for first and second-order averages. The mapping here is for probability distributions that are even with respect to $\theta$: we denote $r_1 = \expectt{\theta}{\mathrm{cos}(\theta)}$ and $r_2 = \expectt{\theta}{\mathrm{cos}(2\theta)}$. The coefficients $p_i$ are the probabilities dictating how the corresponding Clifford circuits are sampled.}
    \label{fig:vqc_to_clifford_mapping_rules}
\end{figure*}

\subsection{Exact mapping and efficient sampling}

As mentioned earlier, we will focus on the class of variational quantum circuits composed of fixed Clifford gates alternated with single qubit parameterized rotations along the $X,Y$ or $Z$ directions, such as the one depicted in Fig.~\ref{fig:vqc_to_clifford_mapping}. As explained in Sec.~\ref{section:variational_problem}, we will restrict our study to rotations along $Z$, as we can obtain the cases of rotations along $Y$ and $X$ by adding extra Clifford gates to the different fixed layers of the considered ansatz. Let us consider a rotation along the $Z$-axis with a distribution that is symmetric about the $\theta = 0$ angle \footnote{This encompasses distributions that are symmetric about the angle $k\pi/2$ for $k\in\mathbb{Z}$. In this case the bias can be factored out in the form of an extra fixed Clifford gate.}. We show in App.~\ref{appendix:k-fold_channel_rz} that the $1$-fold channel corresponding to a first order average can be written as a convex sum of the unitary channels associated to the identity and the Pauli $Z$ gates, as schematically represented on the upper part of Fig.~\ref{fig:vqc_to_clifford_mapping_rules}. Note that this result has been derived and used in \cite{zhao2021} in the case of a uniform probability distribution for $\theta$ in order to analyze a variational ansatz through the lens of $ZX$-calculus. 

To compute the $1$-fold channel for the randomly initialized ansatz of the form given in Eq.~\eqref{eq:layered_ansatz} with independent rotation parameters, one can simply compose the $1$-fold channels associated to each rotation, interwined with the unitary channels associated to the fixed gates $\hat{W}_k$. We find that the $1$-fold channel of the ansatz is a convex sum of $2^M$ Clifford unitary channels, where $M$ is the number of rotations. One can view this convex sum as an average over a finite ensemble of Clifford approximant circuits. Examples of such circuits are provided in App.~\ref{appendix:example_of_Clifford_approximants} for a simple architecture similar to the one in Fig.~\ref{fig:vqc_to_clifford_mapping}. Although the number of Clifford approximant circuits in this ensemble is exponential in the number of parameters, we show in App.~\ref{appendix:sampling_proof} that a number of samples
polynomial in $M/\epsilon^2$ is sufficient to approximate the average of an observable expectation value (or more generally of any first-order quantity) to any desired precision $\epsilon$. This result relies on a classical concentration argument, and is schematically represented in Fig.~\ref{fig:vqc_to_clifford_mapping} for a simple circuit at the first order. From this, one can estimate the expectation value of the gradient, as it suffices to replace $\U(\thetav)$ by $\U_{\pm}(\thetav)$ (as defined in Sec.~\ref{section:variational_problem}) in the $1$-fold channel definition to obtain the expectation of $C(\thetav\pm\pi/2)$. This gives the expectation of the gradient thanks to the parameter-shift rule.

We prove in App.~\ref{appendix:k-fold_channel_rz} that the $2$-fold channel associated to a random $Z$-rotation is also a linear combination of Clifford channels, provided that the probability distribution is an even function of $\theta$. This result is depicted in Fig.~\ref{fig:vqc_to_clifford_mapping_rules}(b). To obtain this mapping, we use the Choi representation of
quantum channels \cite{watrous2018}. The Choi operators representing unitary quantum
channels given by tensor products of Z-rotations gates are diagonal. Hence, one can
represent these channels by the diagonal coefficients of their associated Choi operators.
Using this representation and the linearity of the expectation, we obtain a tractable
representation of the average two-fold channel of a Z-rotation. The decomposition
is then derived by solving a linear system of equations obtained by identifying the
coefficients of the previous representation for the different channels involved. When the inequalities $\expectt{\theta}{f_{+}(\theta)}\geq0$ and $\expectt{\theta}{f_{-}(\theta)}\geq0$ with $f_{\pm}(\theta) = \cos{\theta}(\cos{\theta}\pm 1)$ are satisfied, the previous linear combination is in fact a convex sum. Equivalently, the $2$-fold channel of a $Z$-rotation with an even angular probability distribution is a Clifford mixed-unitary channel if
\bea
\expectt{\theta}{\cos^2{\theta}}\geq\abs{\expectt{\theta}{\cos{\theta}}}.
\label{eq:inequality_condition}
\eea

The zeros of $f_{+}$, $f_{-}$ are the angles $\{k\pi/2, k\in\mathbb{Z}\}$ for which $\hat{R}_Z(\theta)$ matches a Clifford gate (see App.~\ref{appendix:k-fold_channel_rz}). Indeed, if the distribution of $\theta$ is a convex sum of Dirac distributions at these angles, the average over $\theta$ becomes a discrete average over the corresponding Clifford unitaries. Hence, the associated $2$-fold channel is indeed a convex sum of Clifford channels. One can also verify that the previous conditions are satisfied for distributions that are both even with respect to the angle $\theta$ and $\pi$-periodic. For example, the uniform distribution is included. In the case of a centered Gaussian distribution, the previous conditions are satisfied if and only if the corresponding width is large enough. 

Provided the distributions of the rotation angles satisfy the conditions discussed above, the scheme can be extended to the second order, allowing to  approximate second-order quantities, such as the average of the squared cost function $\expectt{\thetav}{C(\thetav)^2}$, using a set of Clifford approximant circuits. By the parameter-shift rule, the expectation of the squared gradient can be calculated from the knowledge of four quantities of the form $\expectt{\thetav}{C(\thetav\pm(\pi/2)\vb*{e}_k)C(\thetav\pm(\pi/2)\vb*{e}_k)}$. The latter can be estimated with Clifford approximants by replacing the $\U^{\otimes 2}$ term in the definition of the $2$-fold channel by $\U_{\pm}\otimes\U_{\pm}$. Hence the scheme covers the estimation of the gradient variance. Note that at the second order, the approximant circuits are obtained by replacing the rotation $2$-fold channels by one of the four 2-qubit Clifford gates depicted on Fig.~\ref{fig:vqc_to_clifford_mapping_rules}, yielding an ensemble of $4^M$ possible Clifford circuits. As for first-order quantities, a number of samples scaling linearly in $M$ is enough to guarantee convergence. These rigorous results are summarized in the following theorem, whose detailed proof is shown in Appendices \ref{appendix:k-fold_channel_rz} and \ref{appendix:sampling_efficiency}.
\begin{theorem}
For a variational ansatz composed of fixed Clifford gates and of $M$ parameterized rotations along the X,Y or Z direction, if the random variational parameters $\(\theta_1,\dots,\theta_M\)$ are independent and symmetric with respect to one of the Clifford angles, i.e. $\in \{0,\frac{\pi}{2},\pi,\frac{3\pi}{2}\}$,
then for any error $\epsilon>0$ and a probability $1-\delta$ to meet such accuracy, any first-order quantity can be computed using 
$$ K \geq O\(\frac{M}{\epsilon^{2}}\log\(\frac{2}{\delta}\)\) $$
Clifford approximant circuits.
Moreover, if the distribution of $\theta_i$ satisfies the inequality
\begin{equation*}
\begin{aligned}
\expectt{\theta_i}{\cos^2(\theta_i-\expectt{\theta_i}{\theta_i})}&\geq\abs{\expectt{\theta_i}{\cos(\theta_i-\expectt{\theta_i}{\theta_i})}}\, ,\\
\forall i\in&\{1,\dots,M\}
\end{aligned}
\end{equation*}
then the same holds for any second-order quantity.
\label{theorem:main_theorem}
\end{theorem}
Finally one makes use of the Gottesman-Knill theorem, which states that for a Clifford unitary $\U$ and an observable $\O$ acting non-trivially on $N_O$ qubits, the expectation value $\tr{\dyad{0}^{\otimes n}\Ud\O\U}$ can be classically computed with a polynomial complexity in both $\nqubits$ and $N_{O}$. Our method inherits this complexity, and in particular we can classically estimate the gradient expectation and variance with a polynomial complexity in $\nqubits$, $N_{O}$ and $M$. 

In App.~\ref{appendix:sampling_efficiency_in_general_case_section} we extend the scheme presented above for the $2$-fold channels to the case where the distribution of the random angle does not satisfy the convex condition of Eq.~\eqref{eq:inequality_condition}. In that case, it is still possible to use Clifford approximant circuits to estimate second-order quantities, but this comes at the price of an exponential complexity in the number of variational parameters $M$. This result is based on a sampling scheme proposed by the authors of \cite{piveteauQuasiprobabilityDecompositionsReduced2022}. In that context, the method allows to trade an exponential complexity in the system size for an exponential complexity in the number of variational parameters $M$. 

In App.~\ref{appendix:N-fold_channel_generalization}, we also present the extension of our scheme to the general case of $N$-fold channels. We prove  that the $N$-fold channel associated to random Z-rotations can be decomposed as a real sum of Clifford unitary channels. From this decomposition, we derive a condition for the $N$-fold channel to be a convex sum of Clifford unitaries by imposing the coefficients of the combination to be positive. However we show that the obtained decomposition is not unique, so that the derived condition is sufficient but not necessary. Finding a sufficient and necessary condition on the distribution of a random angle that guarantees that the corresponding $N$-fold channels are Clifford mixed unitary channels remains an open problem. 

\subsection{Numerical simulations}

\begin{figure}
    \includegraphics[width=\columnwidth]{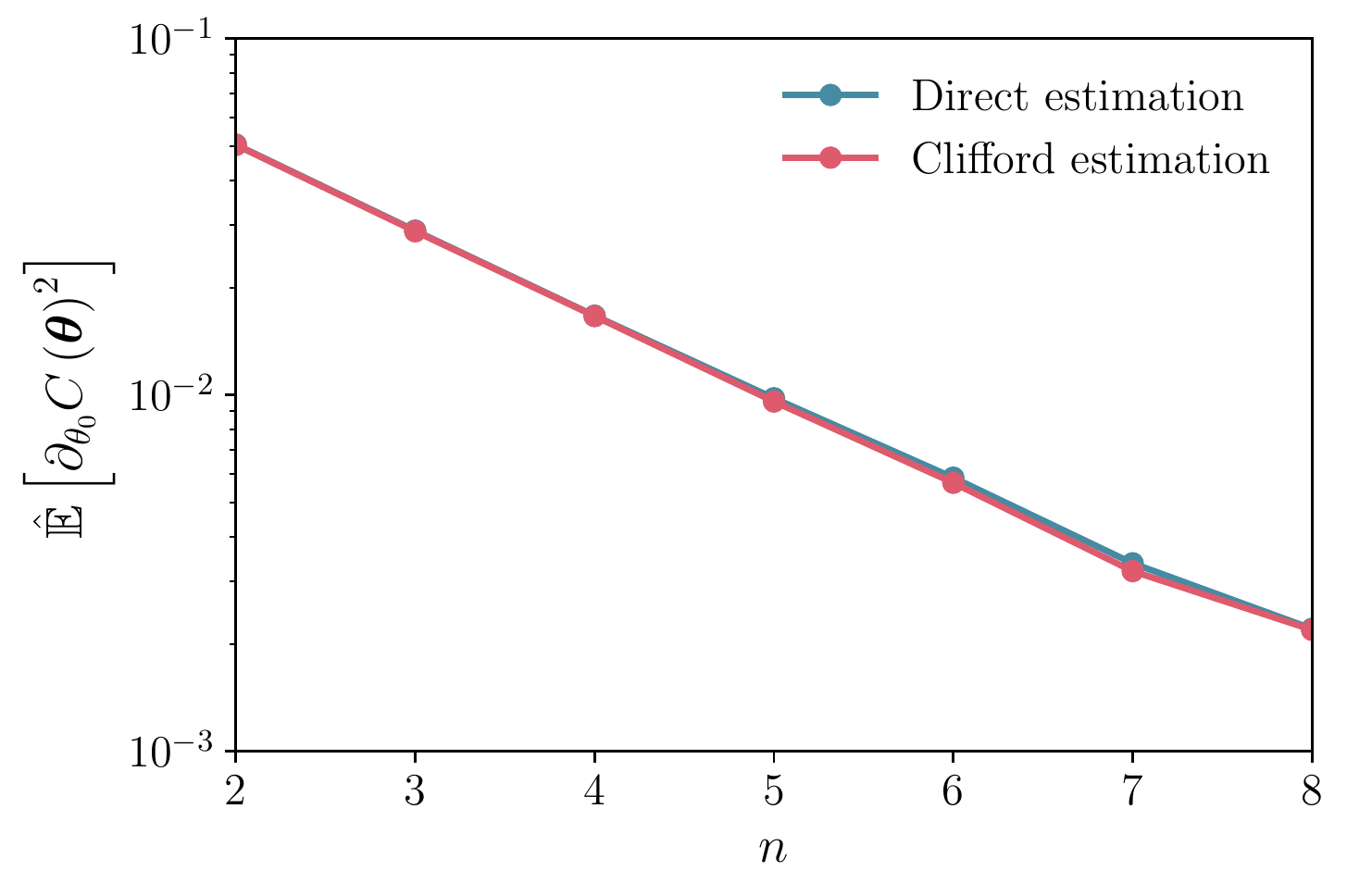}
    \caption{Estimated average of the squared gradient of the cost function with respect to the first variational parameter versus the number $n$ of qubits. We emphasize that derivatives with respect to the other angles $\theta_k$ give similar results (not shown). The results are for random circuits composed of a single layer of gates, with one rotation per qubit. Such rotations are randomly chosen among ${R_X,R_Y,R_Z}$. The rotation layer is followed by a layer of alternated $CZ$ gates (note that this is the same type of architecture as that represented on Fig.~\ref{fig:vqc_to_clifford_mapping}). The random rotation angles are independent and follow the uniform probability distribution on the interval $\[0,2\pi\]$. In order to get the estimation, we have randomly sampled $500$ different circuit architectures. For each gates architecture,  we have computed the average of the squared gradient assuming a uniform distribution of the rotation angles, using both a direct estimation and our method based on the mapping to Clifford approximant circuits. In particular, we have  sampled $500$ vectors of angle parameters for the direct estimation and $500$ Clifford circuits for our method. Note that for the uniform distribution the average gradient vanishes, thus estimating the squared gradient is equivalent to estimate the gradient variance.} 
    \label{fig:simple_circuit_bp}
\end{figure}
\begin{figure}
    \includegraphics[width=\columnwidth]{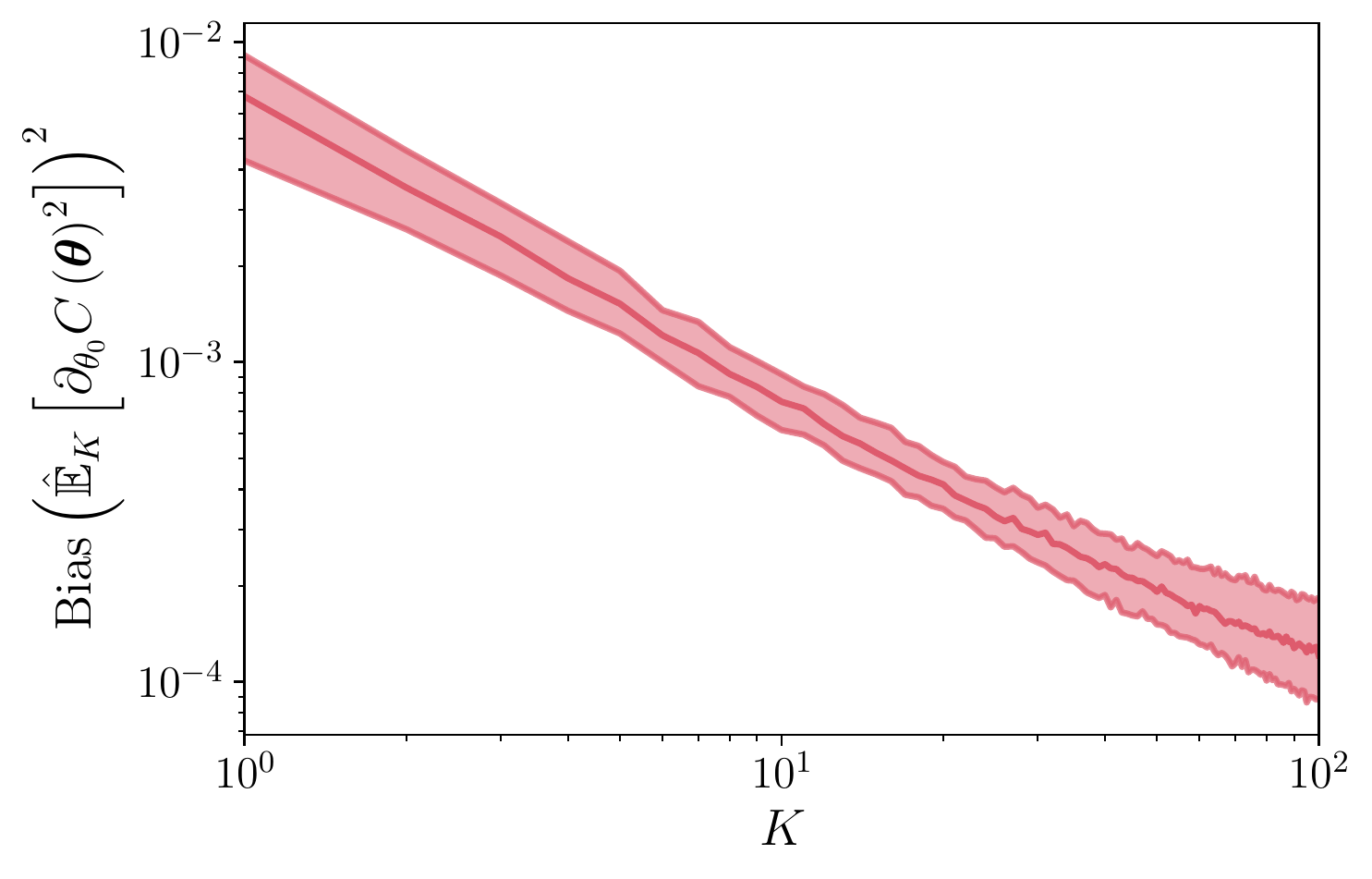}  
    \caption{Squared statistical bias of the estimator considered in Fig. \ref{fig:simple_circuit_bp} for random circuits with $\nqubits=5$ qubits versus the number $K$ of Clifford approximant circuits. The results have been obtained with $500$ randomly drawn circuit architectures.  For each sample size $K$, we consider a bootstrap batch of $100$ estimators (each estimator is obtained by sampling $K$ circuits from a set of $2000$ Clifford approximant circuits for each choice of the rotation directions). Then for each $K$, the statistical bias is derived from the bootstrap batch. The estimator true expected value is provided by the direct estimation of the average squared gradient with $4000$ samples. The shaded area corresponds to the interval between the $20$ and $80$ percentiles of the estimated biases for the $500$ random architectures.}
    \label{fig:simple_circuit_bp_est_bias}
\end{figure}
\begin{figure}
    \includegraphics[width=\columnwidth]{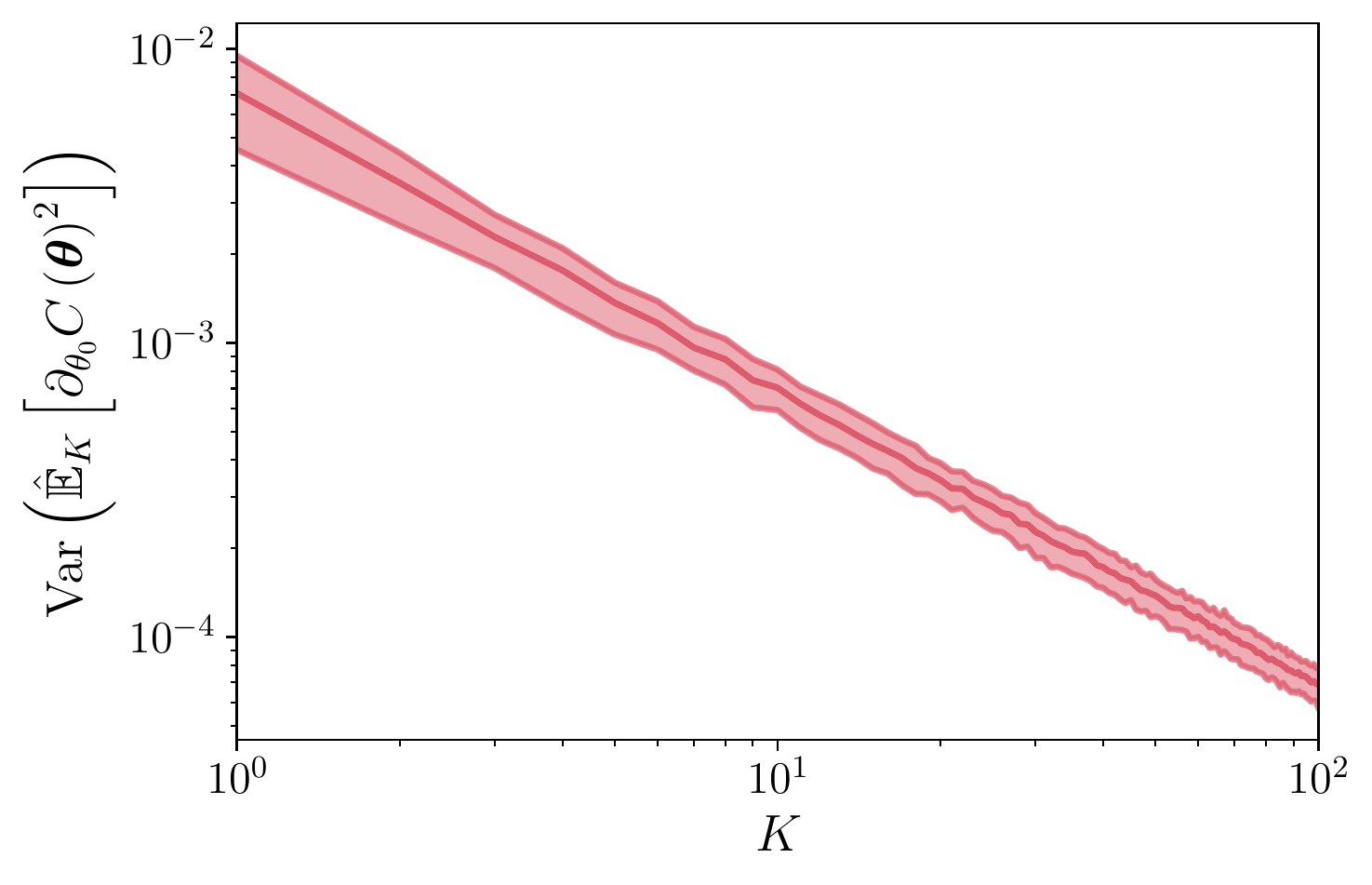}
    \caption{Variance of the estimator of the expected squared gradient with respect to the first parameter $\theta_0$ versus the number $K$ of Clifford approximant circuits. Same type of random circuits as in Fig. \ref{fig:simple_circuit_bp} with $\nqubits=5$ qubits. We have used the same bootstrap procedure as in Fig.~\ref{fig:simple_circuit_bp_est_bias}. The shaded area corresponds to the interval between the $20$ and $80$ percentiles of the estimated biases for $500$ random architectures.}
    \label{fig:simple_circuit_bp_est_var}
\end{figure}

To illustrate the applications of our exact mapping and the ensuing estimation method, we have performed numerical experiments on concrete examples. Let us consider a simple variational quantum circuit composed of layers of single-qubit rotations along either the $X$, $Y$ or $Z$ axes, alternated with fixed layers of Control-Z gates. Such an ansatz is shown for three qubits in Fig.~\ref{fig:vqc_to_clifford_mapping}. We further assume that the rotation angles are independent and identically distributed according to the uniform law over $\left[0,2\pi\right]$. Moreover, we assume that the cost function is of the form in Eq.~\eqref{eq:cost_function} with $\O=\dyad{0}^{\otimes \nqubits}$. 

We consider these architectures with random directions of the rotation gates. Up to a different fixed first layer, such random circuits have been showed to exhibit barren plateaus in Ref.~\cite{mcclean2016}. Note that in this particular case the averaging was done on both the rotations angles and the rotations directions. Here we reproduce this result using Clifford approximants. To do so we sample both the exact circuit architecture by randomly selecting the rotation directions uniformly from $\{X,Y,Z\}$, and then we either sample the rotation angles directly or we sample a Clifford approximant circuit. For a uniform distribution we have $\forall k\in\mathbb{Z},\;\expectt{\theta}{\cos{k\theta}} = 0$ so the sampling of the replacement Clifford gates is uniform (as represented on Fig.~\ref{fig:vqc_to_clifford_mapping_rules} for $r_1=r_2=0$). Moreover, by the parameter-shift rule (Eq. \ref{eq:parameter_shift_rule}) it is clear that for uniformly distributed rotations the average gradient is analytically zero, thus it suffices to estimate the average of the squared gradient as $\variance{\thetav}{\partial_k C(\thetav)} = \expectt{\thetav}{\partial_k C(\thetav)^2}$. 

In Fig.~\ref{fig:simple_circuit_bp} the estimations of the average squared gradient using either direct evaluations or by sampling Clifford approximants are shown. Note that the average is taken over both the random rotation angles and the variable architecture (i.e. the random direction of the rotation gates). The estimation obtained from Clifford approximants accurately matches the direct estimation and the average squared gradient vanishes exponentially with the number of qubits, as expected. In addition, the evolution of the bias of the Clifford estimation with the number of approximant circuits $K$ is shown in Fig.~\ref{fig:simple_circuit_bp_est_bias}. The bias decreases polynomially with $K$. As appears in Fig.~\ref{fig:simple_circuit_bp_est_var}, the same trend holds for the variance of the Clifford estimators. These results are in agreement with the analytical scaling derived in App. \ref{appendix:sampling_efficiency}.

\begin{figure}
    \includegraphics[width=\columnwidth]{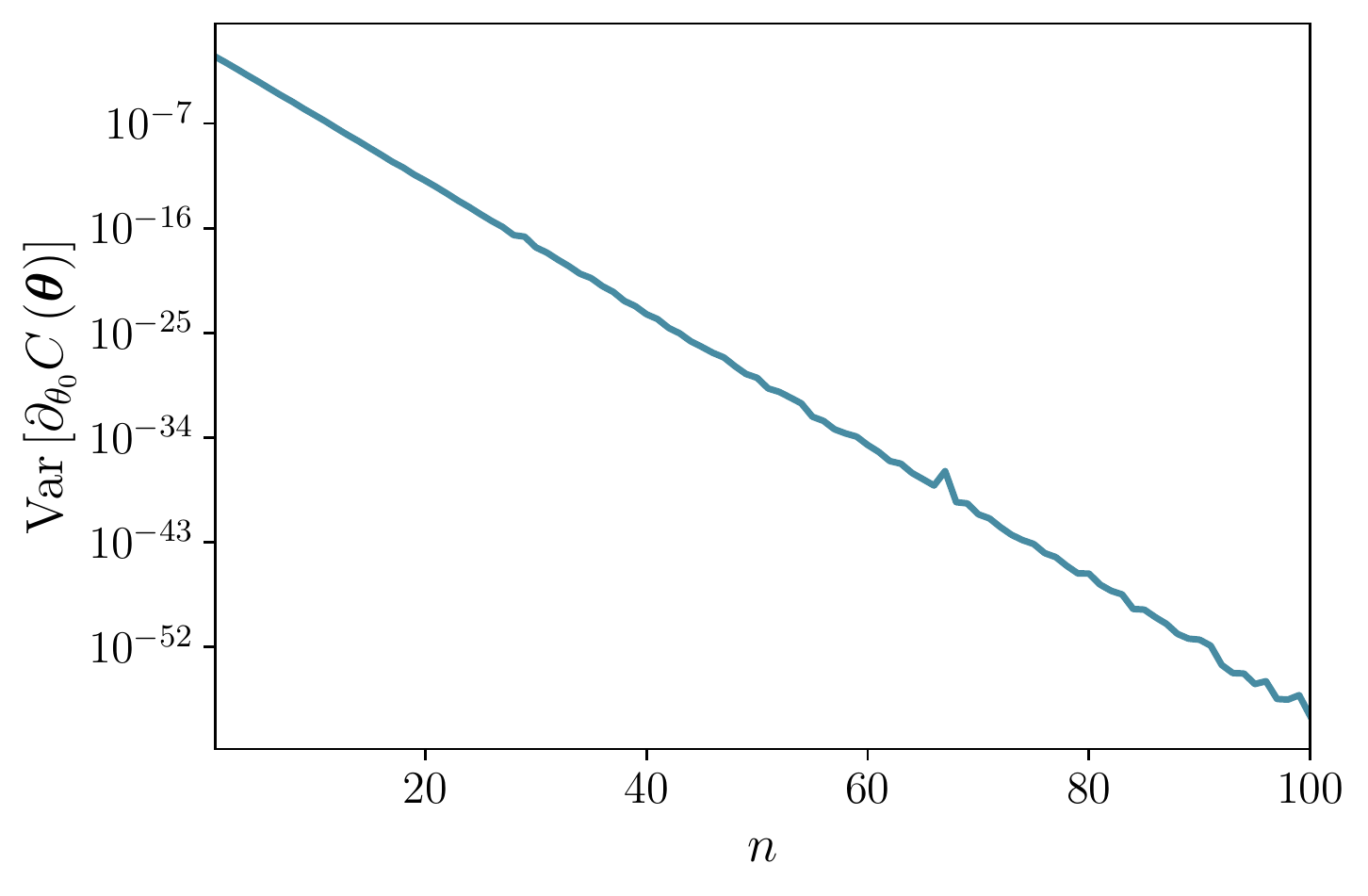}
    
    \caption{Estimated variance of the  gradient of the cost function with respect to the first variational parameter versus the number $n$ of qubits. Each variance is estimated using $10^5$ Clifford circuits. The circuits and cost function used are the same as the for Fig.~\ref{fig:simple_circuit_bp}.}
    \label{fig:barren_plateaus_reply}
\end{figure}

Using our method, we also reproduced the results showing the exponential suppresion of the gradient variance presented in Ref.~\cite{mcclean2016} for up to 100 qubits. The results are presented on Fig.~\ref{fig:barren_plateaus_reply}. 

\begin{figure}
    \includegraphics[width=\columnwidth]{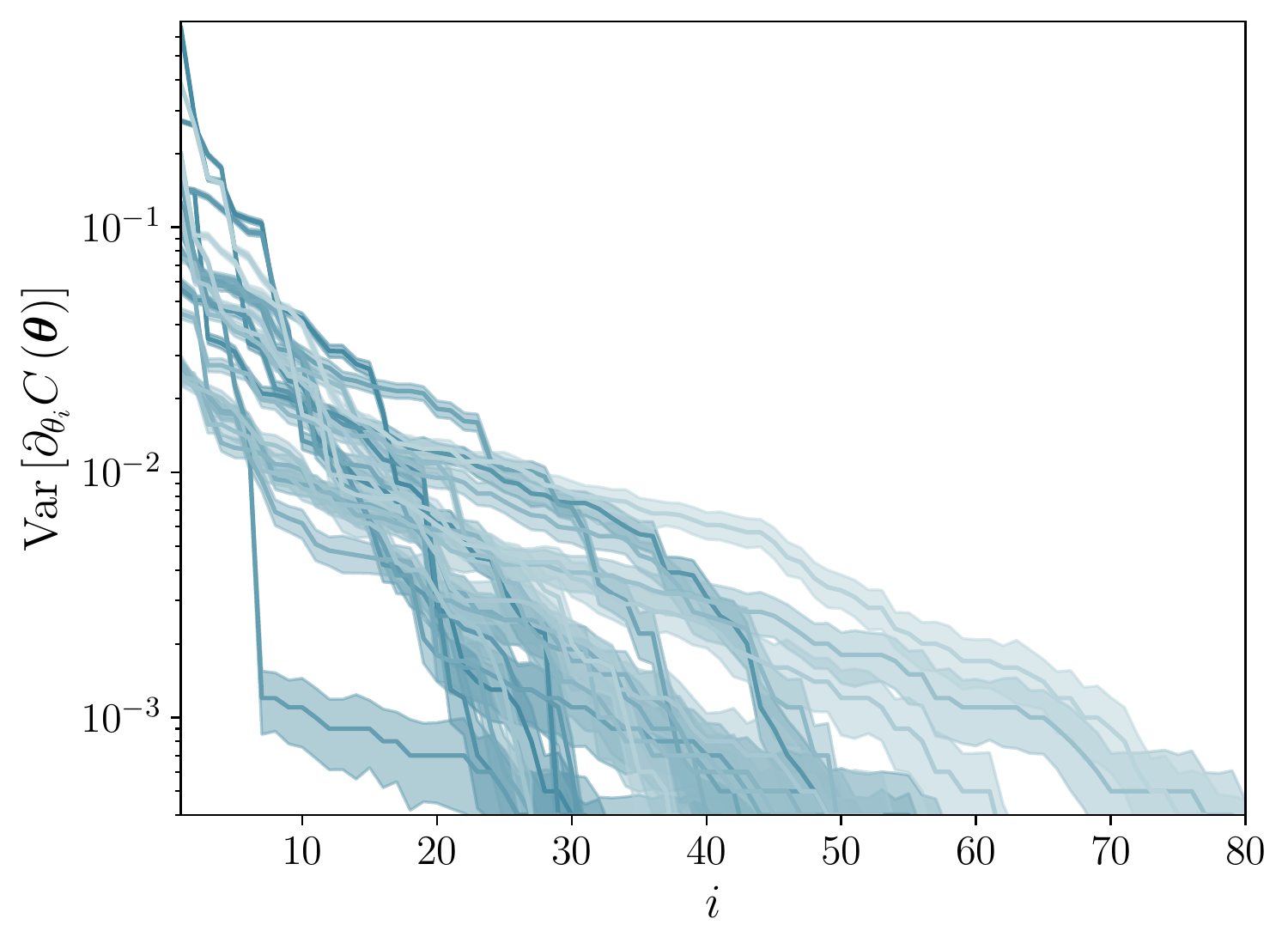}

    \caption{Variance of the cost function gradient with respect to the variational parameters in decreasing order and for 20 random circuits. The circuits are acting on $n=40$ qubits. Each circuit is composed of 10 layers. For each layer $l$, a number $m_l$ is randomly chosen in the interval $[0,n]$, and the layer is then built by first applying $m_l$ random single qubit rotation gates to randomly chosen qubits, and then applying a serie of entangling gates of the same type and with the same arrangement as the ones considered in Fig.~\ref{fig:simple_circuit_bp}. For each circuit and each parameter $\theta_i$, the gradients are estimated using $10^4$ Clifford approximant circuits. The Hamiltonian of the cost function is a sum of $10$ random Pauli strings. Note that each curve corresponds to a different architecture.}
    \label{fig:random_circuits_vars_reply}
    
\end{figure}

\begin{figure}
    \includegraphics[width=\columnwidth]{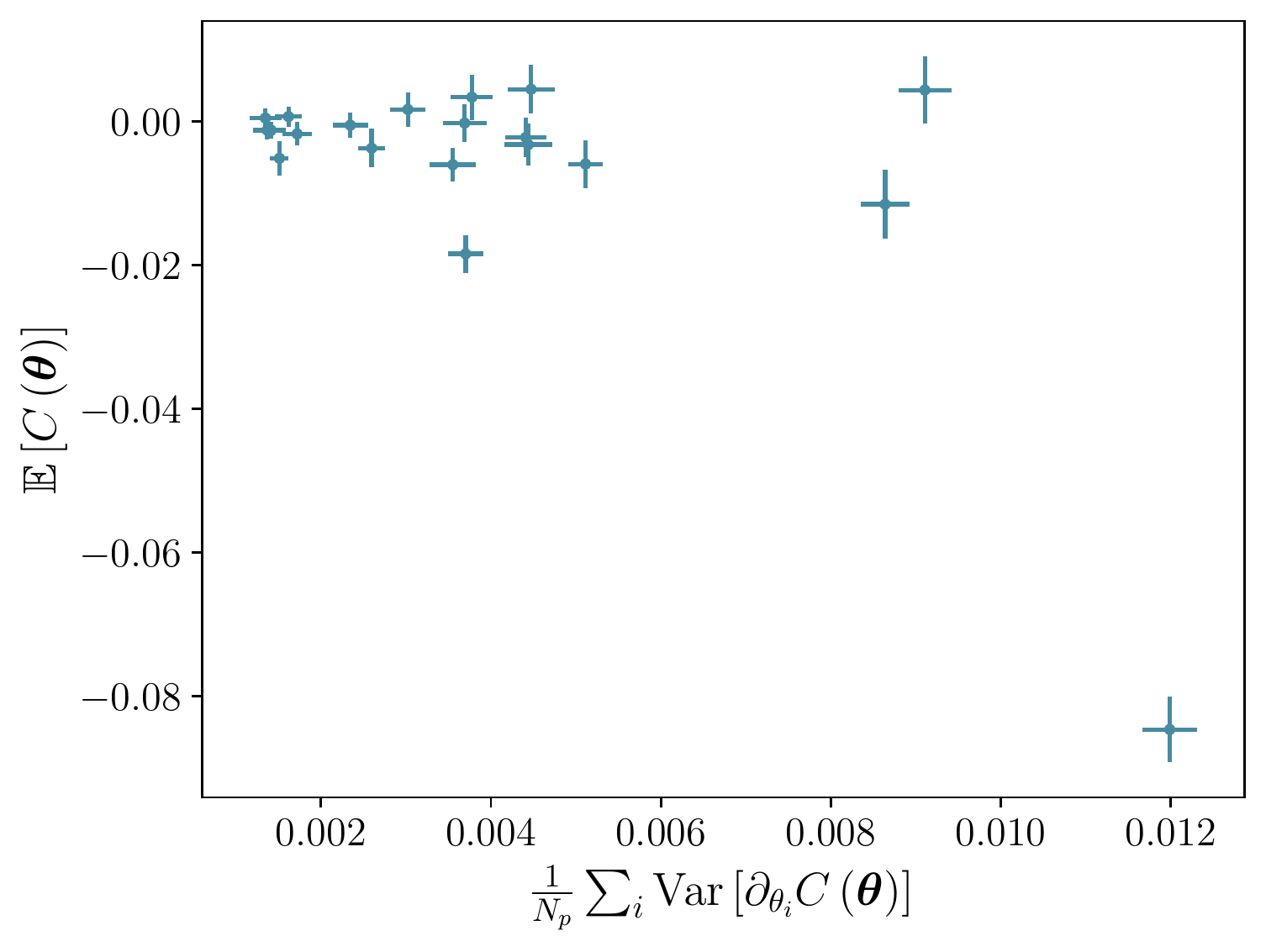}

    \caption{Mean value of the cost function vs. variance of the cost function gradient, for $20$ random circuits and $40$ qubits. Each point represents a circuit, and the circuits considered are the same as the ones of Fig.~\ref{fig:random_circuits_vars_reply}. The x-axis is the average of the variances of the gradients with respect to the $N_p$ circuits rotation parameters $\theta_i,\, i = 1,\dots,N_p$.}
    \label{fig:random_circuits_vars_vs_out_reply}
    
\end{figure}

Finally, we illustrate the impact of an ansatz architecture on its trainability by evaluating the gradients variances for a set of randomly drawn ansatze and for a given cost function Hamiltonian. We consider random circuits acting on 40 qubits, and a Hamiltonian composed of a sum of 10 randomly chosen Pauli strings. As for the results presented in Fig.~\ref{fig:simple_circuit_bp}, the considered variational circuits are composed of layers of rotations alternated with fixed entangling layers. Here we considered circuits with 10 layers. For each rotation layer, a random subset of rotations is replaced by identity gates (see Fig.~\ref{fig:random_circuits_vars_reply} for details). The variances of the cost function partial derivatives with respect to the ansatz parameters are shown on Fig.~\ref{fig:random_circuits_vars_reply}, and Fig.~\ref{fig:random_circuits_vars_vs_out_reply} shows the average variance of the gradients versus the the mean value of the cost function for the different random circuits. These results show that for a given cost function, modifying the ansatz architecture has a strong effect on the trainability. We therefore believe that our method may be used to systematically examine such effects for large systems with a reasonable cost, hence guiding the design of better variational ansätze.

\section{Conclusions and perspectives}
In this paper we presented a classically efficient method to estimate first and second-order expectation values for a large class of randomly initialized variational quantum circuits. This includes estimating the average gradient of the cost function and its variance, which can be used to estimate the trainability. Our method applies to the large class of circuits whose architecture is composed of fixed Clifford gates and single-qubit parameterized rotations, provided that the rotation angles are independent and that their distributions are symmetric with respect to an angle $\theta_0 \in \{k\pi/2,\;k\in\mathbb{Z}\}$ and satisfy $\expectt{\theta}{\cos^2{(\theta-\theta_{0})}}\geq \abs{\expectt{\theta}{\cos{(\theta-\theta_0)}}}$. The method relies on an exact mapping of randomly initialized variational quantum circuits to ensembles of Clifford circuits and on the Gottesman-Knill theorem. We provide rigorous convergence guarantees, and in particular we show that the complexity of the method scales polynomially in both the system size and the number of parameters of the considered ansatz. We investigated the generalization of the proposed scheme to the case of $N$-fold channels, and showed that the $N$-fold average of random Z-rotations can be expressed as a real combination of Clifford unitaries. However, such a decomposition is not unique, and finding a sufficient and necessary condition for the considered $N$-fold channel to be a Clifford mixed-unitary channel remains an open problem. Solving this problem is of great interest as it could allow to generalize the scheme presented in this work to ansätze with correlated variational parameters.

We believe that such a tool will prove very useful in future applications, as it could be employed to conduct classical optimization of architectures and initialization of large scale variational quantum circuits. 
As the absence of barren plateaus can be guaranteed by a large enough variance of the gradient, regardless of the exact origin of the potential barren plateaus, this method could be used to certify trainability for system with a very large number of qubits.
\begin{acknowledgments}
This work was supported by Region Île-de-France in the framework of the Domaine d'Intérêt Majeur (DIM) Science et Ingénierie en Région Île-de-France pour les Technologies Quantiques (SIRTEQ). This work was granted access to the High Performance Computing (HPC) resources of Très Grand Centre de Calcul
(TGCC) under Allocation No. 2022-A0120512462 made by
Grand Equipement National de Calcul Intensif (GENCI). We would like to thank Zakari Denis for helpful discussions during the early stages of this work.
\end{acknowledgments}
\appendix
\section{\texorpdfstring{$\boldsymbol{1}$}{1}-fold and \texorpdfstring{$\boldsymbol{2}$}{2}-fold channels of a random Z-rotation}
\label{appendix:k-fold_channel_rz}

\subsection{\texorpdfstring{$\boldsymbol{1}$}{1}-fold channel}

Here we give the expression of the $1$-fold channel for a single-qubit rotation around the Z axis. The rotations around the X and Y axis can then be obtained by combination with Hadamard and phase gates. Let us define $\hat{\Pi}_0 := \dyad{0}$ and $\hat{\Pi}_1 := \dyad{1}$:

\bea
\hat{R}_{Z}(\theta) =&\; e^{-\rmi\frac{\theta}{2}}\hat{\Pi}_0+e^{\rmi\frac{\theta}{2}}\hat{\Pi}_1\\
\hat{R}_{Z}(\theta)\rhohat \hat{R}^{\dagger}_{Z}(\theta) =&\; \hat{\Pi}_0 \rhohat \hat{\Pi}_0 + \hat{\Pi}_1 \rhohat \hat{\Pi}_1 \\&+ e^{\rmi\theta}\hat{\Pi}_1\rhohat\hat{\Pi}_0+e^{-\rmi\theta}\hat{\Pi}_0\rhohat\hat{\Pi}_1.
\eea
Thus
\bea\label{eq:one-fold-raw-sum}
\expectt{\theta}{\hat{R}_{Z}(\theta)\rhohat \hat{R}^{\dagger}_{Z}(\theta)} =&\; \hat{\Pi}_0 \rhohat \hat{\Pi}_0 + \hat{\Pi}_1 \rhohat \hat{\Pi}_1 \\&+ \expectt{\theta}{e^{\rmi\theta}}\hat{\Pi}_1\rhohat\hat{\Pi}_0\\&+\expectt{\theta}{e^{-\rmi\theta}}\hat{\Pi}_0\rhohat\hat{\Pi}_1.
\eea
We recognize the characteristic function of the distribution of $\theta$, namely $$\phi(t):= \expectt{\theta}{e^{\rmi t\theta}}.$$Assuming this probability distribution is even in $\theta$, we have $\phi(t)\in\mathbb{R},\: \forall t$ and we can define $r_1=\phi(1)=\phi(1)^{\ast} = \phi(-1)$. 
As we have $\id= \hat{\Pi}_0+\hat{\Pi}_1$ and $\hat{Z} = \hat{\Pi}_0-\hat{\Pi}_1$, we get
\bea\label{eq:eq-1fold-terms}
\rhohat = \(\hat{\Pi}_0\rhohat\hat{\Pi}_0+\hat{\Pi}_1\rhohat\hat{\Pi}_1\) + \(\hat{\Pi}_1\rhohat\hat{\Pi}_0+\hat{\Pi}_0\rhohat\hat{\Pi}_1\), \\
\hat{Z}\rhohat \hat{Z} =  \(\hat{\Pi}_0\rhohat\hat{\Pi}_0+\hat{\Pi}_1\rhohat\hat{\Pi}_1\) - \(\hat{\Pi}_1\rhohat\hat{\Pi}_0+\hat{\Pi}_0\rhohat\hat{\Pi}_1\),
\eea
and hence
\bea\label{eq:one-fold-convex-sum}
\frac{1+r_1}{2}\rhohat + \frac{1-r_1}{2}\hat{Z}\rhohat \hat{Z} =&\; \expectt{\theta}{R_{Z}(\theta)\rhohat R^{\dagger}_{Z}(\theta)}.
\eea
This is indeed a convex sum of Clifford channels under the condition that $r_1\in\[-1,1\]$, which is always satisfied. For distributions that are symmetric with respect to a Clifford angles $\in \{k\pi/2, k\in\{0,1,2,3\}\}$, we can factor out the corresponding rotation, which is (up to a phase) a Clifford gate. This way we can fall back to the case of an unbiased even distribution, i.e. symmetric with respect to the zero angle. Note that in the particular case of the uniform distribution over $\[0,2\pi\]$, we have $r_1=0$.
\subsection{\texorpdfstring{$\boldsymbol{2}$}{2}-fold channel}

In this section we will make use of the Choi representation of quantum channels, which allows to represent channels acting on two-qubits states by $16\times16$ matrices. For a quantum channel (i.e. a completely positive trace preserving or CPTP map) $\Ecal$, the Choi operator is defined by:
\bea
    \Lambda(\Ecal) = \sum_{i,j,k,l=0}^{1}\dyad{ij}{kl}\otimes\Ecal\(\dyad{ij}{kl}\).
\eea
Its corresponding matrix entries are:
\bea
    \Lambda(\Ecal)_{(ijkl),(mnpq)} =& \tr{\Lambda(\Ecal)^{\dagger}\left(\dyad{ij}{kl}\otimes\dyad{mn}{pq}\right)}\\
    =&\tr{\Ecal\(\dyad{ij}{kl}\)^{\dagger}\dyad{mn}{pq}}
\eea
In the following, we will write 
\bea
\Ecal[\hat{U}](\rhohat):=\hat{U}\rhohat\hat{U}^\dagger
\eea
for the quantum channel associated to a unitary transformation $\hat{U}$. We assume that $\U$ is diagonal in the computational basis, so that we can write 
\bea
\U=\sum_{i,j=0}^1\lambda_{ij}\hat{\Pi}_{ij},
\label{eq:unitary_diag_in_comp_basis}
\eea
where we define the projectors $\hat{\Pi}_{ij} := \hat{\Pi}_{i}\otimes\hat{\Pi}_{j}$. For $\U$ unitary, we have $\U\Ud=\id=\sum_{i,j}\lambda_{ij}\lambda_{ij}^{*}\hat{\Pi}_{ij}$, and hence $\lambda_{ij} = e^{\rmi\theta_{ij}},\;\forall i,j$. Therefore we have
\bea
\Ecal[\U](\dyad{ij}{kl}) =& \sum_{m,n,p,q}\lambda_{mn}\lambda_{pq}^{*}\hat{\Pi}_{mn}\dyad{ij}{kl}\hat{\Pi}_{pq}\\
&= \lambda_{ij}\lambda_{kl}^{*}\dyad{ij}{kl}\\
&=e^{\rmi(\theta_{ij}-\theta_{kl})}\dyad{ij}{kl}.
\label{eq:choi_matrix_diag_unitary}
\eea
Thus the Choi matrix of $\Ecal[\U]$ is diagonal whenever $\U$ is of the form given in Eq.~\eqref{eq:unitary_diag_in_comp_basis}. We can represent it by a $4\times4$ matrix $M$, whose entries are defined by 
\bea
M_{(ij),(kl)} := \Lambda_{(ijkl),(ijkl)}.
\label{eq:entries_of_M}
\eea
Note that the matrix $M$ is Hermitian and that its diagonal entries are always equal to one, due to Eq.~\eqref{eq:choi_matrix_diag_unitary}.
In the following we will represent each channel by its associated matrix $M$ in the basis $(00),(01),(10),(11)$.

As done earlier, we will focus on rotations around the Z axis. We have
\bea
\Phi_{Z}^{(2)}(\rhohat) := \expectt{\theta}{(\hat{R}_{Z}(\theta)\otimes \hat{R}_{Z}(\theta))\rhohat (\hat{R}^{\dagger}_{Z}(\theta)\otimes \hat{R}^{\dagger}_{Z}(\theta))}\\
\eea
and
\bea
\hat{R}_{Z}(\theta)\otimes \hat{R}_{Z}(\theta) =&\; \(e^{-\rmi\theta}\hat{\Pi}_0\otimes\hat{\Pi}_0 + e^{\rmi\theta}\hat{\Pi}_1\otimes\hat{\Pi}_1\)\\&+\(\hat{\Pi}_0\otimes\hat{\Pi}_1+\hat{\Pi}_1\otimes\hat{\Pi}_0\).
\eea
Defining
\bea
    \Gamma_{\theta} =&\; \(e^{-\rmi\theta}\hat{\Pi}_{00} + e^{\rmi\theta}\hat{\Pi}_{11}\),\\
    \Xi =&\; \hat{\Pi}_{01}+\hat{\Pi}_{10},
\eea
we can write
\bea\label{eq:two-fold-raw-sum}
\Phi_{Z}^{(2)}(\rhohat) =&\; \expectt{\theta}{\Xi\rhohat\Xi^\dagger}+\expectt{\theta}{\Gamma_{\theta}\rhohat\Gamma_{\theta}^\dagger}\\&+\expectt{\theta}{\Gamma_{\theta}\rhohat\Xi^\dagger}+\expectt{\theta}{\Xi\rhohat\Gamma_{\theta}^\dagger}.
\eea
\subsubsection{Uniform distribution}
For the uniform distribution of $\theta$ in $\[0,2\pi\]$, we have $\expectt{\theta}{e^{\pm\rmi\theta}}=\expectt{\theta}{e^{\pm2\rmi\theta}}=0$, and thus:
\bea
\expectt{\theta}{\Gamma_{\theta}\rhohat\Xi^\dagger} =&\; 0,\\
\expectt{\theta}{\Gamma_{\theta}\rhohat\Gamma_{\theta}^\dagger} =&\; \hat{\Pi}_{00}\rhohat\hat{\Pi}_{00}+\hat{\Pi}_{11}\rhohat\hat{\Pi}_{11},\\
\expectt{\theta}{\Xi\rhohat\Xi^\dagger} =&\; \hat{\Pi}_{01}\rhohat\hat{\Pi}_{01}+\hat{\Pi}_{10}\rhohat\hat{\Pi}_{10},\\&+\hat{\Pi}_{01}\rhohat\hat{\Pi}_{10}+\hat{\Pi}_{10}\rhohat\hat{\Pi}_{01}.
\eea
Finally, we get
\bea
\Phi_{Z}^{(2)}(\rhohat) =&\; \hat{\Pi}_{00}\rhohat\hat{\Pi}_{00}+\hat{\Pi}_{11}\rhohat\hat{\Pi}_{11}+\hat{\Pi}_{01}\rhohat\hat{\Pi}_{01}\\&+\hat{\Pi}_{10}\rhohat\hat{\Pi}_{10}+\hat{\Pi}_{01}\rhohat\hat{\Pi}_{10}+\hat{\Pi}_{10}\rhohat\hat{\Pi}_{01}.
\eea
We can represent $\Phi_{Z}^{(2)}$ by its associated matrix 
\bea
M(\Phi_{Z}^{(2)}) = 
    \begin{pmatrix}
    1 & 0 & 0 & 0 \\
    0 & 1 & 1 & 0 \\
    0 & 1 & 1 & 0 \\
    0 & 0 & 0 & 1
    \end{pmatrix}.
\eea
One can verify that the following channels also have a diagonal Choi matrix, and we can use the same representation of their diagonals, giving
\bea
M(\Ecal[\id]) =&\; 
    \begin{pmatrix}
    1 & 1 & 1 & 1 \\
    1 & 1 & 1 & 1 \\
    1 & 1 & 1 & 1 \\
    1 & 1 & 1 & 1
    \end{pmatrix},\\
M(\Ecal[\hat{Z}\otimes \hat{Z}]) =&\; 
    \begin{pmatrix}
    1 & -1 & -1 & 1 \\
    -1 & 1 & 1 & -1 \\
    -1 & 1 & 1 & -1 \\
    1 & -1 & -1 & 1
    \end{pmatrix},\\
M(\Ecal[\hat{S}\otimes\hat{S}]) =&\; 
    \begin{pmatrix}
    1 & \rmi & \rmi & -1 \\
    -\rmi & 1 & 1 & \rmi \\
    -\rmi & 1 & 1 & \rmi \\
    -1 & -\rmi & -\rmi & 1
    \end{pmatrix},\\
M(\Ecal[\hat{S}^{\dagger}\otimes\hat{S}^{\dagger}]) =&\; 
    \begin{pmatrix}
    1 & -\rmi & -\rmi & -1 \\
    \rmi & 1 & 1 & -\rmi \\
    \rmi & 1 & 1 & -\rmi \\
    -1 & \rmi & \rmi & 1
    \end{pmatrix},\\
    \label{eq:M_matrices_gates}
\eea
with $\hat{S} = \Pi_0+\rmi\Pi_1$ the phase gate.
Gathering all together, we have
\bea
4M(\Phi_{Z}^{(2)}) =&\; M(\Ecal[\id])+M(\Ecal[\hat{Z}\otimes \hat{Z}]) \\&+M(\Ecal[\hat{S}\otimes\hat{S}])+M(\Ecal[\hat{S}^{\dagger}\otimes\hat{S}^{\dagger}]).
\eea
The final result in the main text then follows by linearity and uniqueness of the Choi matrix.
\subsubsection{Even distribution}
Let us consider an even probability distribution of $\theta$ (i.e. a distribution for which $\theta$ has the same law as $-\theta$). For such distributions we again have that $\phi_{\theta}(t)=\phi_{\theta}(-t)\in \[-1,1\]\subset\mathbb{R}$ for all $t\in\mathbb{R}$ 
and thus $$\phi_{\theta}(t)=\frac{1}{2}\(\phi_{\theta}(t)+\phi_{\theta}(-t)\) = \expectt{\theta}{\cos(t\theta)}.$$ Defining $r_1 = \phi_{\theta}(1)$ and $r_2 = \phi_{\theta}(2)$, we can write
\bea
    \expectt{\theta}{\Gamma_{\theta}\rhohat\Xi^\dagger} =&\; r_1\(\hat{\Pi}_{00}+\hat{\Pi}_{11}\)\rhohat\(\hat{\Pi}_{01}+\hat{\Pi}_{10}\),\\
    \expectt{\theta}{\Gamma_{\theta}\rhohat\Gamma_{\theta}^\dagger} =&\; \hat{\Pi}_{00}\rhohat\hat{\Pi}_{00}+\hat{\Pi}_{11}\rhohat\hat{\Pi}_{11}\\&+r_2\(\hat{\Pi}_{00}\rhohat\hat{\Pi}_{11}+\hat{\Pi}_{11}\rhohat\hat{\Pi}_{00}\),\\
    \expectt{\theta}{\Xi\rhohat\Xi^\dagger} =&\; \hat{\Pi}_{01}\rhohat\hat{\Pi}_{01}+\hat{\Pi}_{10}\rhohat\hat{\Pi}_{10}\\&+\hat{\Pi}_{01}\rhohat\hat{\Pi}_{10}+\hat{\Pi}_{10}\rhohat\hat{\Pi}_{01}.
\eea
Hence we obtain:
\bea
M(\Phi_{Z}^{(2)}) = 
    \begin{pmatrix}
    1 & r_1 & r_1 & r_2 \\
    r_1 & 1 & 1 & r_1 \\
    r_1 & 1 & 1 & r_1 \\
    r_2 & r_1 & r_1 & 1
    \end{pmatrix}.
\eea
We can express $M(\Phi_{Z}^{(2)})$ as a linear combination of the matrices of Eq.~\eqref{eq:M_matrices_gates}, giving
\bea
M(\Phi_{Z}^{(2)}) =& aM(\Ecal[\id] )+bM(\Ecal[\hat{Z}\otimes \hat{Z}])\\&+\frac{c}{2}\left(M(\Ecal[\hat{S}\otimes\hat{S}])+M(\Ecal[\hat{S}^{\dagger}\otimes\hat{S}^{\dagger}])\right).
\eea
The coefficients $a,b,c$ can be found by solving the linear system
\bea
\begin{cases} a+b+c = 1 \\ a-b = r_1 \\ a+b-c = r_2 \end{cases} \,,
\eea
and one finds
\bea
 M(\Phi_{Z}^{(2)}) =&\; \frac{1}{4}\(1+r_2+2r_1\)M(\Ecal[\id])\\&+\frac{1}{4}\(1+r_2-2r_1\)M(\Ecal[\hat{Z}\otimes \hat{Z}])\\&+\frac{1}{4}\(1-r_2\)M(\Ecal[\hat{S}\otimes\hat{S}])\\&+\frac{1}{4}\(1-r_2\)M(\Ecal[\hat{S}^{\dagger}\otimes\hat{S}^{\dagger}]).
\eea
Therefore, the associated channel is
\bea\label{eq:two-fold-sum}
\Phi_{Z}^{(2)}\(\rhohat\) =&\; \frac{1}{4}\(1+r_2+2r_1\)\rhohat\\&+\frac{1}{4}\(1+r_2-2r_1\)\(\hat{Z}\otimes \hat{Z}\)\rhohat \(\hat{Z}\otimes \hat{Z}\) \\&+\frac{1}{4}\(1-r_2\)\(\hat{S}\otimes\hat{S}\)\rhohat \(\hat{S}^{\dagger}\otimes\hat{S}^{\dagger}\)\\
&+\frac{1}{4}\(1-r_2\)\(\hat{S}^{\dagger}\otimes\hat{S}^{\dagger}\) \rhohat \(\hat{S}\otimes\hat{S})\).
\eea
\begin{remark}
Defining $CZ = \hat{\Pi}_0\otimes\id + \hat{\Pi}_1\otimes \hat{Z}$ the control-Z gate and $CZ_X = (\hat{X}\otimes \hat{X}) CZ (\hat{X}\otimes \hat{X})$, we have
\bea
M(\Ecal[CZ]) =&\; 
    \begin{pmatrix}
    1 & 1 & 1 & -1 \\
    1 & 1 & 1 & -1 \\
    1 & 1 & 1 & -1 \\
    -1 & -1 & -1 & 1
    \end{pmatrix},\\
M(\Ecal[CZ_X]) =&\; 
    \begin{pmatrix}
    1 & -1 & -1 & -1 \\
    -1 & 1 & 1 & 1 \\
    -1 & 1 & 1 & 1 \\
    -1 & 1 & 1 & 1
    \end{pmatrix},\\
\eea
and thus
\bea
\Ecal[\hat{S}\otimes\hat{S}]+\Ecal[\hat{S}^{\dagger}\otimes\hat{S}^{\dagger}] = \Ecal[CZ]+\Ecal[CZ_X].
\eea
Therefore the decomposition of $\Phi^{(2)}_Z$ into a convex sum of Clifford channels of Eq.~\eqref{eq:two-fold-sum} is not unique.
\end{remark}
\begin{figure}[t]
    \includegraphics[width=\columnwidth]{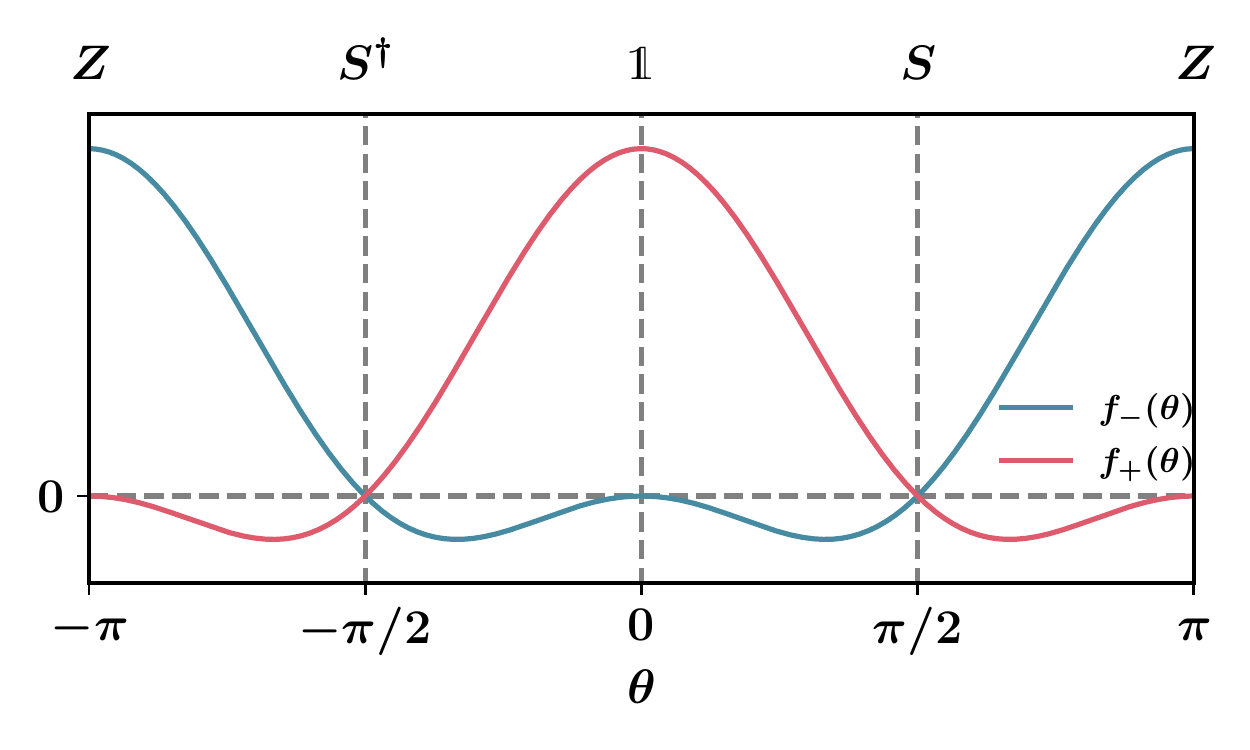}
    
    \caption{Plot of $f_{\pm}(\theta) = \cos{\theta}(\cos{\theta}\pm1)$ versus $\theta$. The condition in Eq.~\eqref{eq:second_order_condition_in_appendix} is fulfilled if and only if $\expectt{\theta}{f_{+}(\theta)}\geq0$ and $\expectt{\theta}{f_{-}(\theta)}\geq0$. $\Shat$ is the phase gate.}
    \label{fig:second_order_condition}
\end{figure}
The decomposition obtained in Eq.~\eqref{eq:two-fold-sum} is a convex sum if one assume that $
\(1+r_2-2r_1\)\geq0$ and $\(1+r_2+2r_1\)\geq0$.
This condition holds if and only if
$$
\expectt{\theta}{\frac{1}{2}(1+\cos{2\theta})\pm\cos{\theta}}\geq 0,
$$
namely if and only if
\bea
\expectt{\theta}{\cos^2\theta}\geq \abs{\expectt{\theta}{\cos{\theta}}}.
\label{eq:second_order_condition_in_appendix}
\eea

This condition is fulfilled for the distributions that are $\pi$-periodic as in that case we have $\expectt{\theta}{\cos{\theta}}=0$. Another example of distribution that satisfy this constraint is a Gaussian distribution with a large enough variance. In fact for a centered Gaussian distribution of variance $\sigma^2$, we have
$r_1 = e^{-\sigma^2/2}$ and $r_2= e^{-2\sigma^2}$, so that the condition becomes
\bea
1+e^{-2\sigma^2}-2e^{-\sigma^2/2}\geq 0.
\eea
One can show that this condition is equivalent to $\sigma^2 \geq \sigma^2_{0}$ for some specific $\sigma_0\in\mathbb{R}$, yielding a requirement on the width of the gaussian.

\section{First and second-order quantities}\label{appendix:first-and-second-order-quantities}

In this appendix, we define the notion of first and second-order quantities as quantities that can be obtained from the knowledge of respectively the $1$-fold and the $2$-fold channels for each of the random rotations appearing in a given ansatz. We also show that the average cost function and the average gradient are first-order quantities, while the average of the squared cost function and of the squared gradient are second-order quantities.
\subsection{First-order quantities}
\label{section:first_order}
Let us consider the ansatz defined by 
\bea
\U (\thetav) = \prod_{i=1}^{M}\U_{i}(\theta_i)\W_i,
\eea
and denote
\bea
\Ucal_i(\theta_i)(\rhohat)=&\;\U_{i}(\theta_i)
\rhohat\Ud_i(\theta_i)\\
\Wcal_i(\rhohat) =&\; \W_i\rhohat \W^{\dagger}_i
\eea
the unitary channels associated to the different layers of the circuit. The whole circuit unitary transformation then reads
\bea
\Ucal(\thetav)(\rhohat) =&\;\Ucal_M(\theta_M)\circ\dots\circ\Wcal_1(\rhohat)\\
=&\; \multicomp_{i=1}^{M}(\Ucal_i(\theta_i)\circ\Wcal_i)(\rhohat).
\eea
The cost function is then given by
\bea
C(\thetav) = \tr{\Ucal(\thetav)(\rhohat)\O}
\eea
and its expectation with respect to $\thetav$ is
\bea
\expectt{\thetav}{C(\thetav)} =&\; \expectt{\thetav}{\tr{\Ucal(\thetav)(\rhohat)\O}}\\
=&\;\tr{\expectt{\thetav}{\Ucal(\thetav)(\rhohat)}\O}\\
=&\; \tr{\expectt{\thetav}{\U(\thetav)\rhohat\Ud(\thetav)}\O}\\
=&\;\tr{\int_{\mathbb{R}^M} \U(\thetav)\rhohat\Ud(\thetav)p(\thetav)\mathrm{d}\thetav\O} \\
=&\;\tr{\Phi^{(1)}_{\thetav}(\rhohat)\O}.
\eea
Here, we used both the linearity of the expectation and the definition of the $1$-fold channel from Eq.~\eqref{eq:$t$-fold-channel}.
The cost function expectation can thus be obtained from the knowledge of the complete $1$-fold channel $\Phi^{(1)}_{\thetav}$. Assuming that the angles $\{\theta_{i}\}$ are independent from each other, the expectation against $\thetav$ can be factored in expectations against the $\theta_{i}$'s, which allows to write:
\bea
\Phi^{(1)}_{\thetav}(\rhohat) =&\; \expectt{\thetav}{\Ucal(\thetav)(\rhohat)} \\
=&\; \multicomp_{i=1}^{M}(\expectt{\theta_{i}}{\Ucal_i(\theta_i)}\circ\Wcal_i)(\rhohat).
\eea
As explained in the main text, we can consider without loss of generality all the rotations to be Z-rotations. Then the channels $\expectt{\theta_{i}}{\Ucal_i(\theta_i)}$ are exactly $1$-fold channels associated to a Z-rotation acting on a single qubit, and hence can be computed from the results of App. \ref{appendix:k-fold_channel_rz}. As stated earlier, we refer to quantities that can be obtained from the knowledge of the $1$-fold channels associated to each rotations of the ansatz as {\it first-order quantities}. Hence the average cost function $\expectt{\thetav}{C(\thetav)}$ is a first-order quantity.

Another example of an interesting first-order quantity is the average of the gradient. From the parameter-shift rule and using the linearity of the expectation, we have:
\bea
\expectt{\thetav}{\partial_{k}C(\thetav)} =&\; \frac{1}{2}\expectt{\thetav}{C\(\thetav + \frac{\pi}{2}\vb*{e}_{k}\)} \\&-\frac{1}{2}\expectt{\thetav}{C\(\thetav - \frac{\pi}{2}\vb*{e}_{k}\)}.
\eea
Here $\vb*{e}_{k}$ is the unit vector along the $k$-th component. 
The $\pm\pi/2$ shifts in the parameter $\theta_k$ can be factored out and seen as an extra Clifford gate added to the fixed layer $\W_k$. In fact, assuming $\hat{P}_k = \hat{Z}$ and denoting $\hat{S}$ the phase gate, we have $\U_k(\theta_k + \pi/2)\W_k = e^{-\mathrm{i}\frac{\theta_k}{2}\P_k}e^{-\mathrm{i}\frac{\pi}{2}\hat{Z}}\W_k = e^{-\mathrm{i}\frac{\pi}{4}} \U_k(\theta_k)\hat{S}\W_k $. Defining $\W_{k,\pm} = e^{\mp\mathrm{i}\frac{\pi}{4}}\hat{S}\W_k$, we get $\U_k(\theta_k + \pi/2)\W_k = \U_k(\theta_k)\W_{k,+}$. We can proceed likewise to define $\W_{k,-}$. In the following, we can write $\;\forall i\neq k \;\V_{i,\pm} = \W_i$ and $\V_{k,\pm} = \W_{k,\pm}$ the modified fixed layers that include the considered shift.
We have
\bea
\expectt{\thetav}{\U_{\pm}(\thetav)\rhohat\Ud_{\pm}(\thetav)}=&\;\expectt{\thetav}{\multicomp_{i=1}^{M}(\Ucal_{i}(\theta_i)\circ\Vcal_{i,\pm})(\rhohat)}\\
=&\; \multicomp_{i=1}^{M}(\expectt{\theta_i}{\Ucal_{i}(\theta_i)}\circ\Vcal_{i,\pm})(\rhohat),
\eea
where $\Vcal_{i,\pm}(\rhohat) = \V_{i,\pm}\rhohat\Vd_{i,\pm}$. The average gradient is therefore a first-order quantity, namely depending on $1$-fold channels only.

\subsection{Second-order quantities}
\label{section:second_order}

We now turn our attention to the mean value of the squared cost function. This is given by
\bea
\expectt{\thetav}{C(\thetav)^2} =&\; \expectt{\thetav}{\tr{\Ucal(\thetav)(\rhohat)\O}^2}\\
=&\; \expectt{\thetav}{\tr{\(\Ucal(\thetav)(\rhohat)\O\)^{\otimes 2}}}\\
=&\;\expectt{\thetav}{\tr{\Ucal^{(2)}(\thetav)(\rhohat^{\otimes 2})\O^{\otimes 2}}}.
\eea
For every state $\rhohat$ of a system of $2\nqubits$ qubits (i.e., a doubled version of the original system where the copy is not connected by gates to the original circuit), we define
\bea
\Ucal^{(2)}(\thetav)(\rhohat) =&\; \U^{\otimes 2}(\thetav)\rhohat\U^{\dagger\otimes 2}(\thetav).
\eea
Likewise we can define the doubled version of the circuit layers as
\bea
\Ucal^{(2)}_i(\theta_i)(\rhohat)=&\;\U^{\otimes 2}_{i}(\theta_i) \rhohat \U^{\otimes 2}_i(\theta_i) \, , \\
\Wcal^{(2)}_i(\rhohat) =&\; \W^{\otimes 2}_{i} \rhohat \W^{\otimes 2}_i,
\eea
giving
\bea
\Ucal^{(2)}(\thetav)(\rhohat) = \multicomp_{i=1}^{M}(\Ucal^{(2)}_{i}(\theta_i)\circ\Wcal^{(2)}_{i})(\rhohat).
\eea
Thus for independent rotations we have
\bea
\Phi^{(2)}_{\thetav}(\rhohat) =&\; \expectt{\thetav}{\Ucal^{(2)}(\thetav)(\rhohat)}\\
=&\; \multicomp_{i=1}^{M}(\expectt{\theta_i}{\Ucal^{(2)}_{i}(\theta_i)}\circ\Wcal^{(2)}_{i})(\rhohat) \, .
\label{eq:$2$-fold-channel-decomposition}
\eea
As for first-order quantities, we refer to quantities that can be obtained from the knowledge of the average $2$-fold channels of the rotations layers $\expectt{\theta_i}{\Ucal^{(2)}_{i}(\theta_i)}$ as second-order quantities. 

The average of the squared cost function is thus a second-order quantity, and as for the first order case, we can show that the squared gradient is also a second-order quantity. In fact, by making use of the parameter-shift rule, we see that to obtain the average of the squared gradient we have to compute the following four terms
\bea
\expectt{\thetav}{C(\thetav+a_1\vb*{e}_k)C(\thetav+a_2\vb*{e}_k)},
\eea
with $a_1,a_2 \in\{\frac{\pi}{2},-\frac{\pi}{2}\}$. 
As done before, it suffices to replace the $\Wcal^{(2)}_i$ in Eq.~\eqref{eq:$2$-fold-channel-decomposition} with 
\bea
\Vcal^{(2)}_{i, a_1, a_2}(\rhohat) = \(\V_{i,a_1}\otimes \V_{i,a_2}\)\rhohat(\V^{\dagger}_{i,a_1}\otimes \Vd_{i,a_2}).
\eea 
Finally, the gradient variance can be computed as
\bea
\variance{\thetav}{\partial_k C(\thetav)} = \expectt{\thetav}{\partial_k C(\thetav)^2} - \expectt{\thetav}{\partial_k C(\thetav)}^2,
\eea
which is the sum of a first and a second-order quantity.
\section{Proof of the sampling efficiency}
\label{appendix:sampling_efficiency}
In this appendix we prove that to obtain an estimation of any first or second-order quantity for a given ansatz up to a precision $\epsilon$ and probability $\delta\in\[0,1\]$ to meet this precision, it suffices to sample a number of Clifford approximant circuits $K\sim \log(2\delta)M/\epsilon^2$. By invoking the Gottesman-Knill theorem, we obtain an estimation of any of the previous quantities with a complexity polynomial in both the size of the system and the number of variational parameters of the considered ansatz.
\subsection{Details on the mapping}
\label{appendix:detail-mapping}

Here we give details on the mapping of the randomly initialized parameterized circuit to Clifford approximants.
\begin{remark}
We use the notations adapted to first-order quantities. The generalization to the second order and the shifted versions is straightforward as it suffices to replace each channel by its doubled and/or shifted version, as done in App.~\ref{appendix:first-and-second-order-quantities}.
\end{remark}
Assuming that the $\theta_i$ are independent from each other, averaging $\Ucal(\thetav)$ over $\thetav$ amount to replace each rotation channel $\Ucal_i(\theta_i)$ by a convex sum of $m$ Clifford unitary channels $\Ucal_{ij}$ with associated weight $p_{ij}$. Thus $\expectt{\thetav}{\Ucal(\thetav)(\rhohat)}$ is replaced by a discrete average over $m^M$ Clifford unitary channels (with $m=2$ for the $1$-fold channel and $m=4$ for the $2$-fold channel):
\bea
\expectt{\thetav}{\Ucal(\thetav)(\rhohat)}
=&\; \multicomp_{i=1}^{M}(\sum_{j=1}^{m}p_{ij}\Ucal_{ij}\circ\Wcal_i)(\rhohat).
\eea
As we want to sample from that sum, we can define for each $i$ a discrete random variable $X_i$ taking values in $\{1,\dots,m\}$ such that $\mathbb{P}(X_i = j) = p_{ij}$. This represents a choice of a given unitary in the previous convex sum. Gathering these for all $k$ we get a random vector $\vb*{X} = (X_1, \dots, X_M) \in \{1, \dots, m\}^M$ that completely defines a unique unitary $\Ucal(\vb*{X})$ through:
\bea
\Ucal(j_1,\dots,j_M) = \multicomp_{i=1}^{M}\Ucal_{i j_i}\circ \Wcal_i.
\eea
Thus we have:
\bea\label{eq:expect-X}
\expectt{\thetav}{\Ucal(\thetav)(\rhohat)} =&\; \expectt{\vb*{X}}{\Ucal(\vb*{X})(\rhohat)}\\
=&\; \multicomp_{i=1}^{M}(\sum_{j=1}^{m}p_{ij}\Ucal_{ij}\circ\Wcal_i)(\rhohat).
\eea
The main idea is now to approximate the $k$-fold channels by an empirical average over $K$ samples of the previous Clifford circuits, namely:
\bea
\hat{\Phi}(\rhohat) := \frac{1}{K}\sum_{i=1}^{K}\Ucal(\vb*{X}_i)(\rhohat).
\eea

\subsection{Sampling efficiency}

\label{appendix:sampling_proof}
Our result relies on classical arguments for the sampling of bounded functions depending on a set of random variables using the McDiarmid's concentration inequality \cite{mcdiarmid1989,mohri2018}, which we remind below.
\begin{definition}[Bounded difference property]
A function $f:\mathcal{X}^M \to\mathbb{R}$ satisfies the bounded difference property if and only if $\exists\{c_1,\dots,c_M\}$ such that $\forall i\in\{1,\dots,M\}, \: \forall \(x_1, \dots, x_M\)$:
$$\sup_{x^{'}_i\in\mathcal{X}}\abs{f(x_1,..\,,x_i,..\,,x_M)-f(x_1,..\,,x^{'}_i,..\,,x_M)}<c_i\,.$$
\end{definition}
\begin{theorem}[McDiarmid's inequality]
Let $f:\mathcal{X}^M \to\mathbb{R}$ satisfy the bounded difference property with bounds $\{c_1,\dots,c_M\}$, and a random vector $\vb*{X} = \(X_1,\dots,X_M\)$ taking values in $\mathcal{X}^M$, then $\forall\epsilon>0$
$$
\mathbb{P}\(\abs{f(\vb*{X})-\expectt{\vb*{X}}{f(\vb*{X})}}\geq \epsilon \)\leq 2\mathrm{exp}\(-\frac{2\epsilon^2}{\sum_{i=1}^M c^{2}_{i}}\).
$$
\end{theorem}
We will show that the quantities we want to estimate satisfy the bounded difference property and apply the McDiarmid's inequality to prove that our previous sampling is efficient.
In the following we define
\bea\label{eq:sample-cost-function}
f(\vb*{x}) = \tr{\Ucal(\vb*{x})(\rhohat)\O},
\eea
where $\O$ is the cost function observable defined in the main text and, as in the previous section, $\mathcal{U}(\vb*{x})$ the unitary channel associated to a given Clifford approximant circuit that is completely specified by a discrete vector $\vb*{x} = (x_1,\dots,x_i,\dots,x_M)\in\{1,\dots,m\}^M$. By the H{\"o}lder inequality~\cite{baumgartner2011, watrous2018}, $f$ is upper-bounded:
\bea\label{eq:fx-bound}
\abs{f(\vb*{x})}\leq \norm{\rhohat}_{1}\norm{\O}_{\infty}
\eea
where $\norm{A}_1, \norm{A}_{\infty}$ are respectively the Schatten-1 norm and the spectral norm~\cite{watrous2018}. Remark that $\norm{\rhohat}_{1}=1$ for $\rhohat$ is a density operator. Defining a second vector for which only the $i$-th component is changed $\vb*{x'}=(x_1,\dots,x^{'}_i,\dots,x_M)$, we get by using the triangle inequality
\bea
\abs{f(\vb*{x}) - f(\vb*{x^{'}})} &\leq \abs{f(\vb*{x})}+\abs{f(\vb*{x'})}\\
&\leq 2\norm{\O}_{\infty}.
\eea
Hence $f$ satisfies the bounded difference property with $c_i = c = 2\norm{\O}_{\infty}$, and we can apply McDiarmid's inequality, which gives almost the desired result. To go further, we define
\bea
f_K(\vb*{x}_1,\dots,\vb*{x}_K) =&\; \sum_{j=1}^K f(x_{j1},\dots,x_{jM})\\
=&\; \sum_{j=1}^K \tr{\Ucal(\vb*{x}_j)(\rhohat)\O}\\
=&\; K\tr{\hat{\Phi}(\rhohat)\O}.
\eea
Clearly, $f_K$ satisfies the bounded difference property with the same bound $c$ [to see this, we take all $x_{ij}$ equal except for $x_{kl}$, and it follows that the difference $f_K(\vb*{x}_1,\dots,\vb*{x}_K)-f_K(\vb*{x}^{'}_1,\dots,\vb*{x}^{'}_K)$ is simply $f(\vb*{x}_k)-f(\vb*{x}^{'}_k)$]. Thus McDiarmid's inequality applies to $f_K$, which is a function of $KM$ parameters:
\bea
    \mathbb{P}(&\abs{f_K(\vb*{X})-\expectt{\vb*{X}}{f_K(\vb*{X})}}\geq K\epsilon )\\
    =&\; \mathbb{P}\(\abs{\frac{1}{K}f_K(\vb*{X})-\expectt{\vb*{X}}{\frac{1}{K}f_K(\vb*{X})}}\geq \epsilon \)\\
    =&\; \mathbb{P}\(\abs{\tr{\hat{\Phi}(\rhohat)\O}-\expectt{\thetav}{\tr{\Ucal(\thetav)(\rhohat)\O}}} \geq \epsilon \)\\
    \leq&\; 2\mathrm{exp}\(-\frac{2 K^2 \epsilon^2}{KM c^{2}}\)\\
    =&\; 2\mathrm{exp}\(-\frac{ K \epsilon^2}{2M\norm{\O}_{\infty}^2}\).
    \label{eq:exponential_concentration}
\eea
Therefore, choosing a precision $\epsilon>0$ and a probability $1-\delta\in\[0,1\]$ to meet this precision, we get
\bea
\mathbb{P}&\(\abs{\tr{\hat{\Phi}(\rhohat)\O}-\expectt{\thetav}{\tr{\Ucal(\thetav)(\rhohat)\O}}} \leq \epsilon  \)\\
&\hspace*{3cm}\geq 1-\delta
\eea
whenever the number of sampled Clifford circuits $K$ is 
\bea\label{eq:sample-number}
K \geq \frac{2}{\epsilon^2}\mathrm{log}\(\frac{2}{\delta}\)M\norm{\O}_{\infty}^2 = O\(M\)\,.
\eea
Note that in Eq.~\eqref{eq:exponential_concentration}, replacing the observable $\O$ by its normalized counterpart $\O/\norm{\O}$ with an associated precision $\tilde{\epsilon}$ gives the same scaling for $K$, as in that case $\tilde{\epsilon}=\epsilon/\norm{\O}$. Hence we can always work with a normalized observable. However, if one is interested in the scaling with the system size $\nqubits$, we have to consider a sequence of observables $\O_{\nqubits}$, whose norms can present a particular scaling in $\nqubits$, so the presence of the norm of $\O$ in Eq.~(\ref{eq:sample-number}) allows to keep track of this effect. In many situations of interest, the observables considered scale polynomialy in the system size, and so does $K$.
Finally, one can use the Gottesman-Knill theorem which states that for a Clifford unitary $\U$ and an observable $\O$ acting non-trivially on $N_O$ qubits, the expectation value $\tr{\dyad{0}^{\otimes n} \Ud \O\U}$ can be classically computed with a complexity polynomial in both $N_O$ and the number of qubits $\nqubits$~\cite{mitarai2022a}. Our scheme inherits this scaling and we can estimate the gradient variance $\variance{\thetav}{\partial_k C(\thetav)}$ for each $k$ with a classical computer in a complexity in $O\(\nqubits^p N_{O}^q M\)$ with $M$ the number of parameters in the variational quantum circuit.

\section{Sampling efficiency in the general case}
\label{appendix:sampling_efficiency_in_general_case_section}

In this section we extend the previous scheme to more general distributions. We first discuss in App.~\ref{app:sampling-nonconvex} the scaling of the sampling complexity with the convexity condition relaxed, i.e. where we no longer require the decomposition of the $2$-fold channel [Eq.~\eqref{eq:two-fold-sum}] to be a convex sum and only assume that the distribution of $\theta$ is even. Then, we study in~App.\ref{app:sample-eff-general} the case of an arbitrary distribution of the rotation angles, which is not necessarily symmetrically distributed. Finally, we show that our previous scheme still applies at the price of an exponential factor in the number of variational parameters $M$ in the number of Clifford approximant circuits to be sampled. Compared to a brute-force simulation, this method can be used to trade an exponential complexity in the system size for an exponential complexity in the number of variational parameters.

\subsection{Sampling efficiency in the nonconvex case}\label{app:sampling-nonconvex}
Here we consider distributions of rotation angle $\theta$ that are even, but do not satisfy the convexity condition of Eq.~\eqref{eq:second_order_condition_in_appendix}. In this case,  our decomposition of the $1$-fold channel remains convex while the $2$-fold channel becomes a nonconvex sum, hence the coefficients for the Clifford channels can no longer be interpreted as probabilities. We first show how one can still estimate such nonconvex sums via probabilistic sampling~\cite{piveteauQuasiprobabilityDecompositionsReduced2022}. 
Denoting 
\bea\label{eq:sum-quasiproba}
\expectt{\thetav}{\Ucal(\thetav)(\rhohat)} =&\;  \multicomp_{k=1}^{M}(\sum_{j=1}^{m}q_{kj}\Ucal_{kj}\circ\Wcal_k)(\rhohat)\,,
\eea
we hereby assume 
\bea\label{eq:quasiproba}
    q_{kj}\in\mathbb{R}\,,\quad \sum_{j=1}^M q_{kj}=1\,,\quad \forall k\,.
\eea
Defining
\bea\label{eq:def-gamma-k}
\gamma_k:=\Sigma_{j=1}^M\abs{q_{kj}}\,,\quad \ptil_{kj}:=\abs{q_{kj}}/\gamma_k\,,
\eea
Eq.~\eqref{eq:sum-quasiproba} can be rewritten in terms of convex sums: 
\bea
\expectt{\thetav}{\Ucal(\thetav)} =&\;  \multicomp_{k=1}^{M}\sum_{j=1}^{m}\ptil_{kj}\left[\gamma_k\sgn(q_{kj})\right]\Ucal_{kj}\circ\Wcal_k\,.
\eea
Similar to App.~\ref{appendix:detail-mapping}, we now define the random vector $\vbXtil = (\Xtil_1, \dots, \Xtil_M) \in \{1, \dots, m\}^M$, with probabilities $\mathbb{P}(\Xtil_k = j) = \ptil_{kj}$, and the rescaled random unitary channel $\Util(\vbXtil)$ through
\bea
\Util(j_1,\dots,j_M) = \multicomp_{k=1}^{M}\left[\gamma_k\sgn(q_{k{j_k
}})\right]\Ucal_{kj_k}\circ\Wcal_k\,.
\eea
Therefore, we recover the form of an expectation value similar to Eq.~\eqref{eq:expect-X}:
\bea
\expectt{\thetav}{\Ucal(\thetav)(\rhohat)} =&\; \expectt{\vbXtil}{\Util(\vbXtil)(\rhohat)}\,.
\eea
This allows us to apply the same arguments as in App.~\ref{appendix:sampling_proof} by considering the function
\bea
    \ftil(\vb*{x})=\Tr\left[\Util(\vb*{x})(\rhohat)\O\right]
\eea
instead of $f(\vb*{x})$ defined in Eq.~\eqref{eq:sample-cost-function}. The function bound~\eqref{eq:fx-bound} should be rescaled accordingly:
\bea
\abs{\ftil(\vb*{x})}\leq \gamma \norm{\O}_{\infty}\,,
\eea
where the scaling factor is defined as
\bea\label{eq:def-gamma}
\gamma:=\prod_{k=1}^M \gamma_k\,.
\eea
The number of sampled Clifford circuits previously derived in Eq.~\eqref{eq:sample-number} should therefore be scaled with the same factor:
\bea\label{eq:sample-number-nonconvex}
K \geq \frac{2}{\epsilon^2}\mathrm{log}\(\frac{2}{\delta}\)\gamma M\norm{\O}_{\infty}^2\,.
\eea
Note that the factor $\gamma_k\geq 1$ can be regarded as a measure of ``nonconvexity'' in the decomposition of the $k$-th channel. In the case of a convex sum, where $q_{kj}>0,~\forall k,j$, the scaling factor is simply $\gamma=1^M=1$ and we recover the previous results. 

We now show that $\gamma_k$ is upper-bounded. Following our discussion in App.~\ref{appendix:k-fold_channel_rz}, it suffices to consider the $2$-fold channel for a single-qubit Z-rotation, where the decomposition can be possibly nonconvex. Without loss of generality, let us rewrite Eq.~\eqref{eq:two-fold-sum} as
\bea\label{eq:two-fold-sym-wlog}
\Phi_{Z}^{(2)}\(\rhohat\) =&\; q_{k1}\rhohat + q_{k2}\(Z\otimes Z\)\rhohat \(Z\otimes Z\) \\&+q_{k3}CZ \rhohat CZ+q_{k4}CZ_X \rhohat CZ_X
\eea
for some $k$, where 
\bea
    q_{k1}=&\;\expectt{\theta}{\frac{1}{4}\(1+\cos{2\theta}+2\cos{\theta}\)}\,,\\
    q_{k2}=&\;\expectt{\theta}{\frac{1}{4}\(1+\cos{2\theta}-2\cos{\theta}\)}\,,\\
    q_{k3}=&\;q_{k4}=\expectt{\theta}{\frac{1}{4}\(1-\cos{2\theta}\)}\,.
\eea
Defining the non-negative function
\bea
    \varphi(\theta):=&\;~\abs{\frac{1}{4}\(1+\cos{2\theta}+2\cos{\theta}\)} \\&+ \abs{\frac{1}{4}\(1+\cos{2\theta}-2\cos{\theta}\)} \\&+ 2\times \abs{\frac{1}{4}\(1-\cos{2\theta}\)}\,,
\eea
We then get
\bea\label{eq:two-fold-sym-gamma-k}
    \gamma_k&=\abs{q_{k1}}+\abs{q_{k2}}+\abs{q_{k3}}+\abs{q_{k4}}\\
        &\leq \expectt{\theta}{\varphi(\theta)}\\
        &\leq \expectt{\theta}{\sup_{\theta'}\varphi(\theta')}\\
        &=\sup_{\theta'}\varphi(\theta')\\
        &=\dfrac{5}{4}\,.
\eea
Here the function $\varphi(\theta)$ reaches its maximum for $\theta=\pm\frac{\pi}{3},\pm\frac{2\pi}{3}$. Therefore, the factor $\gamma_k$ reaches its upper bound $\frac{5}{4}$ if the distribution of $\theta$ is a sum of Dirac-delta distributions peaked at $\theta=\pm\frac{\pi}{3}$ and/or $\theta=\pm\frac{2\pi}{3}$, in which case we obtain the worst-case scaling of the number of sampled Clifford circuits~\eqref{eq:sample-number-nonconvex}:
\bea\label{eq:sample-number-nonconvex-bound}
    K &\geq \frac{2}{\epsilon^2}\mathrm{log}\(\frac{2}{\delta}\)\gamma M\norm{\O}_{\infty}^2=O(\gamma M)\,, \\
    \gamma &\leq \left(\dfrac{5}{4}\right)^M\,.
\eea
Combining the above result with the Gottesman-Knill theorem, for a cost-function observable $\hat{O}$ acting non-trivially on $N_O$ qubits, our scheme implies a complexity of at most $O\(\nqubits^p N_O^q (\frac{5}{4})^M M\)$ for the estimation of gradient variance $\variance{\thetav}{\partial_k C(\thetav)}$ for each $k$ on a classical computer in the general scenario, where $\nqubits$ is the number of qubits, $M$ is the number of parameters in the variational quantum circuit and $p,q$ are some constants inherited from the Gottesman-Knill theorem.

\subsection{Sampling efficiency for the general case}\label{app:sample-eff-general}

In this section, we extend our scheme to the most generic case, by considering an arbitrary probability distribution for the rotation angles $\thetav$, and derive the corresponding sampling complexity. As before, one needs only to consider the one- and two-fold channels for a single-qubit $Z$-rotation gate. In what follows, let us denote $\expectt{\theta}{\rme^{\rmi\theta}}:=r_1+\rmi s_1$ and $\expectt{\theta}{\rme^{2\rmi\theta}}:=r_2+\rmi s_2$. Note that $r_1=\expectt{\theta}{\cos\theta}$ and $r_2=\expectt{\theta}{\cos{2\theta}}$ are defined in the same way as for the symmetric case before, while $s_1=\expectt{\theta}{\sin\theta}$ and $s_2=\expectt{\theta}{\sin{2\theta}}$ are in general nonzero since we no longer assume the distribution of $\theta$ to be even.

\subsubsection{\texorpdfstring{$1$}{1}-fold channel}
The expression of the $1$-fold channel for a single-qubit $Z$-rotation is given by Eq.~\eqref{eq:one-fold-raw-sum}, which we develop below \textit{without} assuming an even distribution in $\theta$. We get:
\bea\label{eq:one-fold-asym}
\,&\expectt{\theta}{\hat{R}_{Z}(\theta)\rhohat \hat{R}^{\dagger}_{Z}(\theta)}\\ =&\; \hat{\Pi}_0 \rhohat \hat{\Pi}_0 + \hat{\Pi}_1 \rhohat \hat{\Pi}_1\\ &+ \expectt{\theta}{e^{\rmi\theta}}\hat{\Pi}_1\rhohat\hat{\Pi}_0+\expectt{\theta}{e^{-\rmi\theta}}\hat{\Pi}_0\rhohat\hat{\Pi}_1\\
=&\; \hat{\Pi}_0 \rhohat \hat{\Pi}_0 + \hat{\Pi}_1 \rhohat \hat{\Pi}_1 \\&+ (r_1+\rmi s_1)\hat{\Pi}_1\rhohat\hat{\Pi}_0+(r_1-\rmi s_1)\hat{\Pi}_0\rhohat\hat{\Pi}_1\\
=&\; \dfrac{1+r_1}{2}\Ecal[\id](\rhohat)+\dfrac{1-r_1}{2}\Ecal[\Zhat](\rhohat)\\
&+\dfrac{s_1}{2}\Ecal[\Shat^\dagger](\rhohat)-\dfrac{s_1}{2}\Ecal[\Shat](\rhohat)\,,
\eea
where $\Shat = \hat{\Pi}_0+\rmi\hat{\Pi}_1$ is the phase gate, and one can use this definition together with Eq.~\eqref{eq:eq-1fold-terms} to verify the equation above.

Here, the parameter $s_1$ can be understood as a measure of asymmetry in the probability distribution of $\theta$. In the symmetric case, we have $s_1=0$ and the sum reduces to the convex one given by Eq.~\eqref{eq:one-fold-convex-sum}. Following the same procedure as in App.~\ref{app:sampling-nonconvex}, this (possibly nonconvex) linear combination of Clifford channels can be estimated via sampling, and the number of required samples should be scaled, according to the nonconvexity of the sum, by a factor $\gamma=\Pi_{k=1}^M \gamma_k$ [see definition in Eqs.~\eqref{eq:sum-quasiproba}-\eqref{eq:def-gamma-k} and Eq.~\eqref{eq:def-gamma}]. We now derive an upper bound for $\gamma^{(1)}_k$, the scaling factor associated to a single (the $k$-th) $1$-fold $Z$-rotation channel that can be decomposed in the form of Eq.~\eqref{eq:one-fold-asym} in general. We proceed by applying the same argument as in Eqs.~\eqref{eq:two-fold-sym-wlog}-\eqref{eq:two-fold-sym-gamma-k}:
\bea
    \gamma^{(1)}_k =&\; \abs{\dfrac{1+r_1}{2}}+\abs{\dfrac{1-r_1}{2}}+\abs{\dfrac{s_1}{2}}+\abs{-\dfrac{s_1}{2}}\\
    =&\;\abs{\expectt{\theta}{\dfrac{1+\cos{\theta}}{2}}}+\abs{\expectt{\theta}{\dfrac{1-\cos{\theta}}{2}}}\\&+\abs{\expectt{\theta}{\sin\theta}}\\
    \leq&\; \expectt{\theta}{\abs{\dfrac{1+\cos\theta}{2}}+\abs{\dfrac{1-\cos\theta}{2}}+\abs{\sin\theta}}\\
    \leq&\; \sup_{\theta}\left\{\abs{\dfrac{1+\cos\theta}{2}}+\abs{\dfrac{1-\cos\theta}{2}}+\abs{\sin\theta}\right\}\\
    =&\; 2\,.
\eea
This implies that the number of samples $K^{(1)}$ required for the estimation of the generic $1$-fold channel [see Eq.~\eqref{eq:sample-number-nonconvex-bound}] scales as
\bea\label{eq:sample-number-one-fold-asym-bound}
    K^{(1)} &\sim O(\gamma^{(1)}M)\,,\\
    \gamma^{(1)} &= \prod_{k=1}^M\gamma^{(1)}_k\leq 2^M\,.
\eea
Note that the bound derived above depends on the specific choice of the Clifford channels in the decomposition. As the Clifford group does not form a linearly independent set, it should be possible to find a different decomposition that yields a different upper bound and further optimize the complexity.
\subsubsection{\texorpdfstring{$2$}{2}-fold channel}
The $2$-fold channel for a single-qubit $Z$-rotation is given by Eq.~\eqref{eq:two-fold-raw-sum}:
\bea
\Phi_{Z}^{(2)}(\rhohat) =&\; \expectt{\theta}{\Xi\rhohat\Xi^\dagger}+\expectt{\theta}{\Gamma_{\theta}\rhohat\Gamma_{\theta}^\dagger}\\&+\expectt{\theta}{\Gamma_{\theta}\rhohat\Xi^\dagger}+\expectt{\theta}{\Xi\rhohat\Gamma_{\theta}^\dagger}\,.
\eea
For a generic probability distribution of $\theta$, we have
\bea
     \expectt{\theta}{\Xi\rhohat\Xi^\dagger} =&\; \hat{\Pi}_{01}\rhohat\hat{\Pi}_{01}+\hat{\Pi}_{10}\rhohat\hat{\Pi}_{10}\\&+\hat{\Pi}_{01}\rhohat\hat{\Pi}_{10}+\hat{\Pi}_{10}\rhohat\hat{\Pi}_{01}\, ,\\
    \expectt{\theta}{\Gamma_{\theta}\rhohat\Gamma_{\theta}^\dagger} =&\; \hat{\Pi}_{00}\rhohat\hat{\Pi}_{00}+\hat{\Pi}_{11}\rhohat\hat{\Pi}_{11}\\&+r_2\(\hat{\Pi}_{00}\rhohat\hat{\Pi}_{11}+\hat{\Pi}_{11}\rhohat\hat{\Pi}_{00}\)\\
    &+\rmi s_2\(\hat{\Pi}_{11}\rhohat\hat{\Pi}_{00}-\hat{\Pi}_{00}\rhohat\hat{\Pi}_{11}\)\,,\\
    \expectt{\theta}{\Gamma_{\theta}\rhohat\Xi^\dagger} =&\; r_1\(\hat{\Pi}_{00}+\hat{\Pi}_{11}\)\rhohat\(\hat{\Pi}_{01}+\hat{\Pi}_{10}\)\\ &+ \rmi s_1\(\hat{\Pi}_{11}-\hat{\Pi}_{00}\)\rhohat\(\hat{\Pi}_{01}+\hat{\Pi}_{10}\)\,,\\
     \expectt{\theta}{\Xi\rhohat\Gamma_{\theta}^\dagger} =&\; \expectt{\theta}{\Gamma_{\theta}\rhohat\Xi^\dagger}^\dagger\,.
\eea
As one can verify, the Choi representation of the above terms are all diagonal, so that their sum can be represented via the $M$ matrix as before:
\bea
M&(\Phi_{Z}^{(2)})\\ =& 
    \left(
\begin{array}{cccc}
 1 & r_1+\rmi s_1 & r_1+\rmi s_1 & r_2+\rmi s_2 \\
 r_1-\rmi s_1 & 1 & 1 & r_1+\rmi s_1 \\
 r_1-\rmi s_1 & 1 & 1 & r_1+\rmi s_1 \\
 r_2-\rmi s_2 & r_1-\rmi s_1 & r_1-\rmi s_1 & 1 \\
\end{array}
\right)\,.
\eea
This can again be decomposed as a weighted sum of the channels $\Ecal[\id]$, $\Ecal[\Zhat\otimes\Zhat]$, $\Ecal[\Shat\otimes\Shat]$ and $\Ecal[\Shat^\dagger\otimes\Shat^\dagger]$ given in Eq.~\eqref{eq:M_matrices_gates} and of the following Clifford channels:
\bea
M&\left(\Ecal\left[\id\otimes\Shat\right]\right)=
\left(
\begin{array}{cccc}
 1 & \rmi & 1 & \rmi\\
 -\rmi & 1 & -\rmi & 1 \\
 1 & \rmi& 1 & \rmi \\
 -\rmi & 1 & -\rmi & 1 \\
\end{array}
\right)\,,\\
M&\left(\Ecal\left[\Shat\otimes\id\right]\right)=
\left(
\begin{array}{cccc}
 1 & 1 & \rmi & \rmi\\
 1 & 1 & \rmi & \rmi \\
 -\rmi & -\rmi& 1 & 1 \\
 -\rmi & -\rmi & 1 & 1 \\
\end{array}
\right)\,,\\
M&\left(\Ecal\left[\id\otimes\Shat^{\dagger}\right]\right)=
\left(
\begin{array}{cccc}
 1 & -\rmi & 1 & -\rmi\\
 \rmi & 1 & \rmi & 1 \\
 1 & -\rmi& 1 & -\rmi \\
 \rmi & 1 & \rmi & 1 \\
\end{array}
\right)\,,\\
M&\left(\Ecal\left[\Shat^{\dagger}\otimes\id\right]\right)=
\left(
\begin{array}{cccc}
 1 & 1 & -\rmi & -\rmi\\
 1 & 1 & -\rmi & -\rmi \\
 \rmi & \rmi& 1 & 1 \\
 \rmi & \rmi & 1 & 1 \\
\end{array}
\right)\,,\\
M&\left(\Ecal\left[\Zhat\otimes\Shat\right]\right)=
\left(
\begin{array}{cccc}
 1 & \rmi & -1 & -\rmi\\
 -\rmi& 1 & \rmi & -1 \\
 -1 & -\rmi& 1 & \rmi \\
 \rmi & -1 & -\rmi& 1 \\
\end{array}
\right)\,,\\
M&\left(\Ecal\left[\Shat\otimes\Zhat\right]\right)=
\left(
\begin{array}{cccc}
 1 & -1 & \rmi & -\rmi\\
 -1 & 1 & -\rmi & \rmi \\
 -\rmi & \rmi & 1 & -1 \\
 \rmi & -\rmi & -1 & 1 \\
\end{array}
\right)\,,\\
M&\left(\Ecal\left[\Zhat\otimes\Shat^{\dagger}\right]\right)=
\left(
\begin{array}{cccc}
 1 & -\rmi & -1 & \rmi\\
 \rmi& 1 & -\rmi & -1 \\
 -1 & \rmi& 1 & -\rmi \\
 -\rmi & -1 & \rmi& 1 \\
\end{array}
\right)\,,\\
M&\left(\Ecal\left[\Shat^{\dagger}\otimes\Zhat\right]\right)=
\left(
\begin{array}{cccc}
 1 & -1 & -\rmi & \rmi\\
 -1 & 1 & \rmi & -\rmi \\
 \rmi & -\rmi & 1 & -1 \\
 -\rmi & \rmi & -1 & 1 \\
\end{array}
\right)\,.
\eea
Note that the channels listed above are all diagonal in the Choi representation and hence the $M$ matrices capture all their nonzero entries. Following the same reasoning as in App.\ref{appendix:k-fold_channel_rz}, we solve a linear system to obtain the following decomposition:
\bea
\Phi_{Z}^{(2)}(\rhohat)=&\; \dfrac{s_2}{8}\left(\Ecal\left[\Shat\otimes\id\right](\rhohat)+\Ecal\left[\id\otimes\Shat\right](\rhohat)\right)\\&+\dfrac{s_2}{8}\left(\Ecal\left[\Zhat\otimes\Shat^{\dagger}\right](\rhohat)+\Ecal\left[\Shat^{\dagger}\otimes\Zhat\right](\rhohat)\right)\\ &- \dfrac{s_2}{8}\left(\Ecal\left[\Shat^{\dagger}\otimes\id\right](\rhohat)+\Ecal\left[\id\otimes\Shat^{\dagger}\right](\rhohat)\right)\\ &-\dfrac{s_2}{8}\left(\Ecal\left[\Zhat\otimes\Shat\right](\rhohat)+\Ecal\left[\Shat\otimes\Zhat\right](\rhohat)\right)\\
&+ \dfrac{1+r_2+2r_1}{4}\Ecal\left[\id\right](\rhohat)\\
&+ \dfrac{1+r_2-2r_1}{4}\Ecal\left[ \Zhat\otimes\Zhat \right](\rhohat)\\
&+ \dfrac{1-r_2+2s_1}{4}\Ecal\left[ \Shat\otimes\Shat \right](\rhohat)\\
&+ \dfrac{1-r_2-2s_1}{4}\Ecal\left[\Shat^{\dagger}\otimes\Shat^{\dagger}\right](\rhohat)\,.
\eea
%\vfill\null
\begin{remark}
Denoting $\CN=\hat{\Pi}_0\otimes\id+\hat{\Pi}_1\otimes\Xhat$ the CNOT gate and $\CNx:=(\Xhat\otimes\Xhat)\CN(\Xhat\otimes\Xhat)$ its conjugation by the $\Xhat\otimes\Xhat$ gate, we have
\bea
M&\left(\Ecal\left[\CN(\Shat\otimes\Shat)\CN\right]\right)\\ =&\;
\left(
\begin{array}{cccc}
 1 & \rmi & -1 & \rmi \\
 -\rmi& 1 & \rmi & 1 \\
 -1 & -\rmi& 1 & -\rmi\\
 -\rmi& 1 & \rmi & 1 \\
\end{array}
\right)\,,\\
M&\left(\Ecal\left[\CNx(\Shat\otimes\Shat)\CNx\right]\right) \\=&\;
\left(
\begin{array}{cccc}
 1 & -\rmi& 1 & \rmi \\
 \rmi & 1 & \rmi & -1 \\
 1 & -\rmi& 1 & \rmi \\
 -\rmi& -1 & -\rmi& 1 \\
\end{array}
\right).
\eea
Again, by solving a linear system one finds another decomposition of the two-fold channel that involves the above channels, namely:
\bea
\Phi_{Z}^{(2)}(\rhohat)=&\; \dfrac{s_2}{4}\Ecal\left[\CN(\Shat\otimes\Shat)\CN\right](\rhohat)\\ &- \dfrac{s_2}{4}\Ecal\left[\Zhat\otimes\Shat\right](\rhohat)\\ &+ \dfrac{s_2}{4}\Ecal\left[ \CNx(\Shat\otimes\Shat)\CNx \right](\rhohat)\\ &-\dfrac{s_2}{4}\Ecal\left[ \id\otimes\Shat^{\dagger} \right](\rhohat) \\
&+ \dfrac{1+r_2+2r_1}{4}\Ecal\left[\id\right](\rhohat)\\
&+ \dfrac{1+r_2-2r_1}{4}\Ecal\left[ \Zhat\otimes\Zhat \right](\rhohat)\\
&+ \dfrac{1-r_2+2s_1}{4}\Ecal\left[ \Shat\otimes\Shat \right](\rhohat)\\
&+ \dfrac{1-r_2-2s_1}{4}\Ecal\left[\Shat^{\dagger}\otimes\Shat^{\dagger}\right](\rhohat)\,.
\eea
\end{remark}

Similar to our treatment with the $1$-fold channel, let us derive an upper bound for $\gamma^{(2)}_k$, the scaling factor for the number of samples required for the estimation of the generic $2$\nobreakdash-fold $k$\nobreakdash-th $Z$\nobreakdash-rotation channel:
\bea
    \gamma^{(2)}_k =&\; 4\abs{\dfrac{s_2}{8}}+ 4\abs{-\dfrac{s_2}{8}}\\&+ \abs{\dfrac{1+r_2+2r_1}{4}} + \abs{\dfrac{1+r_2-2r_1}{4}}\\
    &+ \abs{\dfrac{1-r_2+2s_1}{4}}+ \abs{\dfrac{1-r_2-2s_1}{4}}\\
    =&\; \abs{\expectt{\theta}{\sin 2\theta}} \\
    &+ \abs{\expectt{\theta}{\dfrac{1+\cos 2\theta + 2 \cos\theta}{4}}} \\
    &+ \abs{\expectt{\theta}{\dfrac{1+\cos 2\theta - 2 \cos\theta}{4}}} \\
    &+  \abs{\expectt{\theta}{\dfrac{1-\cos 2\theta + 2 \sin\theta}{4}}} \\
    &+  \abs{\expectt{\theta}{\dfrac{1-\cos 2\theta - 2 \sin\theta}{4}}}.\\
    \leq&\; \sup_{\theta}\left\{ \abs{\sin 2\theta} \right. \\
    &+ \abs{\dfrac{1+\cos 2\theta + 2 \cos\theta}{4}} \\
    &+ \abs{\dfrac{1+\cos 2\theta - 2 \cos\theta}{4}} \\
    &+ \abs{\dfrac{1-\cos 2\theta + 2 \sin\theta}{4}} \\
    &+\left. \abs{\dfrac{1-\cos 2\theta - 2 \sin\theta}{4}} \right\}\\
    =&\; 1+\sqrt{2}\,.
\eea
This implies that the number of samples $K^{(2)}$ required for the estimation of the generic $2$-fold channel scales as
\bea
    K^{(2)}&\sim O(\gamma^{(2)}M)\,,\\
    \gamma^{(2)} &=\prod_{k=1}^M\gamma^{(2)}_k \leq (1+\sqrt{2})^M\,,
\eea
which is dominant over the complexity of the estimation of the $1$-fold channel [Eq.~\eqref{eq:sample-number-one-fold-asym-bound}] since $1+\sqrt{2}>2$. 

Again, combining the above result with the Gottesman-Knill theorem, for a cost-function observable $\hat{O}$ acting non-trivially on $N_O$ qubits, our scheme implies a complexity of no more than $O\(\nqubits^p N_O^q (1+\sqrt{2})^M M\)$ for the estimation of gradient variance $\variance{\thetav}{\partial_k C(\thetav)}$ for each $k$ on a classical computer in the most generic case, where $\nqubits$ is the number of qubits, $M$ is the number of parameters in the variational ansatz and $p,q$ are some constants inherited from the Gottesman-Knill theorem.

\section{\texorpdfstring{$\boldsymbol{N}$}{N}-fold channel for a random Z rotation}
\label{appendix:N-fold_channel_generalization}
In this section we give a decomposition of the $N$-fold channel as a real sum of Clifford unitary channels. This allows us to extend our scheme to the estimation of $N-th$ order quantities with a complexity scaling polynomialy in both the number of variational parameter $M$ and the system size $\nqubits$ when the decomposition is convex, and exponential in $M$ otherwise. We give a \emph{sufficient} condition on the distribution of the random angle $\theta$ for the decomposition to be a convex one.

Recall that for any unitary $\U$ we defined $\Ecal\left[\U\right](\rhohat) := \U\rhohat\Ud$. In Eq.~\eqref{eq:one-fold-asym} we obtained a decomposition of the $1$-fold channel of a Z-rotation in terms of Clifford unitary channels for a generic distribution of the random angle, namely
\bea\label{eq:one-fold-asym-again}
\,&\expectt{\theta}{\hat{R}_{Z}(\theta)\rhohat \hat{R}^{\dagger}_{Z}(\theta)}\\
=& \dfrac{1+r_1}{2}\Ecal[\id](\rhohat)+\dfrac{1-r_1}{2}\Ecal[\Zhat](\rhohat)\\
&+\dfrac{s_1}{2}\Ecal[\Shat^\dagger](\rhohat)-\dfrac{s_1}{2}\Ecal[\Shat](\rhohat).
\eea
More generally, we have that
\bea
\hat{R}_{Z}(\theta)\rhohat \hat{R}^{\dagger}_{Z}(\theta)=&\; \dfrac{1+\cos{\theta}}{2}\Ecal[\id](\rhohat)\\&+\dfrac{1-\cos{\theta}}{2}\Ecal[\Zhat](\rhohat)\\
&+\dfrac{\sin{\theta}}{2}\Ecal[\Shat^\dagger](\rhohat)\\&-\dfrac{\sin{\theta}}{2}\Ecal[\Shat](\rhohat),
\eea
for any $\theta\in\mathbb{R}$. This can be seen as a consequence of Eq.~(\ref{eq:one-fold-asym-again}) for a Dirac probability measure centered at $\theta$. On can directly generalize this equation to obtain an expression of the $N$-fold channel as a real sum of Clifford unitary channels, as
\bea
\hat{R}^{\otimes N}_{Z}(\theta)\rhohat \hat{R}^{\otimes N \dagger}_{Z}(\theta) =&\;\hspace{-0.5cm}\sum_{I=(i_1,\dots,i_n)}\hspace{-0.3cm}\lambda_I(\theta)\Ecal\[\otimes_{j=1}^{N}\U_{i_j}\](\rhohat)
\label{eq:decomp_in_product_channels}
\eea
where the sum goes over all the multi-indices $I=\(i_1,\dots,i_N\)\in\{0,1,2,3\}$, and $\U_0=\id, \U_1=\Zhat, \U_2 = \Shat$ and $\U_3=\Shat^{\dagger}$. The coefficient $\lambda_I(\theta)_I$ for a multi-index $I$ representing a product of numbers $m_i$ of the $\U_i$ gates is given by
\bea
\lambda_{I}(\theta) = \frac{1}{2^N}&\(1+\cos\theta\)^{m_0}\(1-\cos\theta\)^{m_1}\\&\sin^{m_2}\(-\theta\)\sin^{m_3}\(\theta\),
\eea
with $m_0+m_1+m_2+m_3 = N$.
As a result, the $N$-fold channel is given by a real combination of $4^N$ unitary Clifford channels that are composed of products of the gates $\id,\Zhat,\Shat$ and $\Shat^{\dagger}$. This gives us a sufficient condition for the $N$-fold channel to be a convex sum of Clifford unitary channels, namely it suffices that the expectation values of all the coefficient $\expectt{\theta}{\lambda_I(\theta)}$ be positive. 

Although this condition is sufficient, it is \emph{not necessary}. In particular, in the case of the $2$-fold channel, the expectation of coefficients associated to the multi-indices $(2,3)$ and $(3,2)$ is given by $\expectt{\theta}{-\sin^2{\theta}}$, which is always negative. However, we proved that a convex decomposition exists for the uniform distribution. This is due to the fact that the decomposition of Eq.~\eqref{eq:decomp_in_product_channels} is not unique. In fact, the family of channels 
\bea
\mathcal{P} := \{\Ecal[\U\otimes\V]\,:\; \U,\V\in\{\id,\Zhat,\Shat,\Shat^{\dagger}\}\}
\eea
is not linearly independent. Consider two single qubit unitaries $\U$ and $\V$ that are diagonal in the computational basis. As we are free to chose the global phase of these unitaries, we can always write them as $\U = e^{\rmi\theta_U/2}\hat{\Pi}_0+e^{-\rmi\theta_U/2}\hat{\Pi}_1$ and $\V = e^{\rmi\theta_V/2}\hat{\Pi}_0+e^{-\rmi\theta_V/2}\hat{\Pi}_1$. We saw in App.~\ref{appendix:k-fold_channel_rz} that the product unitary $\U\otimes\V$ can be represented by the diagonal of the associated Choi matrix, written as a 4-by-4 matrix M:
\bea
M(\Ecal[\U\otimes\V])=\hspace{4.5cm}&\\ 
    \begin{pmatrix}
    1 & e^{-\rmi\theta_V} & e^{-\rmi\theta_U} & e^{-\rmi(\theta_U+\theta_V)}
    \\
    e^{\rmi\theta_V} & 1 & e^{-\rmi(\theta_U-\theta_V)} & e^{-\rmi\theta_U} \\
    e^{\rmi\theta_U} & e^{\rmi(\theta_U-\theta_V)} & 1 & e^{-\rmi\theta_V} \\
    e^{\rmi(\theta_U+\theta_V)} & e^{\rmi\theta_U} & e^{\rmi\theta_V} & 1
    \end{pmatrix}&.\\
\eea
This shows that for a tensor product of single-qubit unitaries, the matrices M in the basis $((00),(01),(10),(11))$ are symmetric with respect to the anti-diagonal transposition. Therefore, the channels in $\mathcal{P}$ belong to a real vector space of dimension 9 (1 dimension for the diagonal, $2\times3$ dimensions for the complex exponentials of the first row, and 2 dimensions for the third term of the second row). As there are 16 channels in $\mathcal{P}$, the family is not linearly independent. The condition that all the $\expectt{\theta}{\lambda_I(\theta)}$ be positive is clearly too restrictive. One way to extend it to find back the condition we previously derived is to use the fact that
\bea
\,&\Ecal[\id\otimes\id]+\Ecal[\Zhat\otimes\Zhat]\\&+\Ecal[\id\otimes\Zhat]+\Ecal[\Zhat\otimes\id]\\&=\Ecal[\Shat\otimes\Shat]+\Ecal[\Shat^{\dagger}\otimes\Shat^{\dagger}]\\&+\Ecal[\Shat\otimes\Shat^{\dagger}]+\Ecal[\Shat^{\dagger}\otimes\Shat]
\eea
to absorb the $\expectt{\theta}{-\sin^2{\theta}}$ factors into the coefficients associated to other channels. 
\begin{remark}
To obtain the previous relation, we used the following channels:
\bea
M&\left(\Ecal\left[\id\otimes\Zhat\right]\right)=
\left(
\begin{array}{cccc}
 1 & -1 & 1 & -1\\
 -1& 1 & -1 & 1 \\
 1 & -1& 1 & -1 \\
 -1 & 1 & -1& 1 \\
\end{array}
\right)\,,\\
M&\left(\Ecal\left[\Zhat\otimes\id\right]\right)=
\left(
\begin{array}{cccc}
 1 & 1 & -1 & -1\\
 1& 1 & -1 & -1 \\
 -1 & -1& 1 & 1 \\
 -1 & -1 & 1& 1 \\
\end{array}
\right)\,,\\
M&\left(\Ecal\left[\Shat\otimes\Shat^{\dagger}\right]\right)=
\left(
\begin{array}{cccc}
 1 & -\rmi & \rmi & 1\\
 \rmi & 1 & -1 & \rmi \\
 -\rmi & -1& 1 & -\rmi \\
 1 & -\rmi & \rmi & 1 \\
\end{array}
\right)\,,\\
M&\left(\Ecal\left[\Shat^{\dagger}\otimes\Shat\right]\right)=
\left(
\begin{array}{cccc}
 1 & \rmi & -\rmi & 1\\
 -\rmi & 1 & -1 & -\rmi \\
 \rmi & -1& 1 & \rmi \\
 1 & \rmi & -\rmi & 1 \\
\end{array}
\right)\,.
\eea
\end{remark}
We showed that the $N$-fold channel associated to Z-rotations can always be decomposed as a real linear combination of Clifford unitary channels. However, it remains an open problem to find necessary and/or sufficient conditions under which the $N$-fold channel can be decomposed a convex combination of Clifford unitaries, i.e. conditions under which the $N$-fold channel is a Clifford mixed-unitary channel. The knowledge of such conditions could allow to extend the scheme proposed in this work to ansätze with correlated rotation parameters.
\section{Example of first and second order Clifford approximant circuits for a simple ansatz}
\label{appendix:example_of_Clifford_approximants}
\begin{figure}[t]
    \includegraphics[width=1.05\columnwidth]{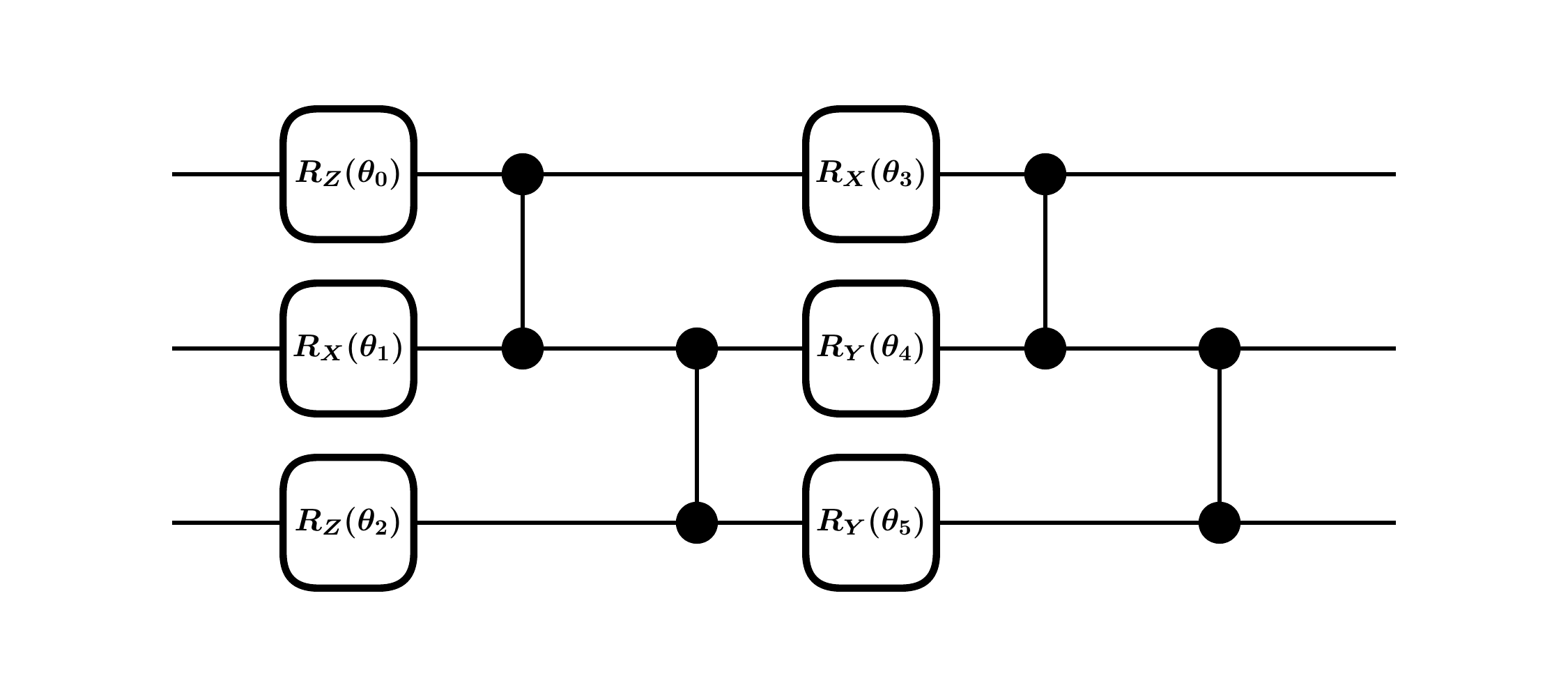}
    \caption{Initial variational circuit with random rotation angles.}
    \label{fig:initial_circuit}
\end{figure}

\begin{figure}[t]
    \centering
    \includegraphics[width=0.9\columnwidth]{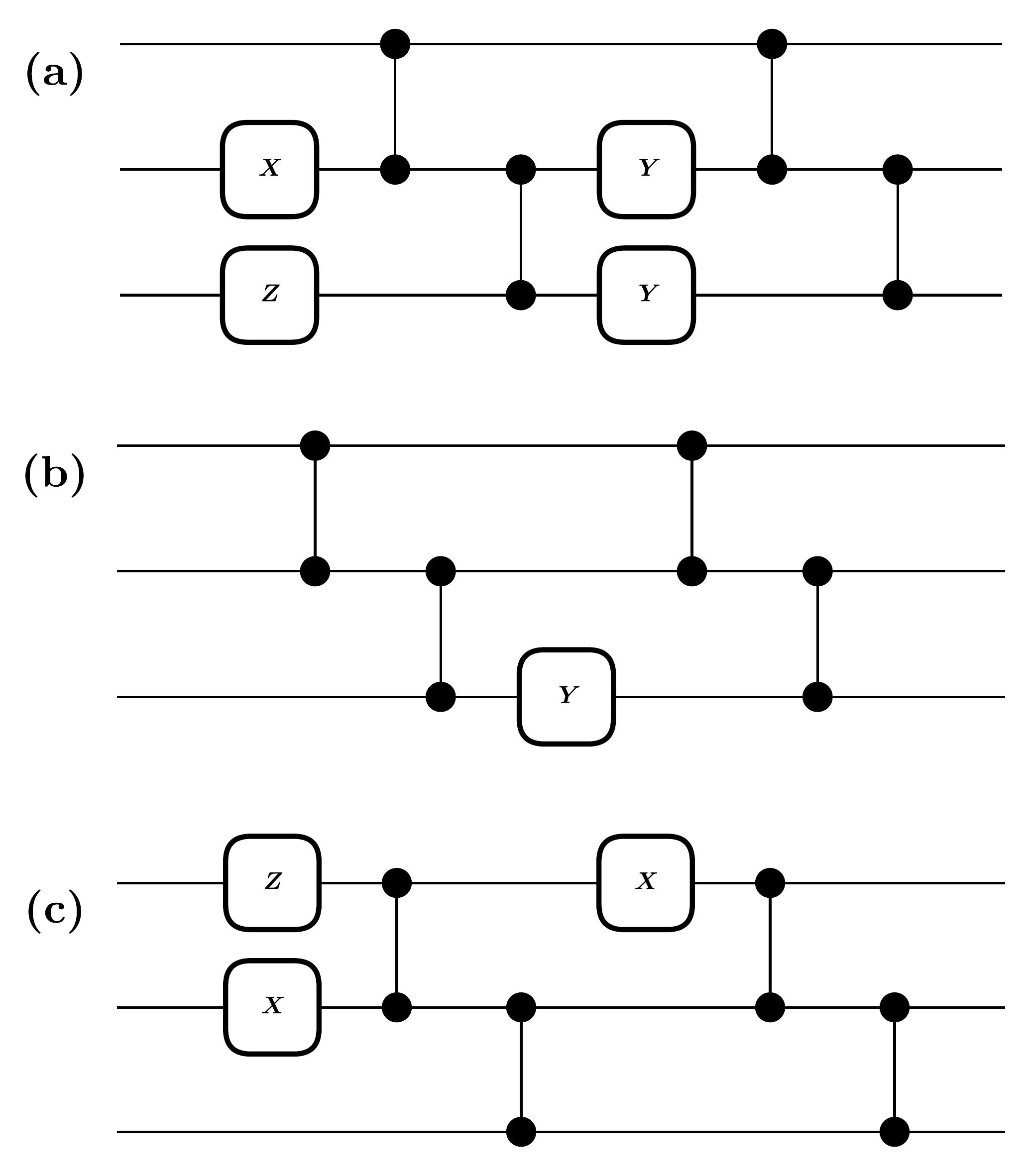}
    \caption{Examples of first-order Clifford approximant circuits for the ansatz of Fig.~\ref{fig:initial_circuit}. Assuming the probability distribution of the angles is even, we replace each rotation by a Clifford gate that is sampled according to Eq.~\eqref{eq:one-fold-convex-sum}.}
    \label{fig:first_order_circuits}
\end{figure}

\begin{figure*}
    \includegraphics[width=0.75\textwidth]{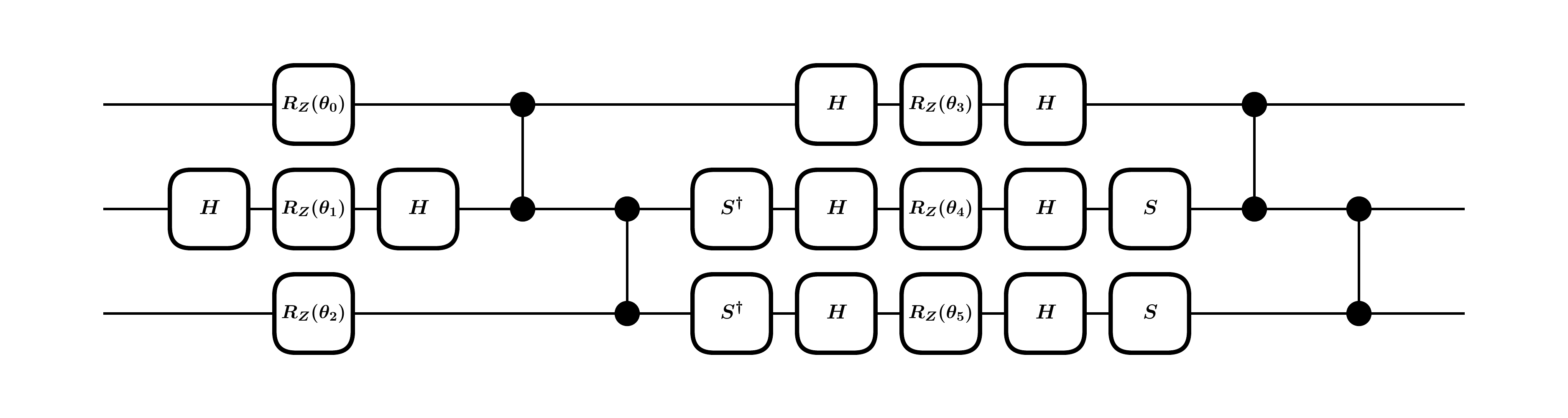}
    \caption{Equivalent form of the initial circuit with Z-rotations only.}
    \label{fig:rz_circuit}
\end{figure*}

In this appendix we provide a sample of Clifford approximant circuits for the estimation of $\expectt{\thetav}{C(\thetav)}$ and $\expectt{\thetav}{C(\thetav)^2}$ for the simple circuit depicted in Fig.~\ref{fig:initial_circuit}. 
The generalisation to Clifford approximants for other quantities, such as the expectation of the squared gradient, can be derived from that example as it suffices to introduce the adequate Clifford gates to the fixed layers to obtain the right estimators (see Sec.~\ref{section:variational_problem} and \ref{appendix:first-and-second-order-quantities}). This circuit acts on three qubits and is composed of two layers of rotations that are alternated with fixed two-qubits Control-Z gates.
To obtain a first order approximant for these circuits it suffices to randomly replace each rotation by either the identity gate (a wire) or the Pauli gate corresponding to the direction of the concerned rotation gate. Three examples of first order Clifford approximant are represented in Fig.~\ref{fig:first_order_circuits}.
The second order approximant are derived by first mapping each rotation along X or Y to a rotation along Z, making use of the identities $\hat{X}=\hat{H}^{\dagger}\hat{Z}\hat{H}$ and $\hat{Y} = (\hat{S}\hat{H})\hat{Z}(\hat{S}\hat{H})^{\dagger}$ where $\hat{H},\hat{S}$ are respectively the Hadamard and phase gates. As a result we get the ansatz with layers of Z rotations alternated with fixed layers composed of Clifford gates represented on Fig.~\ref{fig:rz_circuit}. This circuit is then doubled vertically to give a circuit acting on six qubits. Finally, each pairs of rotations sharing the same angle is randomly replaced by two single-qubit gates according to the scheme of Fig.~\ref{fig:vqc_to_clifford_mapping_rules}.

\begin{figure*}
    \includegraphics[width=0.8\textwidth]{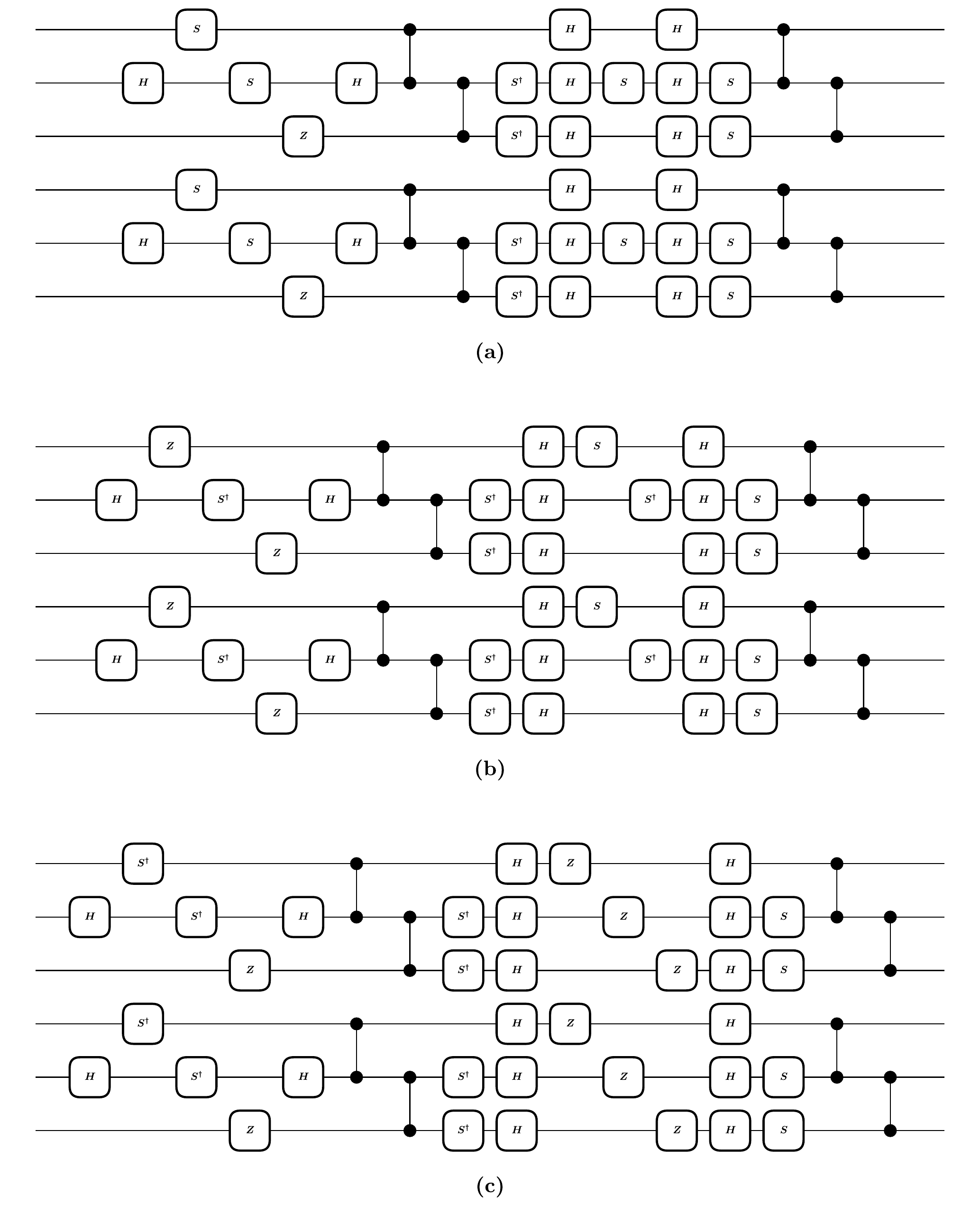}
    \caption{Examples of second-order Clifford approximant circuits for the ansatz of Fig.~\ref{fig:initial_circuit}.}
    \label{fig:second_order_circuits}
\end{figure*}
\clearpage
\bibliography{bibliography.bib}

%apsrev4-2.bst 2019-01-14 (MD) hand-edited version of apsrev4-1.bst
%Control: key (0)
%Control: author (8) initials jnrlst
%Control: editor formatted (1) identically to author
%Control: production of article title (0) allowed
%Control: page (0) single
%Control: year (1) truncated
%Control: production of eprint (0) enabled
\begin{thebibliography}{75}%
\makeatletter
\providecommand \@ifxundefined [1]{%
 \@ifx{#1\undefined}
}%
\providecommand \@ifnum [1]{%
 \ifnum #1\expandafter \@firstoftwo
 \else \expandafter \@secondoftwo
 \fi
}%
\providecommand \@ifx [1]{%
 \ifx #1\expandafter \@firstoftwo
 \else \expandafter \@secondoftwo
 \fi
}%
\providecommand \natexlab [1]{#1}%
\providecommand \enquote  [1]{``#1''}%
\providecommand \bibnamefont  [1]{#1}%
\providecommand \bibfnamefont [1]{#1}%
\providecommand \citenamefont [1]{#1}%
\providecommand \href@noop [0]{\@secondoftwo}%
\providecommand \href [0]{\begingroup \@sanitize@url \@href}%
\providecommand \@href[1]{\@@startlink{#1}\@@href}%
\providecommand \@@href[1]{\endgroup#1\@@endlink}%
\providecommand \@sanitize@url [0]{\catcode `\\12\catcode `\$12\catcode
  `\&12\catcode `\#12\catcode `\^12\catcode `\_12\catcode `\%12\relax}%
\providecommand \@@startlink[1]{}%
\providecommand \@@endlink[0]{}%
\providecommand \url  [0]{\begingroup\@sanitize@url \@url }%
\providecommand \@url [1]{\endgroup\@href {#1}{\urlprefix }}%
\providecommand \urlprefix  [0]{URL }%
\providecommand \Eprint [0]{\href }%
\providecommand \doibase [0]{https://doi.org/}%
\providecommand \selectlanguage [0]{\@gobble}%
\providecommand \bibinfo  [0]{\@secondoftwo}%
\providecommand \bibfield  [0]{\@secondoftwo}%
\providecommand \translation [1]{[#1]}%
\providecommand \BibitemOpen [0]{}%
\providecommand \bibitemStop [0]{}%
\providecommand \bibitemNoStop [0]{.\EOS\space}%
\providecommand \EOS [0]{\spacefactor3000\relax}%
\providecommand \BibitemShut  [1]{\csname bibitem#1\endcsname}%
\let\auto@bib@innerbib\@empty
%</preamble>
\bibitem [{\citenamefont {Cerezo}\ \emph
  {et~al.}(2021{\natexlab{a}})\citenamefont {Cerezo}, \citenamefont
  {Arrasmith}, \citenamefont {Babbush}, \citenamefont {Benjamin}, \citenamefont
  {Endo}, \citenamefont {Fujii}, \citenamefont {McClean}, \citenamefont
  {Mitarai}, \citenamefont {Yuan}, \citenamefont {Cincio},\ and\ \citenamefont
  {Coles}}]{cerezo2021d}%
  \BibitemOpen
  \bibfield  {author} {\bibinfo {author} {\bibfnamefont {M.}~\bibnamefont
  {Cerezo}}, \bibinfo {author} {\bibfnamefont {A.}~\bibnamefont {Arrasmith}},
  \bibinfo {author} {\bibfnamefont {R.}~\bibnamefont {Babbush}}, \bibinfo
  {author} {\bibfnamefont {S.~C.}\ \bibnamefont {Benjamin}}, \bibinfo {author}
  {\bibfnamefont {S.}~\bibnamefont {Endo}}, \bibinfo {author} {\bibfnamefont
  {K.}~\bibnamefont {Fujii}}, \bibinfo {author} {\bibfnamefont {J.~R.}\
  \bibnamefont {McClean}}, \bibinfo {author} {\bibfnamefont {K.}~\bibnamefont
  {Mitarai}}, \bibinfo {author} {\bibfnamefont {X.}~\bibnamefont {Yuan}},
  \bibinfo {author} {\bibfnamefont {L.}~\bibnamefont {Cincio}},\ and\ \bibinfo
  {author} {\bibfnamefont {P.~J.}\ \bibnamefont {Coles}},\ }\bibfield  {title}
  {\bibinfo {title} {Variational quantum algorithms},\ }\href
  {https://doi.org/10.1038/s42254-021-00348-9} {\bibfield  {journal} {\bibinfo
  {journal} {Nature Reviews Physics}\ }\textbf {\bibinfo {volume} {3}},\
  \bibinfo {pages} {625} (\bibinfo {year} {2021}{\natexlab{a}})}\BibitemShut
  {NoStop}%
\bibitem [{\citenamefont {Carleo}\ \emph {et~al.}(2019)\citenamefont {Carleo},
  \citenamefont {Cirac}, \citenamefont {Cranmer}, \citenamefont {Daudet},
  \citenamefont {Schuld}, \citenamefont {Tishby}, \citenamefont
  {{Vogt-Maranto}},\ and\ \citenamefont {Zdeborov{\'a}}}]{carleo2019}%
  \BibitemOpen
  \bibfield  {author} {\bibinfo {author} {\bibfnamefont {G.}~\bibnamefont
  {Carleo}}, \bibinfo {author} {\bibfnamefont {I.}~\bibnamefont {Cirac}},
  \bibinfo {author} {\bibfnamefont {K.}~\bibnamefont {Cranmer}}, \bibinfo
  {author} {\bibfnamefont {L.}~\bibnamefont {Daudet}}, \bibinfo {author}
  {\bibfnamefont {M.}~\bibnamefont {Schuld}}, \bibinfo {author} {\bibfnamefont
  {N.}~\bibnamefont {Tishby}}, \bibinfo {author} {\bibfnamefont
  {L.}~\bibnamefont {{Vogt-Maranto}}},\ and\ \bibinfo {author} {\bibfnamefont
  {L.}~\bibnamefont {Zdeborov{\'a}}},\ }\bibfield  {title} {\bibinfo {title}
  {Machine learning and the physical sciences},\ }\href
  {https://doi.org/10.1103/RevModPhys.91.045002} {\bibfield  {journal}
  {\bibinfo  {journal} {Reviews of Modern Physics}\ }\textbf {\bibinfo {volume}
  {91}},\ \bibinfo {pages} {045002} (\bibinfo {year} {2019})}\BibitemShut
  {NoStop}%
\bibitem [{\citenamefont {Cerezo}\ \emph {et~al.}(2022)\citenamefont {Cerezo},
  \citenamefont {Verdon}, \citenamefont {Huang}, \citenamefont {Cincio},\ and\
  \citenamefont {Coles}}]{cerezo2022a}%
  \BibitemOpen
  \bibfield  {author} {\bibinfo {author} {\bibfnamefont {M.}~\bibnamefont
  {Cerezo}}, \bibinfo {author} {\bibfnamefont {G.}~\bibnamefont {Verdon}},
  \bibinfo {author} {\bibfnamefont {H.-Y.}\ \bibnamefont {Huang}}, \bibinfo
  {author} {\bibfnamefont {L.}~\bibnamefont {Cincio}},\ and\ \bibinfo {author}
  {\bibfnamefont {P.~J.}\ \bibnamefont {Coles}},\ }\bibfield  {title} {\bibinfo
  {title} {Challenges and opportunities in quantum machine learning},\ }\href
  {https://doi.org/10.1038/s43588-022-00311-3} {\bibfield  {journal} {\bibinfo
  {journal} {Nature Computational Science}\ }\textbf {\bibinfo {volume} {2}},\
  \bibinfo {pages} {567} (\bibinfo {year} {2022})}\BibitemShut {NoStop}%
\bibitem [{\citenamefont {Peruzzo}\ \emph {et~al.}(2014)\citenamefont
  {Peruzzo}, \citenamefont {McClean}, \citenamefont {Shadbolt}, \citenamefont
  {Yung}, \citenamefont {Zhou}, \citenamefont {Love}, \citenamefont
  {{Aspuru-Guzik}},\ and\ \citenamefont {O'Brien}}]{peruzzo2014}%
  \BibitemOpen
  \bibfield  {author} {\bibinfo {author} {\bibfnamefont {A.}~\bibnamefont
  {Peruzzo}}, \bibinfo {author} {\bibfnamefont {J.}~\bibnamefont {McClean}},
  \bibinfo {author} {\bibfnamefont {P.}~\bibnamefont {Shadbolt}}, \bibinfo
  {author} {\bibfnamefont {M.-H.}\ \bibnamefont {Yung}}, \bibinfo {author}
  {\bibfnamefont {X.-Q.}\ \bibnamefont {Zhou}}, \bibinfo {author}
  {\bibfnamefont {P.~J.}\ \bibnamefont {Love}}, \bibinfo {author}
  {\bibfnamefont {A.}~\bibnamefont {{Aspuru-Guzik}}},\ and\ \bibinfo {author}
  {\bibfnamefont {J.~L.}\ \bibnamefont {O'Brien}},\ }\bibfield  {title}
  {\bibinfo {title} {A variational eigenvalue solver on a photonic quantum
  processor},\ }\href {https://doi.org/10.1038/ncomms5213} {\bibfield
  {journal} {\bibinfo  {journal} {Nature Communications}\ }\textbf {\bibinfo
  {volume} {5}},\ \bibinfo {pages} {4213} (\bibinfo {year} {2014})}\BibitemShut
  {NoStop}%
\bibitem [{\citenamefont {Kandala}\ \emph {et~al.}(2017)\citenamefont
  {Kandala}, \citenamefont {Mezzacapo}, \citenamefont {Temme}, \citenamefont
  {Takita}, \citenamefont {Brink}, \citenamefont {Chow},\ and\ \citenamefont
  {Gambetta}}]{kandala2017}%
  \BibitemOpen
  \bibfield  {author} {\bibinfo {author} {\bibfnamefont {A.}~\bibnamefont
  {Kandala}}, \bibinfo {author} {\bibfnamefont {A.}~\bibnamefont {Mezzacapo}},
  \bibinfo {author} {\bibfnamefont {K.}~\bibnamefont {Temme}}, \bibinfo
  {author} {\bibfnamefont {M.}~\bibnamefont {Takita}}, \bibinfo {author}
  {\bibfnamefont {M.}~\bibnamefont {Brink}}, \bibinfo {author} {\bibfnamefont
  {J.~M.}\ \bibnamefont {Chow}},\ and\ \bibinfo {author} {\bibfnamefont
  {J.~M.}\ \bibnamefont {Gambetta}},\ }\bibfield  {title} {\bibinfo {title}
  {Hardware-efficient variational quantum eigensolver for small molecules and
  quantum magnets},\ }\href {https://doi.org/10.1038/nature23879} {\bibfield
  {journal} {\bibinfo  {journal} {Nature}\ }\textbf {\bibinfo {volume} {549}},\
  \bibinfo {pages} {242} (\bibinfo {year} {2017})}\BibitemShut {NoStop}%
\bibitem [{\citenamefont {{GOOGLE AI QUANTUM AND COLLABORATORS}}\ \emph
  {et~al.}(2020)\citenamefont {{GOOGLE AI QUANTUM AND COLLABORATORS}},
  \citenamefont {Arute}, \citenamefont {Arya}, \citenamefont {Babbush},
  \citenamefont {Bacon}, \citenamefont {Bardin}, \citenamefont {Barends},
  \citenamefont {Boixo}, \citenamefont {Broughton}, \citenamefont {Buckley},
  \citenamefont {Buell}, \citenamefont {Burkett}, \citenamefont {Bushnell},
  \citenamefont {Chen}, \citenamefont {Chen}, \citenamefont {Chiaro},
  \citenamefont {Collins}, \citenamefont {Courtney}, \citenamefont {Demura},
  \citenamefont {Dunsworth}, \citenamefont {Farhi}, \citenamefont {Fowler},
  \citenamefont {Foxen}, \citenamefont {Gidney}, \citenamefont {Giustina},
  \citenamefont {Graff}, \citenamefont {Habegger}, \citenamefont {Harrigan},
  \citenamefont {Ho}, \citenamefont {Hong}, \citenamefont {Huang},
  \citenamefont {Huggins}, \citenamefont {Ioffe}, \citenamefont {Isakov},
  \citenamefont {Jeffrey}, \citenamefont {Jiang}, \citenamefont {Jones},
  \citenamefont {Kafri}, \citenamefont {Kechedzhi}, \citenamefont {Kelly},
  \citenamefont {Kim}, \citenamefont {Klimov}, \citenamefont {Korotkov},
  \citenamefont {Kostritsa}, \citenamefont {Landhuis}, \citenamefont {Laptev},
  \citenamefont {Lindmark}, \citenamefont {Lucero}, \citenamefont {Martin},
  \citenamefont {Martinis}, \citenamefont {McClean}, \citenamefont {McEwen},
  \citenamefont {Megrant}, \citenamefont {Mi}, \citenamefont {Mohseni},
  \citenamefont {Mruczkiewicz}, \citenamefont {Mutus}, \citenamefont {Naaman},
  \citenamefont {Neeley}, \citenamefont {Neill}, \citenamefont {Neven},
  \citenamefont {Niu}, \citenamefont {O'Brien}, \citenamefont {Ostby},
  \citenamefont {Petukhov}, \citenamefont {Putterman}, \citenamefont
  {Quintana}, \citenamefont {Roushan}, \citenamefont {Rubin}, \citenamefont
  {Sank}, \citenamefont {Satzinger}, \citenamefont {Smelyanskiy}, \citenamefont
  {Strain}, \citenamefont {Sung}, \citenamefont {Szalay}, \citenamefont
  {Takeshita}, \citenamefont {Vainsencher}, \citenamefont {White},
  \citenamefont {Wiebe}, \citenamefont {Yao}, \citenamefont {Yeh},\ and\
  \citenamefont {Zalcman}}]{googleaiquantumandcollaborators2020a}%
  \BibitemOpen
  \bibfield  {author} {\bibinfo {author} {\bibnamefont {{GOOGLE AI QUANTUM AND
  COLLABORATORS}}}, \bibinfo {author} {\bibfnamefont {F.}~\bibnamefont
  {Arute}}, \bibinfo {author} {\bibfnamefont {K.}~\bibnamefont {Arya}},
  \bibinfo {author} {\bibfnamefont {R.}~\bibnamefont {Babbush}}, \bibinfo
  {author} {\bibfnamefont {D.}~\bibnamefont {Bacon}}, \bibinfo {author}
  {\bibfnamefont {J.~C.}\ \bibnamefont {Bardin}}, \bibinfo {author}
  {\bibfnamefont {R.}~\bibnamefont {Barends}}, \bibinfo {author} {\bibfnamefont
  {S.}~\bibnamefont {Boixo}}, \bibinfo {author} {\bibfnamefont
  {M.}~\bibnamefont {Broughton}}, \bibinfo {author} {\bibfnamefont {B.~B.}\
  \bibnamefont {Buckley}}, \bibinfo {author} {\bibfnamefont {D.~A.}\
  \bibnamefont {Buell}}, \bibinfo {author} {\bibfnamefont {B.}~\bibnamefont
  {Burkett}}, \bibinfo {author} {\bibfnamefont {N.}~\bibnamefont {Bushnell}},
  \bibinfo {author} {\bibfnamefont {Y.}~\bibnamefont {Chen}}, \bibinfo {author}
  {\bibfnamefont {Z.}~\bibnamefont {Chen}}, \bibinfo {author} {\bibfnamefont
  {B.}~\bibnamefont {Chiaro}}, \bibinfo {author} {\bibfnamefont
  {R.}~\bibnamefont {Collins}}, \bibinfo {author} {\bibfnamefont
  {W.}~\bibnamefont {Courtney}}, \bibinfo {author} {\bibfnamefont
  {S.}~\bibnamefont {Demura}}, \bibinfo {author} {\bibfnamefont
  {A.}~\bibnamefont {Dunsworth}}, \bibinfo {author} {\bibfnamefont
  {E.}~\bibnamefont {Farhi}}, \bibinfo {author} {\bibfnamefont
  {A.}~\bibnamefont {Fowler}}, \bibinfo {author} {\bibfnamefont
  {B.}~\bibnamefont {Foxen}}, \bibinfo {author} {\bibfnamefont
  {C.}~\bibnamefont {Gidney}}, \bibinfo {author} {\bibfnamefont
  {M.}~\bibnamefont {Giustina}}, \bibinfo {author} {\bibfnamefont
  {R.}~\bibnamefont {Graff}}, \bibinfo {author} {\bibfnamefont
  {S.}~\bibnamefont {Habegger}}, \bibinfo {author} {\bibfnamefont {M.~P.}\
  \bibnamefont {Harrigan}}, \bibinfo {author} {\bibfnamefont {A.}~\bibnamefont
  {Ho}}, \bibinfo {author} {\bibfnamefont {S.}~\bibnamefont {Hong}}, \bibinfo
  {author} {\bibfnamefont {T.}~\bibnamefont {Huang}}, \bibinfo {author}
  {\bibfnamefont {W.~J.}\ \bibnamefont {Huggins}}, \bibinfo {author}
  {\bibfnamefont {L.}~\bibnamefont {Ioffe}}, \bibinfo {author} {\bibfnamefont
  {S.~V.}\ \bibnamefont {Isakov}}, \bibinfo {author} {\bibfnamefont
  {E.}~\bibnamefont {Jeffrey}}, \bibinfo {author} {\bibfnamefont
  {Z.}~\bibnamefont {Jiang}}, \bibinfo {author} {\bibfnamefont
  {C.}~\bibnamefont {Jones}}, \bibinfo {author} {\bibfnamefont
  {D.}~\bibnamefont {Kafri}}, \bibinfo {author} {\bibfnamefont
  {K.}~\bibnamefont {Kechedzhi}}, \bibinfo {author} {\bibfnamefont
  {J.}~\bibnamefont {Kelly}}, \bibinfo {author} {\bibfnamefont
  {S.}~\bibnamefont {Kim}}, \bibinfo {author} {\bibfnamefont {P.~V.}\
  \bibnamefont {Klimov}}, \bibinfo {author} {\bibfnamefont {A.}~\bibnamefont
  {Korotkov}}, \bibinfo {author} {\bibfnamefont {F.}~\bibnamefont {Kostritsa}},
  \bibinfo {author} {\bibfnamefont {D.}~\bibnamefont {Landhuis}}, \bibinfo
  {author} {\bibfnamefont {P.}~\bibnamefont {Laptev}}, \bibinfo {author}
  {\bibfnamefont {M.}~\bibnamefont {Lindmark}}, \bibinfo {author}
  {\bibfnamefont {E.}~\bibnamefont {Lucero}}, \bibinfo {author} {\bibfnamefont
  {O.}~\bibnamefont {Martin}}, \bibinfo {author} {\bibfnamefont {J.~M.}\
  \bibnamefont {Martinis}}, \bibinfo {author} {\bibfnamefont {J.~R.}\
  \bibnamefont {McClean}}, \bibinfo {author} {\bibfnamefont {M.}~\bibnamefont
  {McEwen}}, \bibinfo {author} {\bibfnamefont {A.}~\bibnamefont {Megrant}},
  \bibinfo {author} {\bibfnamefont {X.}~\bibnamefont {Mi}}, \bibinfo {author}
  {\bibfnamefont {M.}~\bibnamefont {Mohseni}}, \bibinfo {author} {\bibfnamefont
  {W.}~\bibnamefont {Mruczkiewicz}}, \bibinfo {author} {\bibfnamefont
  {J.}~\bibnamefont {Mutus}}, \bibinfo {author} {\bibfnamefont
  {O.}~\bibnamefont {Naaman}}, \bibinfo {author} {\bibfnamefont
  {M.}~\bibnamefont {Neeley}}, \bibinfo {author} {\bibfnamefont
  {C.}~\bibnamefont {Neill}}, \bibinfo {author} {\bibfnamefont
  {H.}~\bibnamefont {Neven}}, \bibinfo {author} {\bibfnamefont {M.~Y.}\
  \bibnamefont {Niu}}, \bibinfo {author} {\bibfnamefont {T.~E.}\ \bibnamefont
  {O'Brien}}, \bibinfo {author} {\bibfnamefont {E.}~\bibnamefont {Ostby}},
  \bibinfo {author} {\bibfnamefont {A.}~\bibnamefont {Petukhov}}, \bibinfo
  {author} {\bibfnamefont {H.}~\bibnamefont {Putterman}}, \bibinfo {author}
  {\bibfnamefont {C.}~\bibnamefont {Quintana}}, \bibinfo {author}
  {\bibfnamefont {P.}~\bibnamefont {Roushan}}, \bibinfo {author} {\bibfnamefont
  {N.~C.}\ \bibnamefont {Rubin}}, \bibinfo {author} {\bibfnamefont
  {D.}~\bibnamefont {Sank}}, \bibinfo {author} {\bibfnamefont {K.~J.}\
  \bibnamefont {Satzinger}}, \bibinfo {author} {\bibfnamefont {V.}~\bibnamefont
  {Smelyanskiy}}, \bibinfo {author} {\bibfnamefont {D.}~\bibnamefont {Strain}},
  \bibinfo {author} {\bibfnamefont {K.~J.}\ \bibnamefont {Sung}}, \bibinfo
  {author} {\bibfnamefont {M.}~\bibnamefont {Szalay}}, \bibinfo {author}
  {\bibfnamefont {T.~Y.}\ \bibnamefont {Takeshita}}, \bibinfo {author}
  {\bibfnamefont {A.}~\bibnamefont {Vainsencher}}, \bibinfo {author}
  {\bibfnamefont {T.}~\bibnamefont {White}}, \bibinfo {author} {\bibfnamefont
  {N.}~\bibnamefont {Wiebe}}, \bibinfo {author} {\bibfnamefont {Z.~J.}\
  \bibnamefont {Yao}}, \bibinfo {author} {\bibfnamefont {P.}~\bibnamefont
  {Yeh}},\ and\ \bibinfo {author} {\bibfnamefont {A.}~\bibnamefont {Zalcman}},\
  }\bibfield  {title} {\bibinfo {title} {Hartree-{{Fock}} on a superconducting
  qubit quantum computer},\ }\href {https://doi.org/10.1126/science.abb9811}
  {\bibfield  {journal} {\bibinfo  {journal} {Science}\ }\textbf {\bibinfo
  {volume} {369}},\ \bibinfo {pages} {1084} (\bibinfo {year}
  {2020})}\BibitemShut {NoStop}%
\bibitem [{\citenamefont {Farhi}\ \emph {et~al.}(2014)\citenamefont {Farhi},
  \citenamefont {Goldstone},\ and\ \citenamefont {Gutmann}}]{farhi2014}%
  \BibitemOpen
  \bibfield  {author} {\bibinfo {author} {\bibfnamefont {E.}~\bibnamefont
  {Farhi}}, \bibinfo {author} {\bibfnamefont {J.}~\bibnamefont {Goldstone}},\
  and\ \bibinfo {author} {\bibfnamefont {S.}~\bibnamefont {Gutmann}},\
  }\href@noop {} {\bibinfo {title} {A {{Quantum Approximate Optimization
  Algorithm}}}} (\bibinfo {year} {2014}),\ \Eprint
  {https://arxiv.org/abs/1411.4028} {arXiv:1411.4028} \BibitemShut {NoStop}%
\bibitem [{\citenamefont {Lacroix}\ \emph {et~al.}(2020)\citenamefont
  {Lacroix}, \citenamefont {Hellings}, \citenamefont {Andersen}, \citenamefont
  {Di~Paolo}, \citenamefont {Remm}, \citenamefont {Lazar}, \citenamefont
  {Krinner}, \citenamefont {Norris}, \citenamefont {Gabureac}, \citenamefont
  {Heinsoo}, \citenamefont {Blais}, \citenamefont {Eichler},\ and\
  \citenamefont {Wallraff}}]{lacroix2020}%
  \BibitemOpen
  \bibfield  {author} {\bibinfo {author} {\bibfnamefont {N.}~\bibnamefont
  {Lacroix}}, \bibinfo {author} {\bibfnamefont {C.}~\bibnamefont {Hellings}},
  \bibinfo {author} {\bibfnamefont {C.~K.}\ \bibnamefont {Andersen}}, \bibinfo
  {author} {\bibfnamefont {A.}~\bibnamefont {Di~Paolo}}, \bibinfo {author}
  {\bibfnamefont {A.}~\bibnamefont {Remm}}, \bibinfo {author} {\bibfnamefont
  {S.}~\bibnamefont {Lazar}}, \bibinfo {author} {\bibfnamefont
  {S.}~\bibnamefont {Krinner}}, \bibinfo {author} {\bibfnamefont {G.~J.}\
  \bibnamefont {Norris}}, \bibinfo {author} {\bibfnamefont {M.}~\bibnamefont
  {Gabureac}}, \bibinfo {author} {\bibfnamefont {J.}~\bibnamefont {Heinsoo}},
  \bibinfo {author} {\bibfnamefont {A.}~\bibnamefont {Blais}}, \bibinfo
  {author} {\bibfnamefont {C.}~\bibnamefont {Eichler}},\ and\ \bibinfo {author}
  {\bibfnamefont {A.}~\bibnamefont {Wallraff}},\ }\bibfield  {title} {\bibinfo
  {title} {Improving the {{Performance}} of {{Deep Quantum Optimization
  Algorithms}} with {{Continuous Gate Sets}}},\ }\href
  {https://doi.org/10.1103/PRXQuantum.1.020304} {\bibfield  {journal} {\bibinfo
   {journal} {PRX Quantum}\ }\textbf {\bibinfo {volume} {1}},\ \bibinfo {pages}
  {020304} (\bibinfo {year} {2020})}\BibitemShut {NoStop}%
\bibitem [{\citenamefont {Harrigan}\ \emph {et~al.}(2021)\citenamefont
  {Harrigan}, \citenamefont {Sung}, \citenamefont {Neeley}, \citenamefont
  {Satzinger}, \citenamefont {Arute}, \citenamefont {Arya}, \citenamefont
  {Atalaya}, \citenamefont {Bardin}, \citenamefont {Barends}, \citenamefont
  {Boixo}, \citenamefont {Broughton}, \citenamefont {Buckley}, \citenamefont
  {Buell}, \citenamefont {Burkett}, \citenamefont {Bushnell}, \citenamefont
  {Chen}, \citenamefont {Chen}, \citenamefont {{Ben Chiaro}}, \citenamefont
  {Collins}, \citenamefont {Courtney}, \citenamefont {Demura}, \citenamefont
  {Dunsworth}, \citenamefont {Eppens}, \citenamefont {Fowler}, \citenamefont
  {Foxen}, \citenamefont {Gidney}, \citenamefont {Giustina}, \citenamefont
  {Graff}, \citenamefont {Habegger}, \citenamefont {Ho}, \citenamefont {Hong},
  \citenamefont {Huang}, \citenamefont {Ioffe}, \citenamefont {Isakov},
  \citenamefont {Jeffrey}, \citenamefont {Jiang}, \citenamefont {Jones},
  \citenamefont {Kafri}, \citenamefont {Kechedzhi}, \citenamefont {Kelly},
  \citenamefont {Kim}, \citenamefont {Klimov}, \citenamefont {Korotkov},
  \citenamefont {Kostritsa}, \citenamefont {Landhuis}, \citenamefont {Laptev},
  \citenamefont {Lindmark}, \citenamefont {Leib}, \citenamefont {Martin},
  \citenamefont {Martinis}, \citenamefont {McClean}, \citenamefont {McEwen},
  \citenamefont {Megrant}, \citenamefont {Mi}, \citenamefont {Mohseni},
  \citenamefont {Mruczkiewicz}, \citenamefont {Mutus}, \citenamefont {Naaman},
  \citenamefont {Neill}, \citenamefont {Neukart}, \citenamefont {Niu},
  \citenamefont {O'Brien}, \citenamefont {O'Gorman}, \citenamefont {Ostby},
  \citenamefont {Petukhov}, \citenamefont {Putterman}, \citenamefont
  {Quintana}, \citenamefont {Roushan}, \citenamefont {Rubin}, \citenamefont
  {Sank}, \citenamefont {Skolik}, \citenamefont {Smelyanskiy}, \citenamefont
  {Strain}, \citenamefont {Streif}, \citenamefont {Szalay}, \citenamefont
  {Vainsencher}, \citenamefont {White}, \citenamefont {Yao}, \citenamefont
  {Yeh}, \citenamefont {Zalcman}, \citenamefont {Zhou}, \citenamefont {Neven},
  \citenamefont {Bacon}, \citenamefont {Lucero}, \citenamefont {Farhi},\ and\
  \citenamefont {Babbush}}]{harrigan2021}%
  \BibitemOpen
  \bibfield  {author} {\bibinfo {author} {\bibfnamefont {M.~P.}\ \bibnamefont
  {Harrigan}}, \bibinfo {author} {\bibfnamefont {K.~J.}\ \bibnamefont {Sung}},
  \bibinfo {author} {\bibfnamefont {M.}~\bibnamefont {Neeley}}, \bibinfo
  {author} {\bibfnamefont {K.~J.}\ \bibnamefont {Satzinger}}, \bibinfo {author}
  {\bibfnamefont {F.}~\bibnamefont {Arute}}, \bibinfo {author} {\bibfnamefont
  {K.}~\bibnamefont {Arya}}, \bibinfo {author} {\bibfnamefont {J.}~\bibnamefont
  {Atalaya}}, \bibinfo {author} {\bibfnamefont {J.~C.}\ \bibnamefont {Bardin}},
  \bibinfo {author} {\bibfnamefont {R.}~\bibnamefont {Barends}}, \bibinfo
  {author} {\bibfnamefont {S.}~\bibnamefont {Boixo}}, \bibinfo {author}
  {\bibfnamefont {M.}~\bibnamefont {Broughton}}, \bibinfo {author}
  {\bibfnamefont {B.~B.}\ \bibnamefont {Buckley}}, \bibinfo {author}
  {\bibfnamefont {D.~A.}\ \bibnamefont {Buell}}, \bibinfo {author}
  {\bibfnamefont {B.}~\bibnamefont {Burkett}}, \bibinfo {author} {\bibfnamefont
  {N.}~\bibnamefont {Bushnell}}, \bibinfo {author} {\bibfnamefont
  {Y.}~\bibnamefont {Chen}}, \bibinfo {author} {\bibfnamefont {Z.}~\bibnamefont
  {Chen}}, \bibinfo {author} {\bibnamefont {{Ben Chiaro}}}, \bibinfo {author}
  {\bibfnamefont {R.}~\bibnamefont {Collins}}, \bibinfo {author} {\bibfnamefont
  {W.}~\bibnamefont {Courtney}}, \bibinfo {author} {\bibfnamefont
  {S.}~\bibnamefont {Demura}}, \bibinfo {author} {\bibfnamefont
  {A.}~\bibnamefont {Dunsworth}}, \bibinfo {author} {\bibfnamefont
  {D.}~\bibnamefont {Eppens}}, \bibinfo {author} {\bibfnamefont
  {A.}~\bibnamefont {Fowler}}, \bibinfo {author} {\bibfnamefont
  {B.}~\bibnamefont {Foxen}}, \bibinfo {author} {\bibfnamefont
  {C.}~\bibnamefont {Gidney}}, \bibinfo {author} {\bibfnamefont
  {M.}~\bibnamefont {Giustina}}, \bibinfo {author} {\bibfnamefont
  {R.}~\bibnamefont {Graff}}, \bibinfo {author} {\bibfnamefont
  {S.}~\bibnamefont {Habegger}}, \bibinfo {author} {\bibfnamefont
  {A.}~\bibnamefont {Ho}}, \bibinfo {author} {\bibfnamefont {S.}~\bibnamefont
  {Hong}}, \bibinfo {author} {\bibfnamefont {T.}~\bibnamefont {Huang}},
  \bibinfo {author} {\bibfnamefont {L.~B.}\ \bibnamefont {Ioffe}}, \bibinfo
  {author} {\bibfnamefont {S.~V.}\ \bibnamefont {Isakov}}, \bibinfo {author}
  {\bibfnamefont {E.}~\bibnamefont {Jeffrey}}, \bibinfo {author} {\bibfnamefont
  {Z.}~\bibnamefont {Jiang}}, \bibinfo {author} {\bibfnamefont
  {C.}~\bibnamefont {Jones}}, \bibinfo {author} {\bibfnamefont
  {D.}~\bibnamefont {Kafri}}, \bibinfo {author} {\bibfnamefont
  {K.}~\bibnamefont {Kechedzhi}}, \bibinfo {author} {\bibfnamefont
  {J.}~\bibnamefont {Kelly}}, \bibinfo {author} {\bibfnamefont
  {S.}~\bibnamefont {Kim}}, \bibinfo {author} {\bibfnamefont {P.~V.}\
  \bibnamefont {Klimov}}, \bibinfo {author} {\bibfnamefont {A.~N.}\
  \bibnamefont {Korotkov}}, \bibinfo {author} {\bibfnamefont {F.}~\bibnamefont
  {Kostritsa}}, \bibinfo {author} {\bibfnamefont {D.}~\bibnamefont {Landhuis}},
  \bibinfo {author} {\bibfnamefont {P.}~\bibnamefont {Laptev}}, \bibinfo
  {author} {\bibfnamefont {M.}~\bibnamefont {Lindmark}}, \bibinfo {author}
  {\bibfnamefont {M.}~\bibnamefont {Leib}}, \bibinfo {author} {\bibfnamefont
  {O.}~\bibnamefont {Martin}}, \bibinfo {author} {\bibfnamefont {J.~M.}\
  \bibnamefont {Martinis}}, \bibinfo {author} {\bibfnamefont {J.~R.}\
  \bibnamefont {McClean}}, \bibinfo {author} {\bibfnamefont {M.}~\bibnamefont
  {McEwen}}, \bibinfo {author} {\bibfnamefont {A.}~\bibnamefont {Megrant}},
  \bibinfo {author} {\bibfnamefont {X.}~\bibnamefont {Mi}}, \bibinfo {author}
  {\bibfnamefont {M.}~\bibnamefont {Mohseni}}, \bibinfo {author} {\bibfnamefont
  {W.}~\bibnamefont {Mruczkiewicz}}, \bibinfo {author} {\bibfnamefont
  {J.}~\bibnamefont {Mutus}}, \bibinfo {author} {\bibfnamefont
  {O.}~\bibnamefont {Naaman}}, \bibinfo {author} {\bibfnamefont
  {C.}~\bibnamefont {Neill}}, \bibinfo {author} {\bibfnamefont
  {F.}~\bibnamefont {Neukart}}, \bibinfo {author} {\bibfnamefont {M.~Y.}\
  \bibnamefont {Niu}}, \bibinfo {author} {\bibfnamefont {T.~E.}\ \bibnamefont
  {O'Brien}}, \bibinfo {author} {\bibfnamefont {B.}~\bibnamefont {O'Gorman}},
  \bibinfo {author} {\bibfnamefont {E.}~\bibnamefont {Ostby}}, \bibinfo
  {author} {\bibfnamefont {A.}~\bibnamefont {Petukhov}}, \bibinfo {author}
  {\bibfnamefont {H.}~\bibnamefont {Putterman}}, \bibinfo {author}
  {\bibfnamefont {C.}~\bibnamefont {Quintana}}, \bibinfo {author}
  {\bibfnamefont {P.}~\bibnamefont {Roushan}}, \bibinfo {author} {\bibfnamefont
  {N.~C.}\ \bibnamefont {Rubin}}, \bibinfo {author} {\bibfnamefont
  {D.}~\bibnamefont {Sank}}, \bibinfo {author} {\bibfnamefont {A.}~\bibnamefont
  {Skolik}}, \bibinfo {author} {\bibfnamefont {V.}~\bibnamefont {Smelyanskiy}},
  \bibinfo {author} {\bibfnamefont {D.}~\bibnamefont {Strain}}, \bibinfo
  {author} {\bibfnamefont {M.}~\bibnamefont {Streif}}, \bibinfo {author}
  {\bibfnamefont {M.}~\bibnamefont {Szalay}}, \bibinfo {author} {\bibfnamefont
  {A.}~\bibnamefont {Vainsencher}}, \bibinfo {author} {\bibfnamefont
  {T.}~\bibnamefont {White}}, \bibinfo {author} {\bibfnamefont {Z.~J.}\
  \bibnamefont {Yao}}, \bibinfo {author} {\bibfnamefont {P.}~\bibnamefont
  {Yeh}}, \bibinfo {author} {\bibfnamefont {A.}~\bibnamefont {Zalcman}},
  \bibinfo {author} {\bibfnamefont {L.}~\bibnamefont {Zhou}}, \bibinfo {author}
  {\bibfnamefont {H.}~\bibnamefont {Neven}}, \bibinfo {author} {\bibfnamefont
  {D.}~\bibnamefont {Bacon}}, \bibinfo {author} {\bibfnamefont
  {E.}~\bibnamefont {Lucero}}, \bibinfo {author} {\bibfnamefont
  {E.}~\bibnamefont {Farhi}},\ and\ \bibinfo {author} {\bibfnamefont
  {R.}~\bibnamefont {Babbush}},\ }\bibfield  {title} {\bibinfo {title} {Quantum
  approximate optimization of non-planar graph problems on a planar
  superconducting processor},\ }\href
  {https://doi.org/10.1038/s41567-020-01105-y} {\bibfield  {journal} {\bibinfo
  {journal} {Nature Physics}\ }\textbf {\bibinfo {volume} {17}},\ \bibinfo
  {pages} {332} (\bibinfo {year} {2021})}\BibitemShut {NoStop}%
\bibitem [{\citenamefont {McClean}\ \emph {et~al.}(2018)\citenamefont
  {McClean}, \citenamefont {Boixo}, \citenamefont {Smelyanskiy}, \citenamefont
  {Babbush},\ and\ \citenamefont {Neven}}]{mcclean2018}%
  \BibitemOpen
  \bibfield  {author} {\bibinfo {author} {\bibfnamefont {J.~R.}\ \bibnamefont
  {McClean}}, \bibinfo {author} {\bibfnamefont {S.}~\bibnamefont {Boixo}},
  \bibinfo {author} {\bibfnamefont {V.~N.}\ \bibnamefont {Smelyanskiy}},
  \bibinfo {author} {\bibfnamefont {R.}~\bibnamefont {Babbush}},\ and\ \bibinfo
  {author} {\bibfnamefont {H.}~\bibnamefont {Neven}},\ }\bibfield  {title}
  {\bibinfo {title} {Barren plateaus in quantum neural network training
  landscapes},\ }\href {https://doi.org/10.1038/s41467-018-07090-4} {\bibfield
  {journal} {\bibinfo  {journal} {Nature Communications}\ }\textbf {\bibinfo
  {volume} {9}},\ \bibinfo {pages} {4812} (\bibinfo {year} {2018})}\BibitemShut
  {NoStop}%
\bibitem [{\citenamefont {Holmes}\ \emph {et~al.}(2022)\citenamefont {Holmes},
  \citenamefont {Sharma}, \citenamefont {Cerezo},\ and\ \citenamefont
  {Coles}}]{holmes2022a}%
  \BibitemOpen
  \bibfield  {author} {\bibinfo {author} {\bibfnamefont {Z.}~\bibnamefont
  {Holmes}}, \bibinfo {author} {\bibfnamefont {K.}~\bibnamefont {Sharma}},
  \bibinfo {author} {\bibfnamefont {M.}~\bibnamefont {Cerezo}},\ and\ \bibinfo
  {author} {\bibfnamefont {P.~J.}\ \bibnamefont {Coles}},\ }\bibfield  {title}
  {\bibinfo {title} {Connecting {{Ansatz Expressibility}} to {{Gradient
  Magnitudes}} and {{Barren Plateaus}}},\ }\href
  {https://doi.org/10.1103/PRXQuantum.3.010313} {\bibfield  {journal} {\bibinfo
   {journal} {PRX Quantum}\ }\textbf {\bibinfo {volume} {3}},\ \bibinfo {pages}
  {010313} (\bibinfo {year} {2022})}\BibitemShut {NoStop}%
\bibitem [{\citenamefont {Wang}\ \emph
  {et~al.}(2021{\natexlab{a}})\citenamefont {Wang}, \citenamefont {Fontana},
  \citenamefont {Cerezo}, \citenamefont {Sharma}, \citenamefont {Sone},
  \citenamefont {Cincio},\ and\ \citenamefont {Coles}}]{wang2021c}%
  \BibitemOpen
  \bibfield  {author} {\bibinfo {author} {\bibfnamefont {S.}~\bibnamefont
  {Wang}}, \bibinfo {author} {\bibfnamefont {E.}~\bibnamefont {Fontana}},
  \bibinfo {author} {\bibfnamefont {M.}~\bibnamefont {Cerezo}}, \bibinfo
  {author} {\bibfnamefont {K.}~\bibnamefont {Sharma}}, \bibinfo {author}
  {\bibfnamefont {A.}~\bibnamefont {Sone}}, \bibinfo {author} {\bibfnamefont
  {L.}~\bibnamefont {Cincio}},\ and\ \bibinfo {author} {\bibfnamefont {P.~J.}\
  \bibnamefont {Coles}},\ }\bibfield  {title} {\bibinfo {title} {Noise-induced
  barren plateaus in variational quantum algorithms},\ }\href
  {https://doi.org/10.1038/s41467-021-27045-6} {\bibfield  {journal} {\bibinfo
  {journal} {Nature Communications}\ }\textbf {\bibinfo {volume} {12}},\
  \bibinfo {pages} {6961} (\bibinfo {year} {2021}{\natexlab{a}})}\BibitemShut
  {NoStop}%
\bibitem [{\citenamefont {Ortiz~Marrero}\ \emph {et~al.}(2021)\citenamefont
  {Ortiz~Marrero}, \citenamefont {Kieferov{\'a}},\ and\ \citenamefont
  {Wiebe}}]{ortizmarrero2021a}%
  \BibitemOpen
  \bibfield  {author} {\bibinfo {author} {\bibfnamefont {C.}~\bibnamefont
  {Ortiz~Marrero}}, \bibinfo {author} {\bibfnamefont {M.}~\bibnamefont
  {Kieferov{\'a}}},\ and\ \bibinfo {author} {\bibfnamefont {N.}~\bibnamefont
  {Wiebe}},\ }\bibfield  {title} {\bibinfo {title} {Entanglement-{{Induced
  Barren Plateaus}}},\ }\href {https://doi.org/10.1103/PRXQuantum.2.040316}
  {\bibfield  {journal} {\bibinfo  {journal} {PRX Quantum}\ }\textbf {\bibinfo
  {volume} {2}},\ \bibinfo {pages} {040316} (\bibinfo {year}
  {2021})}\BibitemShut {NoStop}%
\bibitem [{\citenamefont {Uvarov}\ and\ \citenamefont
  {Biamonte}(2021)}]{uvarov2021a}%
  \BibitemOpen
  \bibfield  {author} {\bibinfo {author} {\bibfnamefont {A.~V.}\ \bibnamefont
  {Uvarov}}\ and\ \bibinfo {author} {\bibfnamefont {J.~D.}\ \bibnamefont
  {Biamonte}},\ }\bibfield  {title} {\bibinfo {title} {On barren plateaus and
  cost function locality in variational quantum algorithms},\ }\href
  {https://doi.org/10.1088/1751-8121/abfac7} {\bibfield  {journal} {\bibinfo
  {journal} {Journal of Physics A: Mathematical and Theoretical}\ }\textbf
  {\bibinfo {volume} {54}},\ \bibinfo {pages} {245301} (\bibinfo {year}
  {2021})}\BibitemShut {NoStop}%
\bibitem [{\citenamefont {Cerezo}\ \emph
  {et~al.}(2021{\natexlab{b}})\citenamefont {Cerezo}, \citenamefont {Sone},
  \citenamefont {Volkoff}, \citenamefont {Cincio},\ and\ \citenamefont
  {Coles}}]{cerezo2021c}%
  \BibitemOpen
  \bibfield  {author} {\bibinfo {author} {\bibfnamefont {M.}~\bibnamefont
  {Cerezo}}, \bibinfo {author} {\bibfnamefont {A.}~\bibnamefont {Sone}},
  \bibinfo {author} {\bibfnamefont {T.}~\bibnamefont {Volkoff}}, \bibinfo
  {author} {\bibfnamefont {L.}~\bibnamefont {Cincio}},\ and\ \bibinfo {author}
  {\bibfnamefont {P.~J.}\ \bibnamefont {Coles}},\ }\bibfield  {title} {\bibinfo
  {title} {Cost function dependent barren plateaus in shallow parametrized
  quantum circuits},\ }\href {https://doi.org/10.1038/s41467-021-21728-w}
  {\bibfield  {journal} {\bibinfo  {journal} {Nature Communications}\ }\textbf
  {\bibinfo {volume} {12}},\ \bibinfo {pages} {1791} (\bibinfo {year}
  {2021}{\natexlab{b}})}\BibitemShut {NoStop}%
\bibitem [{\citenamefont {Patti}\ \emph {et~al.}(2021)\citenamefont {Patti},
  \citenamefont {Najafi}, \citenamefont {Gao},\ and\ \citenamefont
  {Yelin}}]{patti2021a}%
  \BibitemOpen
  \bibfield  {author} {\bibinfo {author} {\bibfnamefont {T.~L.}\ \bibnamefont
  {Patti}}, \bibinfo {author} {\bibfnamefont {K.}~\bibnamefont {Najafi}},
  \bibinfo {author} {\bibfnamefont {X.}~\bibnamefont {Gao}},\ and\ \bibinfo
  {author} {\bibfnamefont {S.~F.}\ \bibnamefont {Yelin}},\ }\bibfield  {title}
  {\bibinfo {title} {Entanglement devised barren plateau mitigation},\ }\href
  {https://doi.org/10.1103/PhysRevResearch.3.033090} {\bibfield  {journal}
  {\bibinfo  {journal} {Physical Review Research}\ }\textbf {\bibinfo {volume}
  {3}},\ \bibinfo {pages} {033090} (\bibinfo {year} {2021})}\BibitemShut
  {NoStop}%
\bibitem [{\citenamefont {Wiersema}\ \emph {et~al.}(2021)\citenamefont
  {Wiersema}, \citenamefont {Zhou}, \citenamefont {Carrasquilla},\ and\
  \citenamefont {Kim}}]{wiersema2021}%
  \BibitemOpen
  \bibfield  {author} {\bibinfo {author} {\bibfnamefont {R.}~\bibnamefont
  {Wiersema}}, \bibinfo {author} {\bibfnamefont {C.}~\bibnamefont {Zhou}},
  \bibinfo {author} {\bibfnamefont {J.~F.}\ \bibnamefont {Carrasquilla}},\ and\
  \bibinfo {author} {\bibfnamefont {Y.~B.}\ \bibnamefont {Kim}},\ }\href@noop
  {} {\bibinfo {title} {Measurement-induced entanglement phase transitions in
  variational quantum circuits}} (\bibinfo {year} {2021}),\ \Eprint
  {https://arxiv.org/abs/2111.08035} {arXiv:2111.08035} \BibitemShut {NoStop}%
\bibitem [{\citenamefont {Kim}\ and\ \citenamefont
  {Oz}(2022{\natexlab{a}})}]{kim2022}%
  \BibitemOpen
  \bibfield  {author} {\bibinfo {author} {\bibfnamefont {J.}~\bibnamefont
  {Kim}}\ and\ \bibinfo {author} {\bibfnamefont {Y.}~\bibnamefont {Oz}},\
  }\bibfield  {title} {\bibinfo {title} {Entanglement {{Diagnostics}} for
  {{Efficient Quantum Computation}}},\ }\href
  {https://doi.org/10.1088/1742-5468/ac7791} {\bibfield  {journal} {\bibinfo
  {journal} {Journal of Statistical Mechanics: Theory and Experiment}\ }\textbf
  {\bibinfo {volume} {2022}},\ \bibinfo {pages} {073101} (\bibinfo {year}
  {2022}{\natexlab{a}})},\ \Eprint {https://arxiv.org/abs/2102.12534}
  {arXiv:2102.12534} \BibitemShut {NoStop}%
\bibitem [{\citenamefont {Kim}\ and\ \citenamefont
  {Oz}(2022{\natexlab{b}})}]{kim2022a}%
  \BibitemOpen
  \bibfield  {author} {\bibinfo {author} {\bibfnamefont {J.}~\bibnamefont
  {Kim}}\ and\ \bibinfo {author} {\bibfnamefont {Y.}~\bibnamefont {Oz}},\
  }\bibfield  {title} {\bibinfo {title} {Quantum energy landscape and circuit
  optimization},\ }\href {https://doi.org/10.1103/PhysRevA.106.052424}
  {\bibfield  {journal} {\bibinfo  {journal} {Physical Review A}\ }\textbf
  {\bibinfo {volume} {106}},\ \bibinfo {pages} {052424} (\bibinfo {year}
  {2022}{\natexlab{b}})}\BibitemShut {NoStop}%
\bibitem [{\citenamefont {Sack}\ \emph {et~al.}(2022)\citenamefont {Sack},
  \citenamefont {Medina}, \citenamefont {Michailidis}, \citenamefont {Kueng},\
  and\ \citenamefont {Serbyn}}]{sack2022}%
  \BibitemOpen
  \bibfield  {author} {\bibinfo {author} {\bibfnamefont {S.~H.}\ \bibnamefont
  {Sack}}, \bibinfo {author} {\bibfnamefont {R.~A.}\ \bibnamefont {Medina}},
  \bibinfo {author} {\bibfnamefont {A.~A.}\ \bibnamefont {Michailidis}},
  \bibinfo {author} {\bibfnamefont {R.}~\bibnamefont {Kueng}},\ and\ \bibinfo
  {author} {\bibfnamefont {M.}~\bibnamefont {Serbyn}},\ }\bibfield  {title}
  {\bibinfo {title} {Avoiding {{Barren Plateaus Using Classical Shadows}}},\
  }\href {https://doi.org/10.1103/PRXQuantum.3.020365} {\bibfield  {journal}
  {\bibinfo  {journal} {PRX Quantum}\ }\textbf {\bibinfo {volume} {3}},\
  \bibinfo {pages} {020365} (\bibinfo {year} {2022})}\BibitemShut {NoStop}%
\bibitem [{\citenamefont {Friedrich}\ and\ \citenamefont
  {Maziero}(2022)}]{friedrich2022}%
  \BibitemOpen
  \bibfield  {author} {\bibinfo {author} {\bibfnamefont {L.}~\bibnamefont
  {Friedrich}}\ and\ \bibinfo {author} {\bibfnamefont {J.}~\bibnamefont
  {Maziero}},\ }\bibfield  {title} {\bibinfo {title} {Avoiding barren plateaus
  with classical deep neural networks},\ }\href
  {https://doi.org/10.1103/PhysRevA.106.042433} {\bibfield  {journal} {\bibinfo
   {journal} {Physical Review A}\ }\textbf {\bibinfo {volume} {106}},\ \bibinfo
  {pages} {042433} (\bibinfo {year} {2022})}\BibitemShut {NoStop}%
\bibitem [{\citenamefont {Grant}\ \emph {et~al.}(2019)\citenamefont {Grant},
  \citenamefont {Wossnig}, \citenamefont {Ostaszewski},\ and\ \citenamefont
  {Benedetti}}]{grant2019a}%
  \BibitemOpen
  \bibfield  {author} {\bibinfo {author} {\bibfnamefont {E.}~\bibnamefont
  {Grant}}, \bibinfo {author} {\bibfnamefont {L.}~\bibnamefont {Wossnig}},
  \bibinfo {author} {\bibfnamefont {M.}~\bibnamefont {Ostaszewski}},\ and\
  \bibinfo {author} {\bibfnamefont {M.}~\bibnamefont {Benedetti}},\ }\bibfield
  {title} {\bibinfo {title} {An initialization strategy for addressing barren
  plateaus in parametrized quantum circuits},\ }\href
  {https://doi.org/10.22331/q-2019-12-09-214} {\bibfield  {journal} {\bibinfo
  {journal} {Quantum}\ }\textbf {\bibinfo {volume} {3}},\ \bibinfo {pages}
  {214} (\bibinfo {year} {2019})}\BibitemShut {NoStop}%
\bibitem [{\citenamefont {Liu}\ \emph {et~al.}(2022{\natexlab{a}})\citenamefont
  {Liu}, \citenamefont {Chen}, \citenamefont {Sun}, \citenamefont {Wu},
  \citenamefont {Han},\ and\ \citenamefont {Guo}}]{liu2022}%
  \BibitemOpen
  \bibfield  {author} {\bibinfo {author} {\bibfnamefont {H.-Y.}\ \bibnamefont
  {Liu}}, \bibinfo {author} {\bibfnamefont {Z.-Y.}\ \bibnamefont {Chen}},
  \bibinfo {author} {\bibfnamefont {T.-P.}\ \bibnamefont {Sun}}, \bibinfo
  {author} {\bibfnamefont {Y.-C.}\ \bibnamefont {Wu}}, \bibinfo {author}
  {\bibfnamefont {Y.-J.}\ \bibnamefont {Han}},\ and\ \bibinfo {author}
  {\bibfnamefont {G.-P.}\ \bibnamefont {Guo}},\ }\href@noop {} {\bibinfo
  {title} {Mitigating {{Barren Plateaus}} with {{Transfer-learning-inspired
  Parameter Initializations}}}} (\bibinfo {year} {2022}{\natexlab{a}}),\
  \Eprint {https://arxiv.org/abs/2112.10952} {arXiv:2112.10952} \BibitemShut
  {NoStop}%
\bibitem [{\citenamefont {Mitarai}\ \emph {et~al.}(2022)\citenamefont
  {Mitarai}, \citenamefont {Suzuki}, \citenamefont {Mizukami}, \citenamefont
  {Nakagawa},\ and\ \citenamefont {Fujii}}]{mitarai2022a}%
  \BibitemOpen
  \bibfield  {author} {\bibinfo {author} {\bibfnamefont {K.}~\bibnamefont
  {Mitarai}}, \bibinfo {author} {\bibfnamefont {Y.}~\bibnamefont {Suzuki}},
  \bibinfo {author} {\bibfnamefont {W.}~\bibnamefont {Mizukami}}, \bibinfo
  {author} {\bibfnamefont {Y.~O.}\ \bibnamefont {Nakagawa}},\ and\ \bibinfo
  {author} {\bibfnamefont {K.}~\bibnamefont {Fujii}},\ }\bibfield  {title}
  {\bibinfo {title} {Quadratic {{Clifford}} expansion for efficient
  benchmarking and initialization of variational quantum algorithms},\ }\href
  {https://doi.org/10.1103/PhysRevResearch.4.033012} {\bibfield  {journal}
  {\bibinfo  {journal} {Physical Review Research}\ }\textbf {\bibinfo {volume}
  {4}},\ \bibinfo {pages} {033012} (\bibinfo {year} {2022})}\BibitemShut
  {NoStop}%
\bibitem [{\citenamefont {Ravi}\ \emph {et~al.}(2022)\citenamefont {Ravi},
  \citenamefont {Gokhale}, \citenamefont {Ding}, \citenamefont {Kirby},
  \citenamefont {Smith}, \citenamefont {Baker}, \citenamefont {Love},
  \citenamefont {Hoffmann}, \citenamefont {Brown},\ and\ \citenamefont
  {Chong}}]{ravi2022}%
  \BibitemOpen
  \bibfield  {author} {\bibinfo {author} {\bibfnamefont {G.~S.}\ \bibnamefont
  {Ravi}}, \bibinfo {author} {\bibfnamefont {P.}~\bibnamefont {Gokhale}},
  \bibinfo {author} {\bibfnamefont {Y.}~\bibnamefont {Ding}}, \bibinfo {author}
  {\bibfnamefont {W.~M.}\ \bibnamefont {Kirby}}, \bibinfo {author}
  {\bibfnamefont {K.~N.}\ \bibnamefont {Smith}}, \bibinfo {author}
  {\bibfnamefont {J.~M.}\ \bibnamefont {Baker}}, \bibinfo {author}
  {\bibfnamefont {P.~J.}\ \bibnamefont {Love}}, \bibinfo {author}
  {\bibfnamefont {H.}~\bibnamefont {Hoffmann}}, \bibinfo {author}
  {\bibfnamefont {K.~R.}\ \bibnamefont {Brown}},\ and\ \bibinfo {author}
  {\bibfnamefont {F.~T.}\ \bibnamefont {Chong}},\ }\href@noop {} {\bibinfo
  {title} {{{CAFQA}}: {{A}} classical simulation bootstrap for variational
  quantum algorithms}} (\bibinfo {year} {2022}),\ \Eprint
  {https://arxiv.org/abs/2202.12924} {arXiv:2202.12924} \BibitemShut {NoStop}%
\bibitem [{\citenamefont {Kim}\ \emph {et~al.}(2021)\citenamefont {Kim},
  \citenamefont {Kim},\ and\ \citenamefont {Rosa}}]{kim2021}%
  \BibitemOpen
  \bibfield  {author} {\bibinfo {author} {\bibfnamefont {J.}~\bibnamefont
  {Kim}}, \bibinfo {author} {\bibfnamefont {J.}~\bibnamefont {Kim}},\ and\
  \bibinfo {author} {\bibfnamefont {D.}~\bibnamefont {Rosa}},\ }\bibfield
  {title} {\bibinfo {title} {Universal effectiveness of high-depth circuits in
  variational eigenproblems},\ }\href
  {https://doi.org/10.1103/PhysRevResearch.3.023203} {\bibfield  {journal}
  {\bibinfo  {journal} {Physical Review Research}\ }\textbf {\bibinfo {volume}
  {3}},\ \bibinfo {pages} {023203} (\bibinfo {year} {2021})}\BibitemShut
  {NoStop}%
\bibitem [{\citenamefont {Kim}\ \emph {et~al.}(2022)\citenamefont {Kim},
  \citenamefont {Oz},\ and\ \citenamefont {Rosa}}]{kim2022b}%
  \BibitemOpen
  \bibfield  {author} {\bibinfo {author} {\bibfnamefont {J.}~\bibnamefont
  {Kim}}, \bibinfo {author} {\bibfnamefont {Y.}~\bibnamefont {Oz}},\ and\
  \bibinfo {author} {\bibfnamefont {D.}~\bibnamefont {Rosa}},\ }\href@noop {}
  {\bibinfo {title} {Quantum {{Chaos}} and {{Circuit Parameter Optimization}}}}
  (\bibinfo {year} {2022}),\ \Eprint {https://arxiv.org/abs/2201.01452}
  {arXiv:2201.01452} \BibitemShut {NoStop}%
\bibitem [{\citenamefont {Cheng}\ \emph {et~al.}(2022)\citenamefont {Cheng},
  \citenamefont {Khosla}, \citenamefont {Self}, \citenamefont {Lin},
  \citenamefont {Li}, \citenamefont {Medina},\ and\ \citenamefont
  {Kim}}]{cheng2022}%
  \BibitemOpen
  \bibfield  {author} {\bibinfo {author} {\bibfnamefont {M.~H.}\ \bibnamefont
  {Cheng}}, \bibinfo {author} {\bibfnamefont {K.~E.}\ \bibnamefont {Khosla}},
  \bibinfo {author} {\bibfnamefont {C.~N.}\ \bibnamefont {Self}}, \bibinfo
  {author} {\bibfnamefont {M.}~\bibnamefont {Lin}}, \bibinfo {author}
  {\bibfnamefont {B.~X.}\ \bibnamefont {Li}}, \bibinfo {author} {\bibfnamefont
  {A.~C.}\ \bibnamefont {Medina}},\ and\ \bibinfo {author} {\bibfnamefont
  {M.~S.}\ \bibnamefont {Kim}},\ }\href@noop {} {\bibinfo {title} {Clifford
  {{Circuit Initialisation}} for {{Variational Quantum Algorithms}}}} (\bibinfo
  {year} {2022}),\ \Eprint {https://arxiv.org/abs/2207.01539}
  {arXiv:2207.01539} \BibitemShut {NoStop}%
\bibitem [{\citenamefont {Dborin}\ \emph {et~al.}(2022)\citenamefont {Dborin},
  \citenamefont {Barratt}, \citenamefont {Wimalaweera}, \citenamefont
  {Wright},\ and\ \citenamefont {Green}}]{dborin2022}%
  \BibitemOpen
  \bibfield  {author} {\bibinfo {author} {\bibfnamefont {J.}~\bibnamefont
  {Dborin}}, \bibinfo {author} {\bibfnamefont {F.}~\bibnamefont {Barratt}},
  \bibinfo {author} {\bibfnamefont {V.}~\bibnamefont {Wimalaweera}}, \bibinfo
  {author} {\bibfnamefont {L.}~\bibnamefont {Wright}},\ and\ \bibinfo {author}
  {\bibfnamefont {A.~G.}\ \bibnamefont {Green}},\ }\bibfield  {title} {\bibinfo
  {title} {Matrix product state pre-training for quantum machine learning},\
  }\href {https://doi.org/10.1088/2058-9565/ac7073} {\bibfield  {journal}
  {\bibinfo  {journal} {Quantum Science and Technology}\ }\textbf {\bibinfo
  {volume} {7}},\ \bibinfo {pages} {035014} (\bibinfo {year}
  {2022})}\BibitemShut {NoStop}%
\bibitem [{\citenamefont {Pesah}\ \emph {et~al.}(2021)\citenamefont {Pesah},
  \citenamefont {Cerezo}, \citenamefont {Wang}, \citenamefont {Volkoff},
  \citenamefont {Sornborger},\ and\ \citenamefont {Coles}}]{pesah2021}%
  \BibitemOpen
  \bibfield  {author} {\bibinfo {author} {\bibfnamefont {A.}~\bibnamefont
  {Pesah}}, \bibinfo {author} {\bibfnamefont {M.}~\bibnamefont {Cerezo}},
  \bibinfo {author} {\bibfnamefont {S.}~\bibnamefont {Wang}}, \bibinfo {author}
  {\bibfnamefont {T.}~\bibnamefont {Volkoff}}, \bibinfo {author} {\bibfnamefont
  {A.~T.}\ \bibnamefont {Sornborger}},\ and\ \bibinfo {author} {\bibfnamefont
  {P.~J.}\ \bibnamefont {Coles}},\ }\bibfield  {title} {\bibinfo {title}
  {Absence of {{Barren Plateaus}} in {{Quantum Convolutional Neural
  Networks}}},\ }\href {https://doi.org/10.1103/PhysRevX.11.041011} {\bibfield
  {journal} {\bibinfo  {journal} {Physical Review X}\ }\textbf {\bibinfo
  {volume} {11}},\ \bibinfo {pages} {041011} (\bibinfo {year}
  {2021})}\BibitemShut {NoStop}%
\bibitem [{\citenamefont {Schatzki}\ \emph {et~al.}(2022)\citenamefont
  {Schatzki}, \citenamefont {Larocca}, \citenamefont {Nguyen}, \citenamefont
  {Sauvage},\ and\ \citenamefont {Cerezo}}]{schatzki2022}%
  \BibitemOpen
  \bibfield  {author} {\bibinfo {author} {\bibfnamefont {L.}~\bibnamefont
  {Schatzki}}, \bibinfo {author} {\bibfnamefont {M.}~\bibnamefont {Larocca}},
  \bibinfo {author} {\bibfnamefont {Q.~T.}\ \bibnamefont {Nguyen}}, \bibinfo
  {author} {\bibfnamefont {F.}~\bibnamefont {Sauvage}},\ and\ \bibinfo {author}
  {\bibfnamefont {M.}~\bibnamefont {Cerezo}},\ }\href@noop {} {\bibinfo {title}
  {Theoretical {{Guarantees}} for {{Permutation-Equivariant Quantum Neural
  Networks}}}} (\bibinfo {year} {2022}),\ \Eprint
  {https://arxiv.org/abs/2210.09974} {arXiv:2210.09974} \BibitemShut {NoStop}%
\bibitem [{\citenamefont {Holmes}\ \emph {et~al.}(2021)\citenamefont {Holmes},
  \citenamefont {Arrasmith}, \citenamefont {Yan}, \citenamefont {Coles},
  \citenamefont {Albrecht},\ and\ \citenamefont {Sornborger}}]{holmes2021a}%
  \BibitemOpen
  \bibfield  {author} {\bibinfo {author} {\bibfnamefont {Z.}~\bibnamefont
  {Holmes}}, \bibinfo {author} {\bibfnamefont {A.}~\bibnamefont {Arrasmith}},
  \bibinfo {author} {\bibfnamefont {B.}~\bibnamefont {Yan}}, \bibinfo {author}
  {\bibfnamefont {P.~J.}\ \bibnamefont {Coles}}, \bibinfo {author}
  {\bibfnamefont {A.}~\bibnamefont {Albrecht}},\ and\ \bibinfo {author}
  {\bibfnamefont {A.~T.}\ \bibnamefont {Sornborger}},\ }\bibfield  {title}
  {\bibinfo {title} {Barren {{Plateaus Preclude Learning Scramblers}}},\ }\href
  {https://doi.org/10.1103/PhysRevLett.126.190501} {\bibfield  {journal}
  {\bibinfo  {journal} {Physical Review Letters}\ }\textbf {\bibinfo {volume}
  {126}},\ \bibinfo {pages} {190501} (\bibinfo {year} {2021})}\BibitemShut
  {NoStop}%
\bibitem [{\citenamefont {Arrasmith}\ \emph {et~al.}(2021)\citenamefont
  {Arrasmith}, \citenamefont {Cerezo}, \citenamefont {Czarnik}, \citenamefont
  {Cincio},\ and\ \citenamefont {Coles}}]{arrasmith2021a}%
  \BibitemOpen
  \bibfield  {author} {\bibinfo {author} {\bibfnamefont {A.}~\bibnamefont
  {Arrasmith}}, \bibinfo {author} {\bibfnamefont {M.}~\bibnamefont {Cerezo}},
  \bibinfo {author} {\bibfnamefont {P.}~\bibnamefont {Czarnik}}, \bibinfo
  {author} {\bibfnamefont {L.}~\bibnamefont {Cincio}},\ and\ \bibinfo {author}
  {\bibfnamefont {P.~J.}\ \bibnamefont {Coles}},\ }\bibfield  {title} {\bibinfo
  {title} {Effect of barren plateaus on gradient-free optimization},\ }\href
  {https://doi.org/10.22331/q-2021-10-05-558} {\bibfield  {journal} {\bibinfo
  {journal} {Quantum}\ }\textbf {\bibinfo {volume} {5}},\ \bibinfo {pages}
  {558} (\bibinfo {year} {2021})}\BibitemShut {NoStop}%
\bibitem [{\citenamefont {Arrasmith}\ \emph {et~al.}(2022)\citenamefont
  {Arrasmith}, \citenamefont {Holmes}, \citenamefont {Cerezo},\ and\
  \citenamefont {Coles}}]{arrasmith2022}%
  \BibitemOpen
  \bibfield  {author} {\bibinfo {author} {\bibfnamefont {A.}~\bibnamefont
  {Arrasmith}}, \bibinfo {author} {\bibfnamefont {Z.}~\bibnamefont {Holmes}},
  \bibinfo {author} {\bibfnamefont {M.}~\bibnamefont {Cerezo}},\ and\ \bibinfo
  {author} {\bibfnamefont {P.~J.}\ \bibnamefont {Coles}},\ }\bibfield  {title}
  {\bibinfo {title} {Equivalence of quantum barren plateaus to cost
  concentration and narrow gorges},\ }\href
  {https://doi.org/10.1088/2058-9565/ac7d06} {\bibfield  {journal} {\bibinfo
  {journal} {Quantum Science and Technology}\ }\textbf {\bibinfo {volume}
  {7}},\ \bibinfo {pages} {045015} (\bibinfo {year} {2022})}\BibitemShut
  {NoStop}%
\bibitem [{\citenamefont {Wang}\ \emph
  {et~al.}(2021{\natexlab{b}})\citenamefont {Wang}, \citenamefont {Czarnik},
  \citenamefont {Arrasmith}, \citenamefont {Cerezo}, \citenamefont {Cincio},\
  and\ \citenamefont {Coles}}]{wang2021d}%
  \BibitemOpen
  \bibfield  {author} {\bibinfo {author} {\bibfnamefont {S.}~\bibnamefont
  {Wang}}, \bibinfo {author} {\bibfnamefont {P.}~\bibnamefont {Czarnik}},
  \bibinfo {author} {\bibfnamefont {A.}~\bibnamefont {Arrasmith}}, \bibinfo
  {author} {\bibfnamefont {M.}~\bibnamefont {Cerezo}}, \bibinfo {author}
  {\bibfnamefont {L.}~\bibnamefont {Cincio}},\ and\ \bibinfo {author}
  {\bibfnamefont {P.~J.}\ \bibnamefont {Coles}},\ }\href@noop {} {\bibinfo
  {title} {Can {{Error Mitigation Improve Trainability}} of {{Noisy Variational
  Quantum Algorithms}}?}} (\bibinfo {year} {2021}{\natexlab{b}}),\ \Eprint
  {https://arxiv.org/abs/2109.01051} {arXiv:2109.01051} \BibitemShut {NoStop}%
\bibitem [{\citenamefont {Du}\ \emph {et~al.}(2022)\citenamefont {Du},
  \citenamefont {Huang}, \citenamefont {You}, \citenamefont {Hsieh},\ and\
  \citenamefont {Tao}}]{du2022a}%
  \BibitemOpen
  \bibfield  {author} {\bibinfo {author} {\bibfnamefont {Y.}~\bibnamefont
  {Du}}, \bibinfo {author} {\bibfnamefont {T.}~\bibnamefont {Huang}}, \bibinfo
  {author} {\bibfnamefont {S.}~\bibnamefont {You}}, \bibinfo {author}
  {\bibfnamefont {M.-H.}\ \bibnamefont {Hsieh}},\ and\ \bibinfo {author}
  {\bibfnamefont {D.}~\bibnamefont {Tao}},\ }\bibfield  {title} {\bibinfo
  {title} {Quantum circuit architecture search for variational quantum
  algorithms},\ }\href {https://doi.org/10.1038/s41534-022-00570-y} {\bibfield
  {journal} {\bibinfo  {journal} {npj Quantum Information}\ }\textbf {\bibinfo
  {volume} {8}},\ \bibinfo {pages} {1} (\bibinfo {year} {2022})}\BibitemShut
  {NoStop}%
\bibitem [{\citenamefont {Sharma}\ \emph {et~al.}(2022)\citenamefont {Sharma},
  \citenamefont {Cerezo}, \citenamefont {Cincio},\ and\ \citenamefont
  {Coles}}]{sharma2022}%
  \BibitemOpen
  \bibfield  {author} {\bibinfo {author} {\bibfnamefont {K.}~\bibnamefont
  {Sharma}}, \bibinfo {author} {\bibfnamefont {M.}~\bibnamefont {Cerezo}},
  \bibinfo {author} {\bibfnamefont {L.}~\bibnamefont {Cincio}},\ and\ \bibinfo
  {author} {\bibfnamefont {P.~J.}\ \bibnamefont {Coles}},\ }\bibfield  {title}
  {\bibinfo {title} {Trainability of {{Dissipative Perceptron-Based Quantum
  Neural Networks}}},\ }\href {https://doi.org/10.1103/PhysRevLett.128.180505}
  {\bibfield  {journal} {\bibinfo  {journal} {Physical Review Letters}\
  }\textbf {\bibinfo {volume} {128}},\ \bibinfo {pages} {180505} (\bibinfo
  {year} {2022})}\BibitemShut {NoStop}%
\bibitem [{\citenamefont {De~Palma}\ \emph {et~al.}(2023)\citenamefont
  {De~Palma}, \citenamefont {Marvian}, \citenamefont {Rouz{\'e}},\ and\
  \citenamefont {Fran{\c c}a}}]{depalma2023}%
  \BibitemOpen
  \bibfield  {author} {\bibinfo {author} {\bibfnamefont {G.}~\bibnamefont
  {De~Palma}}, \bibinfo {author} {\bibfnamefont {M.}~\bibnamefont {Marvian}},
  \bibinfo {author} {\bibfnamefont {C.}~\bibnamefont {Rouz{\'e}}},\ and\
  \bibinfo {author} {\bibfnamefont {D.~S.}\ \bibnamefont {Fran{\c c}a}},\
  }\bibfield  {title} {\bibinfo {title} {Limitations of {{Variational Quantum
  Algorithms}}: {{A Quantum Optimal Transport Approach}}},\ }\href
  {https://doi.org/10.1103/PRXQuantum.4.010309} {\bibfield  {journal} {\bibinfo
   {journal} {PRX Quantum}\ }\textbf {\bibinfo {volume} {4}},\ \bibinfo {pages}
  {010309} (\bibinfo {year} {2023})}\BibitemShut {NoStop}%
\bibitem [{\citenamefont {Heyraud}\ \emph {et~al.}(2022)\citenamefont
  {Heyraud}, \citenamefont {Li}, \citenamefont {Denis}, \citenamefont
  {Le~Boit{\'e}},\ and\ \citenamefont {Ciuti}}]{heyraud2022}%
  \BibitemOpen
  \bibfield  {author} {\bibinfo {author} {\bibfnamefont {V.}~\bibnamefont
  {Heyraud}}, \bibinfo {author} {\bibfnamefont {Z.}~\bibnamefont {Li}},
  \bibinfo {author} {\bibfnamefont {Z.}~\bibnamefont {Denis}}, \bibinfo
  {author} {\bibfnamefont {A.}~\bibnamefont {Le~Boit{\'e}}},\ and\ \bibinfo
  {author} {\bibfnamefont {C.}~\bibnamefont {Ciuti}},\ }\bibfield  {title}
  {\bibinfo {title} {Noisy quantum kernel machines},\ }\href
  {https://doi.org/10.1103/PhysRevA.106.052421} {\bibfield  {journal} {\bibinfo
   {journal} {Physical Review A}\ }\textbf {\bibinfo {volume} {106}},\ \bibinfo
  {pages} {052421} (\bibinfo {year} {2022})}\BibitemShut {NoStop}%
\bibitem [{\citenamefont {Jerbi}\ \emph {et~al.}(2023)\citenamefont {Jerbi},
  \citenamefont {Fiderer}, \citenamefont {Poulsen~Nautrup}, \citenamefont
  {K{\"u}bler}, \citenamefont {Briegel},\ and\ \citenamefont
  {Dunjko}}]{jerbi2023}%
  \BibitemOpen
  \bibfield  {author} {\bibinfo {author} {\bibfnamefont {S.}~\bibnamefont
  {Jerbi}}, \bibinfo {author} {\bibfnamefont {L.~J.}\ \bibnamefont {Fiderer}},
  \bibinfo {author} {\bibfnamefont {H.}~\bibnamefont {Poulsen~Nautrup}},
  \bibinfo {author} {\bibfnamefont {J.~M.}\ \bibnamefont {K{\"u}bler}},
  \bibinfo {author} {\bibfnamefont {H.~J.}\ \bibnamefont {Briegel}},\ and\
  \bibinfo {author} {\bibfnamefont {V.}~\bibnamefont {Dunjko}},\ }\bibfield
  {title} {\bibinfo {title} {Quantum machine learning beyond kernel methods},\
  }\href {https://doi.org/10.1038/s41467-023-36159-y} {\bibfield  {journal}
  {\bibinfo  {journal} {Nature Communications}\ }\textbf {\bibinfo {volume}
  {14}},\ \bibinfo {pages} {517} (\bibinfo {year} {2023})}\BibitemShut
  {NoStop}%
\bibitem [{\citenamefont {Li}\ \emph {et~al.}(2022)\citenamefont {Li},
  \citenamefont {Heyraud}, \citenamefont {Donatella}, \citenamefont {Denis},\
  and\ \citenamefont {Ciuti}}]{li2022}%
  \BibitemOpen
  \bibfield  {author} {\bibinfo {author} {\bibfnamefont {Z.}~\bibnamefont
  {Li}}, \bibinfo {author} {\bibfnamefont {V.}~\bibnamefont {Heyraud}},
  \bibinfo {author} {\bibfnamefont {K.}~\bibnamefont {Donatella}}, \bibinfo
  {author} {\bibfnamefont {Z.}~\bibnamefont {Denis}},\ and\ \bibinfo {author}
  {\bibfnamefont {C.}~\bibnamefont {Ciuti}},\ }\bibfield  {title} {\bibinfo
  {title} {Machine learning via relativity-inspired quantum dynamics},\ }\href
  {https://doi.org/10.1103/PhysRevA.106.032413} {\bibfield  {journal} {\bibinfo
   {journal} {Physical Review A}\ }\textbf {\bibinfo {volume} {106}},\ \bibinfo
  {pages} {032413} (\bibinfo {year} {2022})}\BibitemShut {NoStop}%
\bibitem [{\citenamefont {Schuld}(2021)}]{schuld2021a}%
  \BibitemOpen
  \bibfield  {author} {\bibinfo {author} {\bibfnamefont {M.}~\bibnamefont
  {Schuld}},\ }\href@noop {} {\bibinfo {title} {Supervised quantum machine
  learning models are kernel methods}} (\bibinfo {year} {2021}),\ \Eprint
  {https://arxiv.org/abs/2101.11020} {arXiv:2101.11020} \BibitemShut {NoStop}%
\bibitem [{\citenamefont {Rebentrost}\ \emph {et~al.}(2014)\citenamefont
  {Rebentrost}, \citenamefont {Mohseni},\ and\ \citenamefont
  {Lloyd}}]{rebentrost2014a}%
  \BibitemOpen
  \bibfield  {author} {\bibinfo {author} {\bibfnamefont {P.}~\bibnamefont
  {Rebentrost}}, \bibinfo {author} {\bibfnamefont {M.}~\bibnamefont
  {Mohseni}},\ and\ \bibinfo {author} {\bibfnamefont {S.}~\bibnamefont
  {Lloyd}},\ }\bibfield  {title} {\bibinfo {title} {Quantum {{Support Vector
  Machine}} for {{Big Data Classification}}},\ }\href
  {https://doi.org/10.1103/PhysRevLett.113.130503} {\bibfield  {journal}
  {\bibinfo  {journal} {Physical Review Letters}\ }\textbf {\bibinfo {volume}
  {113}},\ \bibinfo {pages} {130503} (\bibinfo {year} {2014})}\BibitemShut
  {NoStop}%
\bibitem [{\citenamefont {Mujal}\ \emph {et~al.}(2021)\citenamefont {Mujal},
  \citenamefont {Mart{\'{\i}}nez-Pe{\~{n}}a}, \citenamefont {Nokkala},
  \citenamefont {Garc{\'{\i}}a-Beni}, \citenamefont {Giorgi}, \citenamefont
  {Soriano},\ and\ \citenamefont {Zambrini}}]{mujal2021}%
  \BibitemOpen
  \bibfield  {author} {\bibinfo {author} {\bibfnamefont {P.}~\bibnamefont
  {Mujal}}, \bibinfo {author} {\bibfnamefont {R.}~\bibnamefont
  {Mart{\'{\i}}nez-Pe{\~{n}}a}}, \bibinfo {author} {\bibfnamefont
  {J.}~\bibnamefont {Nokkala}}, \bibinfo {author} {\bibfnamefont
  {J.}~\bibnamefont {Garc{\'{\i}}a-Beni}}, \bibinfo {author} {\bibfnamefont
  {G.~L.}\ \bibnamefont {Giorgi}}, \bibinfo {author} {\bibfnamefont {M.~C.}\
  \bibnamefont {Soriano}},\ and\ \bibinfo {author} {\bibfnamefont
  {R.}~\bibnamefont {Zambrini}},\ }\bibfield  {title} {\bibinfo {title}
  {Opportunities in quantum reservoir computing and extreme learning
  machines},\ }\href {https://doi.org/10.1002/qute.202100027} {\bibfield
  {journal} {\bibinfo  {journal} {Advanced Quantum Technologies}\ }\textbf
  {\bibinfo {volume} {4}},\ \bibinfo {pages} {2100027} (\bibinfo {year}
  {2021})}\BibitemShut {NoStop}%
\bibitem [{\citenamefont {Denis}\ \emph {et~al.}(2022)\citenamefont {Denis},
  \citenamefont {Favero},\ and\ \citenamefont {Ciuti}}]{denis2022a}%
  \BibitemOpen
  \bibfield  {author} {\bibinfo {author} {\bibfnamefont {Z.}~\bibnamefont
  {Denis}}, \bibinfo {author} {\bibfnamefont {I.}~\bibnamefont {Favero}},\ and\
  \bibinfo {author} {\bibfnamefont {C.}~\bibnamefont {Ciuti}},\ }\bibfield
  {title} {\bibinfo {title} {Photonic {{Kernel Machine Learning}} for
  {{Ultrafast Spectral Analysis}}},\ }\href
  {https://doi.org/10.1103/PhysRevApplied.17.034077} {\bibfield  {journal}
  {\bibinfo  {journal} {Physical Review Applied}\ }\textbf {\bibinfo {volume}
  {17}},\ \bibinfo {pages} {034077} (\bibinfo {year} {2022})}\BibitemShut
  {NoStop}%
\bibitem [{\citenamefont {Marcucci}\ \emph {et~al.}(2020)\citenamefont
  {Marcucci}, \citenamefont {Pierangeli},\ and\ \citenamefont
  {Conti}}]{marcucci2020}%
  \BibitemOpen
  \bibfield  {author} {\bibinfo {author} {\bibfnamefont {G.}~\bibnamefont
  {Marcucci}}, \bibinfo {author} {\bibfnamefont {D.}~\bibnamefont
  {Pierangeli}},\ and\ \bibinfo {author} {\bibfnamefont {C.}~\bibnamefont
  {Conti}},\ }\bibfield  {title} {\bibinfo {title} {Theory of {{Neuromorphic
  Computing}} by {{Waves}}: {{Machine Learning}} by {{Rogue Waves}},
  {{Dispersive Shocks}}, and {{Solitons}}},\ }\href
  {https://doi.org/10.1103/PhysRevLett.125.093901} {\bibfield  {journal}
  {\bibinfo  {journal} {Physical Review Letters}\ }\textbf {\bibinfo {volume}
  {125}},\ \bibinfo {pages} {093901} (\bibinfo {year} {2020})}\BibitemShut
  {NoStop}%
\bibitem [{\citenamefont {Pierangeli}\ \emph {et~al.}(2021)\citenamefont
  {Pierangeli}, \citenamefont {Marcucci},\ and\ \citenamefont
  {Conti}}]{pierangeli2021}%
  \BibitemOpen
  \bibfield  {author} {\bibinfo {author} {\bibfnamefont {D.}~\bibnamefont
  {Pierangeli}}, \bibinfo {author} {\bibfnamefont {G.}~\bibnamefont
  {Marcucci}},\ and\ \bibinfo {author} {\bibfnamefont {C.}~\bibnamefont
  {Conti}},\ }\bibfield  {title} {\bibinfo {title} {Photonic extreme learning
  machine by free-space optical propagation},\ }\href
  {https://doi.org/10.1364/PRJ.423531} {\bibfield  {journal} {\bibinfo
  {journal} {Photonics Research}\ }\textbf {\bibinfo {volume} {9}},\ \bibinfo
  {pages} {1446} (\bibinfo {year} {2021})}\BibitemShut {NoStop}%
\bibitem [{\citenamefont {Thanasilp}\ \emph {et~al.}(2022)\citenamefont
  {Thanasilp}, \citenamefont {Wang}, \citenamefont {Cerezo},\ and\
  \citenamefont {Holmes}}]{thanasilp2022}%
  \BibitemOpen
  \bibfield  {author} {\bibinfo {author} {\bibfnamefont {S.}~\bibnamefont
  {Thanasilp}}, \bibinfo {author} {\bibfnamefont {S.}~\bibnamefont {Wang}},
  \bibinfo {author} {\bibfnamefont {M.}~\bibnamefont {Cerezo}},\ and\ \bibinfo
  {author} {\bibfnamefont {Z.}~\bibnamefont {Holmes}},\ }\href@noop {}
  {\bibinfo {title} {Exponential concentration and untrainability in quantum
  kernel methods}} (\bibinfo {year} {2022}),\ \Eprint
  {https://arxiv.org/abs/2208.11060} {arXiv:2208.11060} \BibitemShut {NoStop}%
\bibitem [{\citenamefont {Watrous}(2018)}]{watrous2018}%
  \BibitemOpen
  \bibfield  {author} {\bibinfo {author} {\bibfnamefont {J.}~\bibnamefont
  {Watrous}},\ }\href@noop {} {\emph {\bibinfo {title} {The {{Theory}} of
  {{Quantum Information}}}}}\ (\bibinfo  {publisher} {{Cambridge University
  Press}},\ \bibinfo {year} {2018})\BibitemShut {NoStop}%
\bibitem [{\citenamefont {Nielsen}\ and\ \citenamefont
  {Chuang}(2010)}]{nielsen2010}%
  \BibitemOpen
  \bibfield  {author} {\bibinfo {author} {\bibfnamefont {M.~A.}\ \bibnamefont
  {Nielsen}}\ and\ \bibinfo {author} {\bibfnamefont {I.~L.}\ \bibnamefont
  {Chuang}},\ }\href@noop {} {\emph {\bibinfo {title} {Quantum {{Computation}}
  and {{Quantum Information}}: 10th {{Anniversary Edition}}}}}\ (\bibinfo
  {publisher} {{Cambridge University Press}},\ \bibinfo {year}
  {2010})\BibitemShut {NoStop}%
\bibitem [{\citenamefont {Gottesman}(1998)}]{gottesman1998}%
  \BibitemOpen
  \bibfield  {author} {\bibinfo {author} {\bibfnamefont {D.}~\bibnamefont
  {Gottesman}},\ }\href@noop {} {\bibinfo {title} {The {{Heisenberg
  Representation}} of {{Quantum Computers}}}} (\bibinfo {year} {1998}),\
  \Eprint {https://arxiv.org/abs/quant-ph/9807006} {arXiv:quant-ph/9807006}
  \BibitemShut {NoStop}%
\bibitem [{\citenamefont {Aaronson}\ and\ \citenamefont
  {Gottesman}(2004)}]{aaronson2004}%
  \BibitemOpen
  \bibfield  {author} {\bibinfo {author} {\bibfnamefont {S.}~\bibnamefont
  {Aaronson}}\ and\ \bibinfo {author} {\bibfnamefont {D.}~\bibnamefont
  {Gottesman}},\ }\bibfield  {title} {\bibinfo {title} {Improved simulation of
  stabilizer circuits},\ }\href {https://doi.org/10.1103/PhysRevA.70.052328}
  {\bibfield  {journal} {\bibinfo  {journal} {Physical Review A}\ }\textbf
  {\bibinfo {volume} {70}},\ \bibinfo {pages} {052328} (\bibinfo {year}
  {2004})}\BibitemShut {NoStop}%
\bibitem [{Note1()}]{Note1}%
  \BibitemOpen
  \bibinfo {note} {We denote $\protect \hat {H}$ the Hadamard gate and
  $\protect \hat {S}$ the phase gate, which both belong to the Clifford group.
  For $X$-rotations we have that $\protect \hat {X} = \protect \hat
  {H}^{\protect \dag }\protect \hat {Z}\protect \hat {H}$ and hence
  $e^{-\protect \mathrm {i}\protect \frac {\theta _i}{2}\protect \hat {X}} =
  \protect \hat {H}^{\protect \dag }e^{-\protect \mathrm {i}\protect \frac
  {\theta _i}{2}\protect \hat {Z}}\protect \hat {H}$ and we can replace
  $\protect \hat {W}_i$ and $\protect \hat {W}_{i+1}$ respectively by $\protect
  \hat {H}\protect \hat {W}_i$ and $\protect \hat {W}_{i+1}\protect \hat {H}$
  to get another ansatz with the same form as the original one and with only
  $Y$ and $Z$ rotations. We proceed likewise for $Y$-rotations using the fact
  that $\protect \hat {Y} = (\protect \hat {S}\protect \hat {H})\protect \hat
  {Z}(\protect \hat {S}\protect \hat {H})^{\protect \dag }$. Note that in the
  case of the last layer one of the extra gates must be absorbed in the cost
  function observable to get the same ansatz structure.}\BibitemShut {Stop}%
\bibitem [{\citenamefont {Mitarai}\ \emph {et~al.}(2018)\citenamefont
  {Mitarai}, \citenamefont {Negoro}, \citenamefont {Kitagawa},\ and\
  \citenamefont {Fujii}}]{mitarai2018}%
  \BibitemOpen
  \bibfield  {author} {\bibinfo {author} {\bibfnamefont {K.}~\bibnamefont
  {Mitarai}}, \bibinfo {author} {\bibfnamefont {M.}~\bibnamefont {Negoro}},
  \bibinfo {author} {\bibfnamefont {M.}~\bibnamefont {Kitagawa}},\ and\
  \bibinfo {author} {\bibfnamefont {K.}~\bibnamefont {Fujii}},\ }\bibfield
  {title} {\bibinfo {title} {Quantum {{Circuit Learning}}},\ }\href
  {https://doi.org/10.1103/PhysRevA.98.032309} {\bibfield  {journal} {\bibinfo
  {journal} {Physical Review A}\ }\textbf {\bibinfo {volume} {98}},\ \bibinfo
  {pages} {032309} (\bibinfo {year} {2018})},\ \Eprint
  {https://arxiv.org/abs/1803.00745} {arXiv:1803.00745} \BibitemShut {NoStop}%
\bibitem [{\citenamefont {Schuld}\ \emph {et~al.}(2019)\citenamefont {Schuld},
  \citenamefont {Bergholm}, \citenamefont {Gogolin}, \citenamefont {Izaac},\
  and\ \citenamefont {Killoran}}]{schuld2019a}%
  \BibitemOpen
  \bibfield  {author} {\bibinfo {author} {\bibfnamefont {M.}~\bibnamefont
  {Schuld}}, \bibinfo {author} {\bibfnamefont {V.}~\bibnamefont {Bergholm}},
  \bibinfo {author} {\bibfnamefont {C.}~\bibnamefont {Gogolin}}, \bibinfo
  {author} {\bibfnamefont {J.}~\bibnamefont {Izaac}},\ and\ \bibinfo {author}
  {\bibfnamefont {N.}~\bibnamefont {Killoran}},\ }\bibfield  {title} {\bibinfo
  {title} {Evaluating analytic gradients on quantum hardware},\ }\href
  {https://doi.org/10.1103/PhysRevA.99.032331} {\bibfield  {journal} {\bibinfo
  {journal} {Physical Review A}\ }\textbf {\bibinfo {volume} {99}},\ \bibinfo
  {pages} {032331} (\bibinfo {year} {2019})}\BibitemShut {NoStop}%
\bibitem [{\citenamefont {Roberts}\ and\ \citenamefont
  {Yoshida}(2017)}]{roberts2017a}%
  \BibitemOpen
  \bibfield  {author} {\bibinfo {author} {\bibfnamefont {D.~A.}\ \bibnamefont
  {Roberts}}\ and\ \bibinfo {author} {\bibfnamefont {B.}~\bibnamefont
  {Yoshida}},\ }\bibfield  {title} {\bibinfo {title} {Chaos and complexity by
  design},\ }\href {https://doi.org/10.1007/JHEP04(2017)121} {\bibfield
  {journal} {\bibinfo  {journal} {Journal of High Energy Physics}\ }\textbf
  {\bibinfo {volume} {2017}},\ \bibinfo {pages} {121} (\bibinfo {year}
  {2017})}\BibitemShut {NoStop}%
\bibitem [{\citenamefont {Sim}\ \emph {et~al.}(2019)\citenamefont {Sim},
  \citenamefont {Johnson},\ and\ \citenamefont {{Aspuru-Guzik}}}]{sim2019a}%
  \BibitemOpen
  \bibfield  {author} {\bibinfo {author} {\bibfnamefont {S.}~\bibnamefont
  {Sim}}, \bibinfo {author} {\bibfnamefont {P.~D.}\ \bibnamefont {Johnson}},\
  and\ \bibinfo {author} {\bibfnamefont {A.}~\bibnamefont {{Aspuru-Guzik}}},\
  }\bibfield  {title} {\bibinfo {title} {Expressibility and {{Entangling
  Capability}} of {{Parameterized Quantum Circuits}} for {{Hybrid
  Quantum-Classical Algorithms}}},\ }\href
  {https://doi.org/10.1002/qute.201900070} {\bibfield  {journal} {\bibinfo
  {journal} {Advanced Quantum Technologies}\ }\textbf {\bibinfo {volume} {2}},\
  \bibinfo {pages} {1900070} (\bibinfo {year} {2019})}\BibitemShut {NoStop}%
\bibitem [{\citenamefont {Nakaji}\ and\ \citenamefont
  {Yamamoto}(2021)}]{nakaji2021b}%
  \BibitemOpen
  \bibfield  {author} {\bibinfo {author} {\bibfnamefont {K.}~\bibnamefont
  {Nakaji}}\ and\ \bibinfo {author} {\bibfnamefont {N.}~\bibnamefont
  {Yamamoto}},\ }\bibfield  {title} {\bibinfo {title} {Expressibility of the
  alternating layered ansatz for quantum computation},\ }\href
  {https://doi.org/10.22331/q-2021-04-19-434} {\bibfield  {journal} {\bibinfo
  {journal} {Quantum}\ }\textbf {\bibinfo {volume} {5}},\ \bibinfo {pages}
  {434} (\bibinfo {year} {2021})}\BibitemShut {NoStop}%
\bibitem [{\citenamefont {Gross}\ \emph {et~al.}(2007)\citenamefont {Gross},
  \citenamefont {Audenaert},\ and\ \citenamefont {Eisert}}]{gross2007}%
  \BibitemOpen
  \bibfield  {author} {\bibinfo {author} {\bibfnamefont {D.}~\bibnamefont
  {Gross}}, \bibinfo {author} {\bibfnamefont {K.}~\bibnamefont {Audenaert}},\
  and\ \bibinfo {author} {\bibfnamefont {J.}~\bibnamefont {Eisert}},\
  }\bibfield  {title} {\bibinfo {title} {Evenly distributed unitaries: {{On}}
  the structure of unitary designs},\ }\href
  {https://doi.org/10.1063/1.2716992} {\bibfield  {journal} {\bibinfo
  {journal} {Journal of Mathematical Physics}\ }\textbf {\bibinfo {volume}
  {48}},\ \bibinfo {pages} {052104} (\bibinfo {year} {2007})}\BibitemShut
  {NoStop}%
\bibitem [{\citenamefont {Iosue}\ \emph {et~al.}(2022)\citenamefont {Iosue},
  \citenamefont {Sharma}, \citenamefont {Gullans},\ and\ \citenamefont
  {Albert}}]{iosue2022a}%
  \BibitemOpen
  \bibfield  {author} {\bibinfo {author} {\bibfnamefont {J.~T.}\ \bibnamefont
  {Iosue}}, \bibinfo {author} {\bibfnamefont {K.}~\bibnamefont {Sharma}},
  \bibinfo {author} {\bibfnamefont {M.~J.}\ \bibnamefont {Gullans}},\ and\
  \bibinfo {author} {\bibfnamefont {V.~V.}\ \bibnamefont {Albert}},\
  }\href@noop {} {\bibinfo {title} {Continuous-variable quantum state designs:
  Theory and applications}} (\bibinfo {year} {2022}),\ \Eprint
  {https://arxiv.org/abs/2211.05127} {arXiv:2211.05127} \BibitemShut {NoStop}%
\bibitem [{\citenamefont {Harrow}\ and\ \citenamefont
  {Low}(2009)}]{harrow2009}%
  \BibitemOpen
  \bibfield  {author} {\bibinfo {author} {\bibfnamefont {A.~W.}\ \bibnamefont
  {Harrow}}\ and\ \bibinfo {author} {\bibfnamefont {R.~A.}\ \bibnamefont
  {Low}},\ }\bibfield  {title} {\bibinfo {title} {Random {{Quantum Circuits}}
  are {{Approximate}} 2-designs},\ }\href
  {https://doi.org/10.1007/s00220-009-0873-6} {\bibfield  {journal} {\bibinfo
  {journal} {Communications in Mathematical Physics}\ }\textbf {\bibinfo
  {volume} {291}},\ \bibinfo {pages} {257} (\bibinfo {year}
  {2009})}\BibitemShut {NoStop}%
\bibitem [{\citenamefont {Brand{\~a}o}\ \emph {et~al.}(2016)\citenamefont
  {Brand{\~a}o}, \citenamefont {Harrow},\ and\ \citenamefont
  {Horodecki}}]{brandao2016}%
  \BibitemOpen
  \bibfield  {author} {\bibinfo {author} {\bibfnamefont {F.~G. S.~L.}\
  \bibnamefont {Brand{\~a}o}}, \bibinfo {author} {\bibfnamefont {A.~W.}\
  \bibnamefont {Harrow}},\ and\ \bibinfo {author} {\bibfnamefont
  {M.}~\bibnamefont {Horodecki}},\ }\bibfield  {title} {\bibinfo {title} {Local
  {{Random Quantum Circuits}} are {{Approximate Polynomial-Designs}}},\ }\href
  {https://doi.org/10.1007/s00220-016-2706-8} {\bibfield  {journal} {\bibinfo
  {journal} {Communications in Mathematical Physics}\ }\textbf {\bibinfo
  {volume} {346}},\ \bibinfo {pages} {397} (\bibinfo {year}
  {2016})}\BibitemShut {NoStop}%
\bibitem [{\citenamefont {Haferkamp}(2022)}]{haferkamp2022}%
  \BibitemOpen
  \bibfield  {author} {\bibinfo {author} {\bibfnamefont {J.}~\bibnamefont
  {Haferkamp}},\ }\bibfield  {title} {\bibinfo {title} {Random quantum circuits
  are approximate unitary $t$-designs in depth ${{O}}\left(n
  t^{5+o(1)}\right)$},\ }\href {https://doi.org/10.22331/q-2022-09-08-795}
  {\bibfield  {journal} {\bibinfo  {journal} {Quantum}\ }\textbf {\bibinfo
  {volume} {6}},\ \bibinfo {pages} {795} (\bibinfo {year} {2022})}\BibitemShut
  {NoStop}%
\bibitem [{\citenamefont {McClean}\ \emph {et~al.}(2016)\citenamefont
  {McClean}, \citenamefont {Romero}, \citenamefont {Babbush},\ and\
  \citenamefont {{Aspuru-Guzik}}}]{mcclean2016}%
  \BibitemOpen
  \bibfield  {author} {\bibinfo {author} {\bibfnamefont {J.~R.}\ \bibnamefont
  {McClean}}, \bibinfo {author} {\bibfnamefont {J.}~\bibnamefont {Romero}},
  \bibinfo {author} {\bibfnamefont {R.}~\bibnamefont {Babbush}},\ and\ \bibinfo
  {author} {\bibfnamefont {A.}~\bibnamefont {{Aspuru-Guzik}}},\ }\bibfield
  {title} {\bibinfo {title} {The theory of variational hybrid quantum-classical
  algorithms},\ }\href {https://doi.org/10.1088/1367-2630/18/2/023023}
  {\bibfield  {journal} {\bibinfo  {journal} {New Journal of Physics}\ }\textbf
  {\bibinfo {volume} {18}},\ \bibinfo {pages} {023023} (\bibinfo {year}
  {2016})}\BibitemShut {NoStop}%
\bibitem [{\citenamefont {Goodfellow}\ \emph {et~al.}(2016)\citenamefont
  {Goodfellow}, \citenamefont {Bengio},\ and\ \citenamefont
  {Courville}}]{goodfellow2016}%
  \BibitemOpen
  \bibfield  {author} {\bibinfo {author} {\bibfnamefont {I.}~\bibnamefont
  {Goodfellow}}, \bibinfo {author} {\bibfnamefont {Y.}~\bibnamefont {Bengio}},\
  and\ \bibinfo {author} {\bibfnamefont {A.}~\bibnamefont {Courville}},\
  }\href@noop {} {\emph {\bibinfo {title} {Deep {{Learning}}}}},\ edited by\
  \bibinfo {editor} {\bibfnamefont {F.}~\bibnamefont {Bach}},\ Adaptive
  {{Computation}} and {{Machine Learning}} Series\ (\bibinfo  {publisher} {{MIT
  Press}},\ \bibinfo {address} {{Cambridge, MA, USA}},\ \bibinfo {year}
  {2016})\BibitemShut {NoStop}%
\bibitem [{\citenamefont {Liu}\ \emph {et~al.}(2022{\natexlab{b}})\citenamefont
  {Liu}, \citenamefont {Lin},\ and\ \citenamefont {Jiang}}]{liu2022d}%
  \BibitemOpen
  \bibfield  {author} {\bibinfo {author} {\bibfnamefont {J.}~\bibnamefont
  {Liu}}, \bibinfo {author} {\bibfnamefont {Z.}~\bibnamefont {Lin}},\ and\
  \bibinfo {author} {\bibfnamefont {L.}~\bibnamefont {Jiang}},\ }\href
  {https://doi.org/10.48550/arXiv.2206.09313} {\bibinfo {title} {Laziness,
  {{Barren Plateau}}, and {{Noise}} in {{Machine Learning}}}} (\bibinfo {year}
  {2022}{\natexlab{b}}),\ \Eprint {https://arxiv.org/abs/2206.09313}
  {arxiv:2206.09313} \BibitemShut {NoStop}%
\bibitem [{\citenamefont {Anschuetz}\ and\ \citenamefont
  {Kiani}(2022)}]{anschuetz2022a}%
  \BibitemOpen
  \bibfield  {author} {\bibinfo {author} {\bibfnamefont {E.~R.}\ \bibnamefont
  {Anschuetz}}\ and\ \bibinfo {author} {\bibfnamefont {B.~T.}\ \bibnamefont
  {Kiani}},\ }\bibfield  {title} {\bibinfo {title} {Quantum variational
  algorithms are swamped with traps},\ }\href
  {https://doi.org/10.1038/s41467-022-35364-5} {\bibfield  {journal} {\bibinfo
  {journal} {Nature Communications}\ }\textbf {\bibinfo {volume} {13}},\
  \bibinfo {pages} {7760} (\bibinfo {year} {2022})}\BibitemShut {NoStop}%
\bibitem [{\citenamefont {Bittel}\ and\ \citenamefont
  {Kliesch}(2021)}]{bittel2021}%
  \BibitemOpen
  \bibfield  {author} {\bibinfo {author} {\bibfnamefont {L.}~\bibnamefont
  {Bittel}}\ and\ \bibinfo {author} {\bibfnamefont {M.}~\bibnamefont
  {Kliesch}},\ }\bibfield  {title} {\bibinfo {title} {Training {{Variational
  Quantum Algorithms Is NP-Hard}}},\ }\href
  {https://doi.org/10.1103/PhysRevLett.127.120502} {\bibfield  {journal}
  {\bibinfo  {journal} {Physical Review Letters}\ }\textbf {\bibinfo {volume}
  {127}},\ \bibinfo {pages} {120502} (\bibinfo {year} {2021})}\BibitemShut
  {NoStop}%
\bibitem [{\citenamefont {Cerezo}\ and\ \citenamefont
  {Coles}(2021)}]{cerezo2021f}%
  \BibitemOpen
  \bibfield  {author} {\bibinfo {author} {\bibfnamefont {M.}~\bibnamefont
  {Cerezo}}\ and\ \bibinfo {author} {\bibfnamefont {P.~J.}\ \bibnamefont
  {Coles}},\ }\bibfield  {title} {\bibinfo {title} {Higher order derivatives of
  quantum neural networks with barren plateaus},\ }\href
  {https://doi.org/10.1088/2058-9565/abf51a} {\bibfield  {journal} {\bibinfo
  {journal} {Quantum Science and Technology}\ }\textbf {\bibinfo {volume}
  {6}},\ \bibinfo {pages} {035006} (\bibinfo {year} {2021})}\BibitemShut
  {NoStop}%
\bibitem [{Note2()}]{Note2}%
  \BibitemOpen
  \bibinfo {note} {This encompasses distributions that are symmetric about the
  angle $k\pi /2$ for $k\in \protect \mathbb {Z}$. In this case the bias can be
  factored out in the form of an extra fixed Clifford gate.}\BibitemShut
  {Stop}%
\bibitem [{\citenamefont {Zhao}\ and\ \citenamefont {Gao}(2021)}]{zhao2021}%
  \BibitemOpen
  \bibfield  {author} {\bibinfo {author} {\bibfnamefont {C.}~\bibnamefont
  {Zhao}}\ and\ \bibinfo {author} {\bibfnamefont {X.-S.}\ \bibnamefont {Gao}},\
  }\bibfield  {title} {\bibinfo {title} {Analyzing the barren plateau
  phenomenon in training quantum neural networks with the {{ZX-calculus}}},\
  }\href {https://doi.org/10.22331/q-2021-06-04-466} {\bibfield  {journal}
  {\bibinfo  {journal} {Quantum}\ }\textbf {\bibinfo {volume} {5}},\ \bibinfo
  {pages} {466} (\bibinfo {year} {2021})}\BibitemShut {NoStop}%
\bibitem [{\citenamefont {Piveteau}\ \emph {et~al.}(2022)\citenamefont
  {Piveteau}, \citenamefont {Sutter},\ and\ \citenamefont
  {Woerner}}]{piveteauQuasiprobabilityDecompositionsReduced2022}%
  \BibitemOpen
  \bibfield  {author} {\bibinfo {author} {\bibfnamefont {C.}~\bibnamefont
  {Piveteau}}, \bibinfo {author} {\bibfnamefont {D.}~\bibnamefont {Sutter}},\
  and\ \bibinfo {author} {\bibfnamefont {S.}~\bibnamefont {Woerner}},\
  }\bibfield  {title} {\bibinfo {title} {Quasiprobability decompositions with
  reduced sampling overhead},\ }\href
  {https://doi.org/10.1038/s41534-022-00517-3} {\bibfield  {journal} {\bibinfo
  {journal} {npj Quantum Information}\ }\textbf {\bibinfo {volume} {8}},\
  \bibinfo {pages} {1} (\bibinfo {year} {2022})}\BibitemShut {NoStop}%
\bibitem [{\citenamefont {McDiarmid}\ \emph {et~al.}(1989)\citenamefont
  {McDiarmid} \emph {et~al.}}]{mcdiarmid1989}%
  \BibitemOpen
  \bibfield  {author} {\bibinfo {author} {\bibfnamefont {C.}~\bibnamefont
  {McDiarmid}} \emph {et~al.},\ }\bibfield  {title} {\bibinfo {title} {On the
  method of bounded differences},\ }\href@noop {} {\bibfield  {journal}
  {\bibinfo  {journal} {Surveys in combinatorics}\ }\textbf {\bibinfo {volume}
  {141}},\ \bibinfo {pages} {148} (\bibinfo {year} {1989})}\BibitemShut
  {NoStop}%
\bibitem [{\citenamefont {Mohri}\ \emph {et~al.}(2018)\citenamefont {Mohri},
  \citenamefont {Rostamizadeh},\ and\ \citenamefont {Talwalkar}}]{mohri2018}%
  \BibitemOpen
  \bibfield  {author} {\bibinfo {author} {\bibfnamefont {M.}~\bibnamefont
  {Mohri}}, \bibinfo {author} {\bibfnamefont {A.}~\bibnamefont
  {Rostamizadeh}},\ and\ \bibinfo {author} {\bibfnamefont {A.}~\bibnamefont
  {Talwalkar}},\ }\href@noop {} {\emph {\bibinfo {title} {Foundations of
  {{Machine Learning}}, Second Edition}}}\ (\bibinfo  {publisher} {{MIT
  Press}},\ \bibinfo {year} {2018})\BibitemShut {NoStop}%
\bibitem [{\citenamefont {Baumgartner}(2011)}]{baumgartner2011}%
  \BibitemOpen
  \bibfield  {author} {\bibinfo {author} {\bibfnamefont {B.}~\bibnamefont
  {Baumgartner}},\ }\href {https://doi.org/10.48550/arXiv.1106.6189} {\bibinfo
  {title} {An inequality for the trace of matrix products, using absolute
  values}} (\bibinfo {year} {2011}),\ \Eprint {https://arxiv.org/abs/1106.6189}
  {arxiv:1106.6189} \BibitemShut {NoStop}%
\end{thebibliography}%
\end{document}